\documentclass[twocolumn]{aastex63}
\usepackage{color}

\usepackage{array, boldline, makecell, booktabs}

\usepackage{graphicx}
\usepackage{longtable}
\usepackage{amsmath}
\usepackage{amsfonts}
\usepackage{amssymb}
\usepackage{amsthm}
\usepackage{soul}
\usepackage{hyperref}
\usepackage{enumitem}
\usepackage[hyphens]{xurl}
\usepackage[T1]{fontenc}

\newcommand{\mkms}{\mathrm{km~s^{-1}}}
\usepackage{natbib}

\def \nHI {N_\text{H\textsc{i}}}
\def \nSiII {N_\text{Si\textsc{ii}}}
\def \nSiIV {N_\text{Si\textsc{iv}}}
\def \nCII {N_\text{C\textsc{ii}}}
\def \nCIV {N_\text{C\textsc{iv}}}
\def \nAlII {N_\text{Al\textsc{ii}}}
\def \nAlIII {N_\text{Al\textsc{iii}}}
\def \nOI {N_\text{O\textsc{i}}}

\begin{document}
\title{On the Metallicities and Kinematics of the Circumgalactic Media of Damped L$\lowercase{\text{y}}\alpha$ Systems at \MakeLowercase{\textit{z}}$\sim$2.5  {\footnote{The data presented herein were obtained at the W.~M.~Keck Observatory, which is operated as a scientific partnership among the California Institute of Technology, the University of California and the National Aeronautics and Space Administration. The Observatory was made possible by the generous financial support of the W.~M.~Keck Foundation.}}}

\author[0000-0001-8169-7249]{Stephanie M. Urbano Stawinski}
\affiliation{Department of Physics and Astronomy, University of California, Irvine, CA 92697, USA}

\author[0000-0001-6248-1864]{Kate H. R. Rubin}
\affiliation{Department of Astronomy, San Diego State University, San Diego, CA 92182 USA}
\affiliation{Center for Astrophysics and Space Sciences, University of California, San Diego, La Jolla, CA 92093, USA}

\author[0000-0002-7738-6875]{J. Xavier Prochaska}
\affiliation{Department of Astronomy \& Astrophysics, University of California, 1156 High Street, Santa Cruz, CA 95064, USA}
\affiliation{University of California Observatories, Lick Observatory, 1156 High Street, Santa Cruz, CA 95064, USA}
\affiliation{Kavli Institute for the Physics and Mathematics of the Universe (WIP), 5-1-5 Kashiwanoha, Kashiwa, 277-8583, Japan}

\author[0000-0002-7054-4332]{Joseph F. Hennawi}
\affiliation{Department of Physics, Broida Hall, University of California, Santa Barbara, CA 93106, USA}
\affiliation{Max-Planck-Institut fuer Astronomie, Koenigstuhl 17, D-69117 Heidelberg, Germany}

\author[0000-0002-1883-4252]{Nicolas Tejos}
\affiliation{Instituto de F\'isica, Pontificia Universidad Cat\'olica de Valpara\'iso, Casilla 4059, Valpara\'iso, Chile}

\author[0000-0001-6676-3842]{Michele Fumagalli}
\affiliation{Dipartimento di Fisica ``G. Occhialini'', Universit\`a degli Studi di Milano Bicocca, Piazza della Scienza 3, 20126 Milano, Italy}
\affiliation{INAF Osservatorio Astronomico di Trieste, via G. Tiepolo 11, Trieste, Italy}

\author[0000-0002-9946-4731]{Marc Rafelski}
\affiliation{Space Telescope Science Institute, 3700 San Martin Drive, Baltimore, MD 21218, USA}
\affiliation{Department of Physics \& Astronomy, Johns Hopkins University, Baltimore, MD 21218, USA}

\author[0000-0001-6196-5162]{Evan N. Kirby}
\affiliation{California Institute of Technology, 1200 East California Boulevard, MC 249-17, Pasadena, CA 91125, USA}
\affiliation{Department of Physics, University of Notre Dame, Notre Dame, IN 46556, USA}

\author[0000-0003-0083-1157]{Elisabeta Lusso}
\affiliation{Dipartimento di Fisica e Astronomia, Universit\`a di Firenze, via G. Sansone 1, 50019 Sesto Fiorentino, Firenze, Italy}
\affiliation{INAF-Osservatorio Astrofisico di Arcetri, Largo E. Fermi 5, I-50125, Firenze, Italy}
  
\author[0000-0001-7326-1736]{Zachary Hafen}
\affiliation{Department of Physics and Astronomy, University of California, Irvine, CA 92697, USA}

\email{ststawinski@gmail.com}

\begin{abstract}
We use medium- and high-resolution spectroscopy of close pairs of quasars to analyze the circumgalactic medium (CGM) surrounding 32 damped Ly$\alpha$ absorption systems (DLAs). The primary quasar sightline in each pair probes an intervening DLA in the redshift range $1.6<z_\text{abs}<3.5$, such that the secondary sightline probes absorption from Ly$\alpha$ and a large suite of metal-line transitions (including \ion{O}{1}, \ion{C}{2}, \ion{C}{4}, \ion{Si}{2}, and \ion{Si}{4}) in the DLA host galaxy's CGM at transverse distances $24~\text{kpc}\le R_\bot\le284\ \text{kpc}$. Analysis of Ly$\alpha$ in the CGM sightlines shows an anti-correlation between $R_\bot$ and \ion{H}{1} column density ($\nHI$) with 99.8$\%$ confidence, similar to that observed around luminous galaxies. The incidences of \ion{C}{2} and \ion{Si}{2} with $N>10^{13}~\rm cm^{-2}$ within 100 kpc of DLAs are larger by $2\sigma$ than those measured in the CGM of Lyman break galaxies (C$_f (\nCII)>0.89$ and C$_f(\nSiII)=0.75_{-0.17}^{+0.12}$). Metallicity constraints derived from ionic ratios for nine CGM systems with negligible ionization corrections and $\nHI>10^{18.5}~\rm cm^{-2}$ show a significant degree of scatter (with metallicities/limits across the range $-2.06\lesssim\log Z/Z_{\odot}\lesssim-0.75$), suggesting inhomogeneity in the metal distribution in these environments. Velocity widths of \ion{C}{4} $\lambda1548$ and low-ionization metal species in the DLA vs.\ CGM sightlines are strongly ($>2\sigma$) correlated, suggesting they trace the potential well of the host halo over $R_\bot\lesssim300$ kpc scales. At the same time, velocity centroids for \ion{C}{4} $\lambda1548$ differ in DLA vs.\ CGM sightlines by $>100~\rm\mkms$ for $\sim50\%$ of velocity components, but few components have velocities that would exceed the escape velocity assuming dark matter host halos of $\ge10^{12}M_\odot$. 
\end{abstract}

\section{Introduction} \label{sec:intro}

The circumgalactic medium (CGM) is defined as the gaseous halo surrounding galaxies that hosts the exchange of gas between  large-scale outflows from the host galaxy interstellar medium (ISM), the ambient halo, and  accretion from the intergalactic medium (IGM; \citealt{Tumlinson_2017_55}). In the last decade it has become evident that studying the CGM is crucial to fully understanding galaxy evolution \citep[e.g.,][]{Peeples2019}. Studies have shown $<25 \%$ of the mass expected in $\sim L^*$ galaxies at $z \sim 0$ is detected in observations of stars or the interstellar medium (ISM; \citealt{Peeples_2014_786}), potentially making the CGM an opportune place to look for these missing baryons. CGM studies at low redshift have detected the majority of these galactic missing baryons, with just the cool gas mass of the CGM around $\sim L^*$ galaxies estimated to be $\sim (3-6) \times 10^{10}~ M_\odot$  (\citealt{Prochaska_2011_740}, \citealt{Werk_2014_792}, \citealt{Prochaska_2017_837}). The physical origin of these diffuse baryons remains unclear; however, it is likely that they are deposited in part by winds launched by star formation or active galactic nuclei in the central galaxy \citep[e.g.,][]{Bordoloi2011,Kacprzak2012,LanMo2018}, as well as by accretion of the IGM or recycled wind material \citep[][]{Keres2009, Oppenheimer_2010_406,Rubin2012}.   Hence, the detailed properties of the CGM can provide essential insight into the processes driving the evolution of the galaxies. 

A critical epoch to study the CGM is at $z \sim 2-3$, during the peak of cosmic star formation (\citealt{Storrie-Lombardi&Wolfe_2000_543}, \citealt{Madau&Dickinson_2014_52}) and supermassive black hole growth (\citealt{Marconi_2006}, \citealt{Richards_2006}). However, the gaseous material that makes up the CGM is diffuse and difficult to detect in emission, especially at higher redshifts (although see, e.g., \citealt{Erb2018,FAB2018,Cai2019,OSullivan2020}). As a result, the vast majority of high-redshift CGM studies have analyzed its \ion{H}{1} and metal content in absorption detected along sightlines to bright background QSOs. For the most part, this work has focused on characterizing the gaseous environments of systems with host galaxies that are bright in the rest-frame UV (i.e., QSOs and Lyman Break Galaxies). These studies have established the masses and extent of the neutral hydrogen overdensities around these systems \citep[e.g.,][]{Rakic2012,Rudie_2012_750,Prochaska_2013_776_136P}, and have likewise assessed the sizes and masses of their metal components  \citep[e.g.,][]{Adelberger2005,Simcoe_2006_637,Turner_2014_445,Prochaska_2014_796_140P,Lau_2016_226,Rudie_2019_arXiv}. With the recent advent of VLT/MUSE, it is now also possible to select large samples of Ly$\alpha$-emitting systems close to background QSO sightlines, enabling similar analyses of the bulk properties of their halos \citep[e.g.,][]{Muzahid2021,Lofthouse2023}.
 
Studies benefiting from sensitive, high-resolution background QSO spectroscopy have gone beyond assessment of these bulk properties to analyze quantities that provide constraints on the origins and ultimate fate of the circumgalactic material. Detailed analyses of metal-line kinematics along sightlines probing QSO host halos have found that low-ionization transitions trace $\approx300~\mkms$ line widths, consistent with material tracing virial motions in these halos, but that robustly-measured asymmetries in these profiles are suggestive of gas tracing large-scale outflows \citep{Lau2018}.  Similarly, \citet{Rudie_2019_arXiv} found that within projected distances ($R_{\perp}$) $< 100$ kpc, the majority of Lyman Break Galaxies (LBGs) exhibit metal-enriched halo gas with velocities which exceed that required to escape the system. In contrast, \citet{Turner_2017_471} found that the absorption kinematics of \ion{H}{1}, \ion{C}{4} and \ion{Si}{4} at larger projected separations (up to $R_{\perp} = 2$ Mpc) from LBGs are best explained by large-scale inflow onto their host halos.

Assessment of the metallicity of QSO host halo gas has revealed high levels of overall enrichment ([M/H]$\gtrsim -0.6$) and significantly $\alpha$-enhanced abundance ratios \citep{Lau_2016_226,Fossati_2021_503}, suggestive that core-collapse supernovae play a dominant role in the enrichment of these environments. A handful of studies have analyzed the metallicities of material both within and well beyond the virial radii of LBGs, uncovering examples of systems as distant as $R_{\perp} = 30-110$ kpc that exhibit large scatter in their enrichment levels (e.g. $Z/Z_{\odot} \gtrsim 0.08-0.3$; \citealt{Simcoe_2006_637,Crighton_2013_776L,Fumagalli_2017_471}). Several studies \citep{Crighton_2013_776L, Crighton_2015_446, Fumagalli_2016_462, Lofthouse_2020_491} have also now provided clear evidence that at least some high-redshift star-forming galaxy halos are not well-mixed, and can give rise to both near-pristine material ($Z/Z_{\odot} \sim 1/100$) and mildly sub-solar gas along the same background QSO sightline (e.g., with velocity offsets of $\approx 200~\mkms$; \citealt{Crighton_2013_776L}).  These latter authors in particular used their measurements to argue for the presence of a cold-accretion stream amidst extended, metal-enriched wind ejecta.  

Due to their use of continuum or Ly$\alpha$ emission for the identification of foreground galaxy samples, the works discussed above have assessed the gaseous environments of halos hosting active galaxies with total halo masses $M_{h} \gtrsim 10^{11-11.5}~ M_{\odot}$ \citep{Adelberger2005,Gawiser2007,Conroy2008,Rakic2013,Wild2008,White2012,Font-Ribera2013,Bielby2016}.  An alternative approach is to instead identify high-redshift galaxies from their absorption-line signatures.  We pioneered this technique in \citet{Rubin_2015_808}, which used spectroscopy of close pairs of QSO sightlines to search for the damped Lyman-$\alpha$ (DLA) absorption profile associated with galaxies in the foreground. DLAs, defined as absorbers having neutral hydrogen column densities $N_\text{H\textsc{i}} \geq 2 \times 10^{20} \ \text{cm}^{-2}$ (\citealt{Wolfe_1986_61}), have been the dominant reservoir of HI gas since at least $z \sim 5$ (\citealt{Wolfe_2005_43}). The spatial relationship between DLAs and high-redshift star formation remained opaque for more than two decades after their discovery; however, studies of DLAs in cosmological simulations have long suggested that they are associated with galaxies spanning a wide range of halo masses ($10^{10} ~M_{\odot} \lesssim M_h \lesssim 10^{12} ~ M_{\odot}$; \citealt{Haehnelt1998,Nagamine2004,Pontzen2008,Razoumov2008,Fumagalli_2011_418,Cen_2012_748,Bird_2014_445,Garratt-Smithson2021}).  

Observational searches for the luminous counterparts of DLAs have recently become successful with the advent of near-infrared IFUs on 8-10m-class telescopes \citep{Peroux2012,Jorgenson&Wolfe2014}, and in programs targeting DLAs with high metallicities \citep{Fynbo2010,Fynbo2013,Krogager2012,Noterdaeme2012,Krogager2017}. A meta-analysis of these latter studies conducted by \citet{Krogager2017} concluded that 
DLAs with detected counterparts typically arise within $R_{\perp}<25$ kpc of galaxies having star formation rates (SFRs) $\approx 1-30~ M_{\odot}~\rm yr^{-1}$, and that lower-metallicity DLAs are likely associated with host galaxies having luminosities that extend $\gtrsim 2$ magnitudes fainter, and with SFRs as low as $\sim 0.01~ M_{\odot}~\rm yr^{-1}$.
More recent follow-up of high-metallicity DLA hosts at $z\sim4$ with ALMA
\citep{Neeleman_2017,Neeleman_2019_870L,Prochaska2019} has identified massive, high-SFR ($\sim 7-110 ~ M_{\odot} ~\rm yr^{-1}$) counterparts with impact parameters of $10-50$ kpc \citep{Neeleman_2019_870L}. Most recently, a Keck Cosmic Web Imager (KCWI) study has mapped two DLAs along multiple lines of sight toward a bright, gravitationally-lensed background galaxy at $z = 2.7$ \citep{Bordoloi_2022}. This work identified Ly$\alpha$ emission that likely arises from the DLA hosts within $R_{\perp}\lesssim 1-2$ kpc of the damped sightlines, and moreover measured the spatial extent of these DLAs to be $\gtrsim$ 238 kpc$^2$ (assuming circular geometry with $d = 17$ kpc), implying neutral gas masses of $\gtrsim  5.5 \times 10^8 - 1.4 \times 10^9 M_\odot$.

Taken together, these studies are suggestive of a scenario in which high-metallicity DLAs arise close to actively star-forming galaxies at high redshift (with SFRs $\gtrsim1~ M_{\odot}~\rm yr^{-1}$ and halo masses $10^{11}~M_{\odot} \lesssim M_h \lesssim 10^{12}~ M_{\odot}$), while lower-metallicity DLAs likely trace halos with lower star formation rates and masses. 
This scenario is further corroborated by analyses of the relation between DLA absorption line widths, metallicities, and host galaxy stellar masses \citep[e.g.,][]{Neeleman_2013_769,Moller_2013_430,Christensen_2014_445}. 
This picture is also broadly consistent with that advocated by analytical work modeling the global distribution function of $\nHI$ for DLAs in tandem with their metallicities \citep{Krogager2020}, line widths, and molecular gas content \citep{Theuns_2021_500}. This implies that DLAs are effective signposts for high-redshift galaxies having a broad range of masses and SFRs, and furthermore that the metallicities of DLAs may provide a rough indication of their relative host halo masses \citep[e.g.,][]{Wolfe_1998_494,Ledoux_2006_457,Neeleman_2013_769}.

In our previous work \citep{Rubin_2015_808}, we searched optical spectroscopy of close pairs of quasars \citep{Findlay_2018_236} for pairs in which at least one line of sight probed an intervening DLA. Our search yielded a sample of 40 pairs with foreground DLAs having redshifts in the range $1.6 < z_{\rm DLA} < 3.6$.  Our quasar spectroscopy was for the most part obtained at low spectral resolution ($\mathcal{R} \lesssim 2000$), but nevertheless permitted assessment of the covering fraction of optically thick \ion{H}{1} and the incidence of strong \ion{Si}{2} $\lambda 1526$ and \ion{C}{4} $\lambda \lambda 1548, 1550$ absorption in DLA environments to projected distances $R_{\perp} < 300$ kpc. Since this first work, we have obtained follow-up spectroscopy of a subset of this sample at medium and high spectral resolution ($\mathcal{R} \gtrsim 4000$), enabling detailed assessment of the column densities and kinematics of several ionic species including \ion{Al}{2}, \ion{Al}{3}, \ion{C}{2}, \ion{C}{4}, \ion{Fe}{2}, \ion{Mg}{2}, \ion{O}{1}, \ion{Si}{2}, and \ion{Si}{4}. We present these measurements here, together with a comparison between these CGM properties and those measured in the denser environments surrounding LBGs and QSO hosts \citep{Rudie_2019_arXiv,Lau_2016_226}. We assume here that our sample DLAs serve as signposts for nearby star formation, and are located either within the interstellar medium of their galaxy hosts or (more likely) in their ``inner'' CGM \citep[e.g.,][]{Theuns_2021_500, Stern2021}. Our dataset thus offers the unique opportunity to constrain metallicities of both the extended CGM {\it and} the ISM/inner CGM material traced by DLAs. Such comparisons are expected to yield important insight into the origins of circumgalactic gas; however, they have only been attempted in relatively low-redshift ($z < 1.5$) systems to date \citep[e.g.,][]{Peroux2016,Prochaska2017,Kacprzak2019,Weng2023}.

Our sample selection and data preparation are described in Section \ref{sec:data}. We then discuss our methods for measuring \ion{H}{1} column densities in the DLA and CGM sightlines in Section \ref{sec:NHI}, and discuss our methods for measuring metal-line column densities and kinematics in Section \ref{sec:analy}. Section \ref{sec:results} presents the resulting column densities, metallicities, and kinematics measured for our DLA and CGM sightlines. Finally, in Section \ref{sec:discuss} we combine our results with those in the literature, and present summaries of the relation between metallicity and $R_{\perp}$ around both DLAs and LBGs, as well as of the relation between metallicity and velocity width. We adopt a Planck $\Lambda$CDM cosmology with $\Omega_{\rm M} = 0.3$, $\Omega_\Lambda = 0.70$, and $H_0 = 68$ km s$^{-1} $ Mpc$^{-1}$ (\citealt{PlanckCollaboration_2016_594A}).

\section{Data and Sample Selection} \label{sec:data}

\subsection{QSO Pair Sample Selection}

Our sample is primarily selected from the Quasars Probing Quasars (QPQ) spectral database (as described in \citealt{Findlay_2018_236}). The QPQ database contains spectra for 5,627 objects with $z >$ 2 which were collected for the purpose of observing pairs of quasars that have close transverse separations on the sky. QPQ targets were initially drawn from low-resolution spectroscopic and photometric surveys that identified sources as quasars, including the SDSS Legacy Survey (2000-2008; \citealt{York_2000_120}; \citealt{Gunn_2006_131}), the Baryon Oscillation Spectroscopic Survey (\citealt{Dawson_2013_145}), and the 2dF QSO Redshift Survey (\citealt{Croom_2004_349}). These targets were supplemented with a photometrically-selected sample of QSO pair candidates, with photometry measured in SDSS, VST ATLAS  (\citealt{Shanks_2015_451}), and WISE imaging (\citealt{Wright_2010_140}). Photometrically-identified pairs were followed up with spectroscopy using $2-4$ meter-class telescopes as described in  \cite{Hennawi_2006_651,Hennawi_2010_719}. A subset of these confirmed candidates that were close on the sky (within $\lesssim 30\arcsec$) and that have $g\lesssim 21.5$ were then observed with medium- or high-resolution spectrographs, including ESI (\citealt{Sheinis_2002_114}) on the Keck II telescope, MagE (\citealt{Marshall_2008_7014E}) and MIKE (\citealt{Bernstein_2003_4841}) on the Magellan Telescopes, and XSHOOTER on the Very Large Telescope \citep{Vernet2011}. The majority of these high-fidelity spectra were obtained for the purpose of studying the CGM of the foreground QSOs \citep{Prochaska_2009_690,Lau_2016_226,Lau2018}.  A subset of this sample was targeted specifically for the present study due to the presence of a foreground DLA discovered in lower-resolution spectroscopy. A full listing of the telescopes and instruments used to obtain data analyzed in this paper, along with the corresponding spectral coverage and resolution of each instrumental setup, is presented in Table \ref{table: instr}.

\begin{deluxetable*}{lcccl}
	\tabletypesize{\footnotesize}
	\tablecolumns{5}
	\tablecaption{ List of instruments  \label{table: instr}}
	\tablehead{
		\colhead{Instrument }  & \colhead{Telescope} &  \colhead{Resolution ($\mathcal{R}$)} & \colhead{$\Delta v$ (km s$^{-1}$) \tablenotemark{a} } &  \colhead{Wavelength Coverage}}
	\startdata
	\vspace{-0.05cm}
	MIKE (Blue+Red) & Magellan Clay & 35714 & 8 & $3350-9500$ \AA \\    
	MIKE-Blue & Magellan Clay & 28000 & 11 & $3350-5000$ \AA \\ 
	XSHOOTER & VLT UT2 & 8000 & 37 & $3000-25000$ \AA \\
	MagE & Magellan Clay & 5857 & 51 & $3100-10000$ \AA \\
	& & 4824 & 62 & $3100-10000$ \AA \\ 
	ESI & Keck II & 4545 &  66 & $3900-10900$ \AA \\ 
	BOSS & Sloan 2.5m Telescope & 2100 & 143 & $3600-10400$ \AA \\ 
	GMOS-N & Gemini Telescope & 1872 & 160 & $3600-9400$ \AA \\
	\enddata
	\tablenotetext{a}{$\Delta v$ is the velocity width of the FWHM resolution element.}
\end{deluxetable*}

In an effort to increase our quasar pair sample, we also searched the IGMspec database. IGMspec is a large database that contains 434,686 spectra in the UV, optical, and near-infrared from 16 different surveys (\citealt{Prochaska_2017_19}). The database includes all the quasars from BOSS DR7 (\citealt{Abazajian_2009_182}) and DR12 (\citealt{Alam_2015_219}). Our search yielded eight pairs using the selection criteria described below; however, none of these sightlines were found to probe foreground DLAs. 

From the QPQ and IGMspec databases, we selected only quasar pairs with a maximum transverse proper distance on the sky of $R_\bot = 300$ kpc (calculated at the redshift of the foreground QSO). This is much larger than the typical virial radius of massive LBGs at $z \sim 2$ ($R_\text{virial}$ $\sim$ 90 kpc), and thus this distance criterion allows us to probe the CGM both within and beyond the virial radii of DLA host galaxies at $z \sim 2 $. We required that the quasars have redshifts 1.58 $< z <$ 4 so that their Ly$\alpha$ transition falls redward of the atmospheric cutoff at 3140~$\text{\AA}$, and so that there is wavelength coverage redward of the Ly$\alpha$ forest. This initial query yielded 411 QSO pairs.

We then required that at least one QSO in the pair have a medium- or high-resolution spectrum (with $\mathcal{R} \geq$ 4000) to enable the metal line analysis described later in Section \ref{sec:analy}. While the majority of these sightlines were targeted 
solely due to the presence of a foreground QSO (i.e., for reasons unrelated to the possible presence of a foreground DLA), a subset were targeted after the discovery of a DLA in low-resolution spectroscopy. This latter subsample may be biased 
toward probing low \ion{H}{1} column density DLAs, due to broadening of DLA absorption profiles in low-resolution spectroscopy. However, this effect is small, as this marginally biased sample comprises a small fraction of the medium-/high-resolution spectra used in this work. The resulting sample consists of 85 pairs. For each pair, we collected all spectra in each database, including low-resolution spectra if they extended the blue wavelength coverage, in order to maximize our wavelength search window for DLA signatures. For two pairs in this sample, we also made use of {\it HST} WFC3/UVIS grism spectroscopy obtained and reduced as described in \citet{Lusso_2018_860}. These data cover $2000~\text{\AA} < \lambda_\mathrm{obs} < 4500~\text{\AA}$ at a FWHM resolution of $\sim 60$ \AA, and therefore can provide useful coverage of the Lyman limit for absorbers discovered at $z<2.5$ along these sightlines. As described below in Section \ref{sec:NHI}, the {\it HST} data was used to improve our constraints on $N_{\text{H}\textsc{i}}$ for sightlines with medium or high-resolution optical spectroscopy.

\subsection{Continuum Fitting} 

We fit each quasar continuum using the function \texttt{fit\_continuum} in the Python package \texttt{linetools}\footnote{\url{https://linetools.readthedocs.io/en/latest/}} (\citealt{Prochaska_2016_linetools}), which allows the user to interactively modify a spline fit to the level of the continuum across the spectrum. The typical uncertainty in the continuum level using this method is $\lesssim 10\%$ in the Ly$\alpha$ forest and $\sim 5\%$ redward of the QSO's Ly$\alpha$ line \citep{Prochaska_2013_776_136P}.

\subsection{Identification of QSO Pairs with Foreground DLAs} \label{subsec:pairs}
We then performed a search for foreground DLAs among these pairs. Initially, we searched each spectrum in a given pair for strong absorption features blueward of the quasar Ly$\alpha$ emission line. We required that these features meet the following criteria:
\begin{enumerate}
\item They appear as a single line with apparent damping wings.
\item The DLA candidate must have a redshift more than $5000~\mkms$ blueward of the foreground QSO. If the system is redward of that limit, it may be associated with the QSO and may not probe the same environment as DLAs that are intervening. 
\item There is metal line absorption present at the same redshift as the putative Ly$\alpha$ in the same sightline. We searched for metal absorption lines within a $\pm 350~\mkms$  window from the transitions \ion{Si}{2} $\lambda$1304, \ion{Si}{4} $\lambda$1526, \ion{O}{1} $\lambda$1302, and \ion{C}{2} $\lambda$1334 (\citealt{Wolfe_2005_43}). We chose this velocity window to ensure we encompass any absorption that could be associated with the DLA.  Assuming DLAs are predominately hosted by halos with masses up to $10^{12}~M_{\odot}$, and that the FWHM of the line-of-sight velocity distribution of virialized halo gas is $v_{\rm FWHM} = 2\sqrt{\ln 2}~ v_{\rm vir}$, we expect $v_{\rm FWHM} \approx 360~\mkms$ at the mean redshift of our sample ($\langle z_\text{abs} \rangle = 2.45$; \citealt{Maller&Bullock_2004_355}). A search window of $\pm 350~\mkms$ therefore fully encompasses the velocity extent of this virialized gas. We note that all absorption features in our sample which satisfy the first two criteria also satisfied this third criterion.
\end{enumerate}

This initial search included absorption from DLAs as well as super Lyman limit systems (SLLS) with $10^{19}~\mathrm {cm^{-2}}\le N_{\text{H}\textsc{i}} \le 10^{20.3}~\mathrm{cm^{-2}}$. Once a DLA candidate was identified, we performed an initial fit of the absorption profile using the \texttt{XSpecGUI} in \texttt{linetools} to determine if the column density satisfies the DLA threshold  ($N_{\text{H}\textsc{i}} \geq 10^{20.3} \  \text{cm}^{-2}$). We assigned a redshift to each DLA that corresponds to the velocity of the peak optical depth of the metal lines. We prioritized lines which arise from low-ionization transitions (i.e., of \ion{Si}{2} or \ion{C}{2}) and which are not saturated. We then refined our measurement of the \ion{H}{1} column density using this redshift as described below in Section \ref{subsec:dlafit}. The resulting sample included 49 DLAs having $N_{\text{H}\textsc{i}} \ge 10^{20.3}$ cm$^{-2}$. 

Our final step was to examine the CGM sightlines associated with each of the confirmed DLAs. To ensure precise metal line analysis, we required the corresponding CGM sightlines to be observed at a resolution $\mathcal{R}>4000$. If a medium- or high-resolution spectrum for the CGM was not available, the pair was removed from our sample. The final sample used throughout this paper includes 32 DLA-CGM pairs. Table \ref{table: obs} lists coordinates and redshifts for the QSOs and DLAs in these pairs, as well as the instruments used for spectroscopy of each sightline.

For each DLA-CGM pair, we searched the CGM spectrum for strong Ly$\alpha$ absorption within $\pm350~\mkms$ of the DLA redshift. We selected the single, strongest Ly$\alpha$ component that is present within this velocity window. We note that there could be multiple components of Ly$\alpha$ absorption in the CGM sightline associated with the DLA, so by choosing a single feature we are setting a lower limit on \ion{H}{1}. Some CGM sightlines have multiple Ly$\alpha$ absorption components of similar strength in this range. In these cases, we included all \ion{H}{1} absorption  within $\pm 350~\mkms$ of the DLA redshift in our column density measurement.

The redshift of the CGM Ly$\alpha$ absorber was found using the same method described above: we adopted the redshift corresponding to the velocity of the peak optical depth of the metal lines in the CGM sightline. In cases where there are no securely-detected CGM metal lines, we estimated the redshift using the Ly$\alpha$ absorption line. For two CGM sightlines, there was no spectral coverage of \ion{H}{1} absorption near the redshift of the DLA. For these systems, we used the DLA redshift as the initial guess to search for associated metal lines. 

With our DLA-CGM pairs selected, we then measured column densities of \ion{H}{1} and column densities and kinematics of several metal ions. Section \ref{sec:NHI} describes in more detail the methodology we use to measure the column densities for \ion{H}{1} in both the DLA and CGM sightlines. Section \ref{sec:metallines} describes the methods we use to assess metal line column densities and kinematics. 
 These measurements are listed in Table \ref{table: metalcolm} and Table \ref{table: kinmeasurements}, respectively.

\section{Determining Column Densities of \ion{H}{1}}\label{sec:NHI}
In this section we describe several complementary approaches we used to measuring the column density of neutral hydrogen present in each DLA and CGM system. Table \ref{table: Zmeasurements} lists our CGM $N_\text{H\textsc{i}}$ measurements and specifies which of the following constraints we used to make this assessment for each sightline pair.

\subsection{Damped Lyman-$\alpha$ Profile Fitting} \label{subsec:dlafit}
\begin{figure*}[ht]
	\centering
	\begin{minipage}{\textwidth}
		\includegraphics[width=\linewidth]{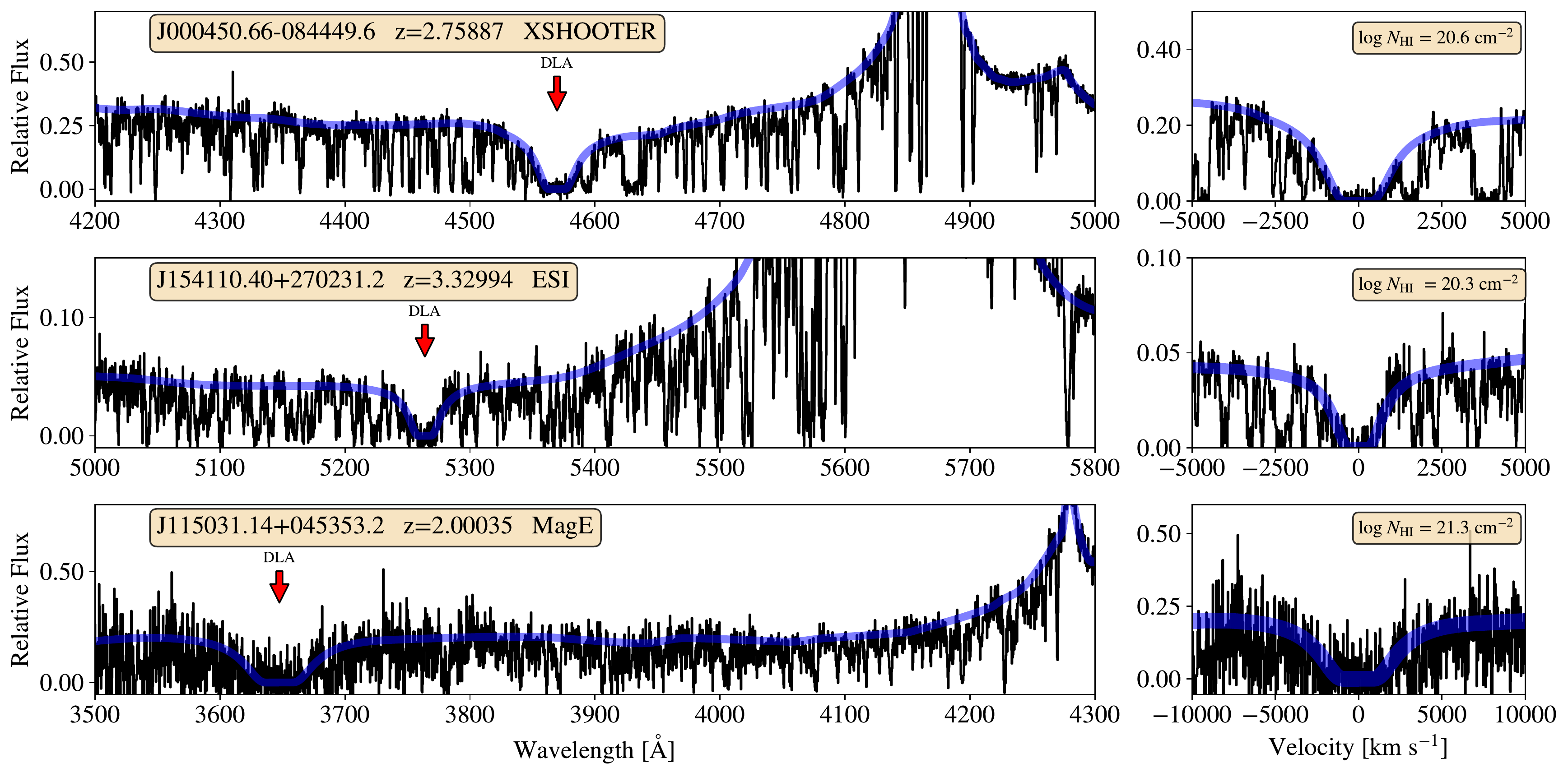}
		\caption{Examples of QSO spectroscopy, the DLA profiles, and continuum fits (in blue) for three systems. The QSO identifier, DLA redshift, and the instrument are given in the top left corner of each panel. The DLA fit and adopted $N_\text{H\textsc{i}}$ value are shown the subsequent right-hand panels.  Velocities are measured relative to the DLA redshift.
\label{fig: dlafit}}
	\end{minipage}
\end{figure*}

In all of our DLA sightlines and about a third of our CGM sightlines, we were able to estimate $N_\text{H\textsc{i}}$ by fitting the characteristic wings of the damped line profile at $\lambda_\text{rest}  =$ 1215.67 \AA \ as in \citet{Wolfe_1986_61}.  To fit the \ion{H}{1} absorption profile, we used the interactive GUI \texttt{XFitDLA} in the Python package \texttt{pyigm} \citep{Prochaska_2017_pyigm}. With the redshift obtained as described in Section \ref{subsec:pairs}, we manually adjusted the $N_\text{H\textsc{i}}$ and broadening parameters while continuously modifying the continuum level around the absorption line to achieve a close match to the data (as assessed by eye). Examples of three best-fit DLA profiles are shown in Figure \ref{fig: dlafit}. 
We adopted a $\pm$0.2 dex error for our $N_\text{H\textsc{i}}$ fits, based on similar analysis from \cite{Prochaska_2003_148}.


There is one case where the CGM \ion{H}{1} absorber has $N_\text{H\textsc{i}}$ $=$ 10$^{20.2}$ cm$^{-2} \pm 0.2$, near the limit of a DLA itself.
The sightline with the highest column density is labeled as the DLA sightline, and the other is treated as the CGM sightline. This choice has implications for the interpretation of our results for this pair, so we highlight it in our discussion below (and refer to it as our double-DLA system). Because there is only one such double-DLA in our sample, this does not significantly affect any general conclusions made later in this analysis. We indicate the specific CGM sightlines for which we constrain $N_\text{H\textsc{i}}$ by fitting damping wings with the number 1 in the Method column in Table \ref{table: Zmeasurements}.

	
\subsection{Lyman Limit Fitting}

For some cases in which there is undamped ($N_\text{H\textsc{i}}$ $\leq$ 10$^{18}$ cm$^{-2}$) but strong Ly$\alpha$ absorption, we have access to either {\it HST} WFC3/UVIS or optical spectroscopic coverage of the flux blueward of the Lyman limit ($\lambda_{\rm rest}= 912$\,\AA). 
For systems with optical spectral coverage of the Lyman limit, we used the interactive GUI \texttt{XFitLLS} from the \texttt{pyigm} package to fit the Lyman limits of these systems as described in \citet{OMeara_2013_765}. The program generates a continuum model of the QSO from \citet{Telfer_2002_565}, and allows the user to 
adjust the normalization and power-law tilt of the template to match the QSO continuum redward of the Ly$\alpha$ forest. Any sharp drops in the flux below the QSO's 912 \AA\ break may then be modeled as LLSs with optical depth $\tau_{912}^{\rm LL} \approx N_\text{H\textsc{i}} / 10^{17.19}~\rm cm^{-2}$. None of the systems in this work have a strong, clean Lyman limit feature that allows for a direct measurement of $N_\text{H\textsc{i}}$. This is due either to strong intervening systems absorbing the continuum close to the Lyman limit at $z_\text{abs}$, or to the amount of \ion{H}{1} in the target absorption system being sufficiently low that it does not produce a detectable Lyman limit break.
Therefore, this method allowed for an estimate of the minimum amount of \ion{H}{1} that is required to account for the decrease in flux blueward of the DLA's Lyman limit. Moreover, because there is strong but undamped Ly$\alpha$ absorption associated with these systems, we also placed an upper bound on their \ion{H}{1} columns of $N_\text{H\textsc{i}} < 10^{18} \ \text{cm}^{-2}$.

Seven CGM sightlines in this work were targeted in the {\it HST} WFC3/UVIS grism survey of paired quasars described in \citet{Lusso_2018_860}\footnote{The {\it HST} WFC3/UVIS data presented in this paper can be found in Mikulski Archive for Space Telescopes (MAST) at the Space Telescope Science Institute. The specific observations analyzed can be accessed via \dataset[10.17909/n7qq-vc30]{https://doi.org/10.17909/n7qq-vc30}.}; however, four of our absorbers were not detected in the {\it HST} spectroscopy. This was due to either weak \ion{H}{1} absorption that limits the detection of a flux decrement at $\lambda_{\rm rest} = 912$\,\AA, or to strong background absorbers that significantly reduce the flux and signal-to-noise ratio (S/N) near the Lyman limit. 
One CGM system, in sightline J105644.88-005933.4, has Ly$\alpha$ damping wings observed in our optical spectroscopy, so we do not make use of the {\it HST} coverage to improve our constraints on $N_\text{H\textsc{i}}$. The Lyman limit coverage of the grism spectroscopy of the remaining two systems (in sightlines J161302.03+080814.3 at $z_\text{abs} = $ 1.617 and J123635.42+522057.3 at $z_\text{abs} = $ 2.39691) were modeled in the  same manner as described above, using the \texttt{XFitLLS} GUI. Uncertainties in the value of $N_\text{H\textsc{i}}$ were determined by perturbing the best-fit value in increments of $0.1$ dex and assessing the degree to which each perturbed value was consistent with the data by visual inspection. Using this method, we estimated the error on each $N_\text{H\textsc{i}}$ measured from these grism spectra to be $\pm 0.2-0.3$ dex. The CGM sightlines for which we found $N_\text{H\textsc{i}}$ from Lyman Limit fitting of either {\it HST} WFC3/UVIS or optical spectroscopy are indicated with the number 2 or 3 in the Method column in Table \ref{table: Zmeasurements}, respectively.

\subsection{Limits on $N_\text{H\textsc{i}}$}

CGM sightlines for which $N_\text{H\textsc{i}}$ could not be constrained using the methods described above (but must have $N_\text{H\textsc{i}} < 10^{18}~\rm cm^{-2}$) were treated in one of two ways described below. 
If there is a single, strong absorption line with some associated metal absorption at the same redshift, we estimated a lower limit on the \ion{H}{1} column density using the apparent optical depth method \citep{Savage&Sembach_1991_379}. We adopted this limit, along with the upper limit $N_\text{H\textsc{i}}$ $<$ 10$^{18}$ cm$^{-2}$, as conservative bounds on the \ion{H}{1} column density assuming the Ly$\alpha$ transitions are within the flat region of the curve of growth. We indicate these CGM sightlines with the number 4 in the Method column in Table \ref{table: Zmeasurements}.
For CGM systems that have many weak absorption features near Ly$\alpha$ with no associated metal lines, we assumed the Ly$\alpha$ is optically thin and calculated the column density of each absorption feature within $\pm$ 350~$\mkms$ of the DLA redshift using the same apparent optical depth method.  We then summed the resulting column densities for these features to use as our final estimate of $N_\text{H\textsc{i}}$. These sightlines are designated with the number 5 in the Method column in Table \ref{table: Zmeasurements}.
Lastly, two CGM sightlines had no spectral coverage of Ly$\alpha$, and therefore are not used in any \ion{H}{1} analysis.

\section{Metal Line Profile Analysis} \label{sec:analy}

\begin{figure*}[ht]
	\centering
	\begin{minipage}{7in}
		\includegraphics[width=0.321\textwidth]{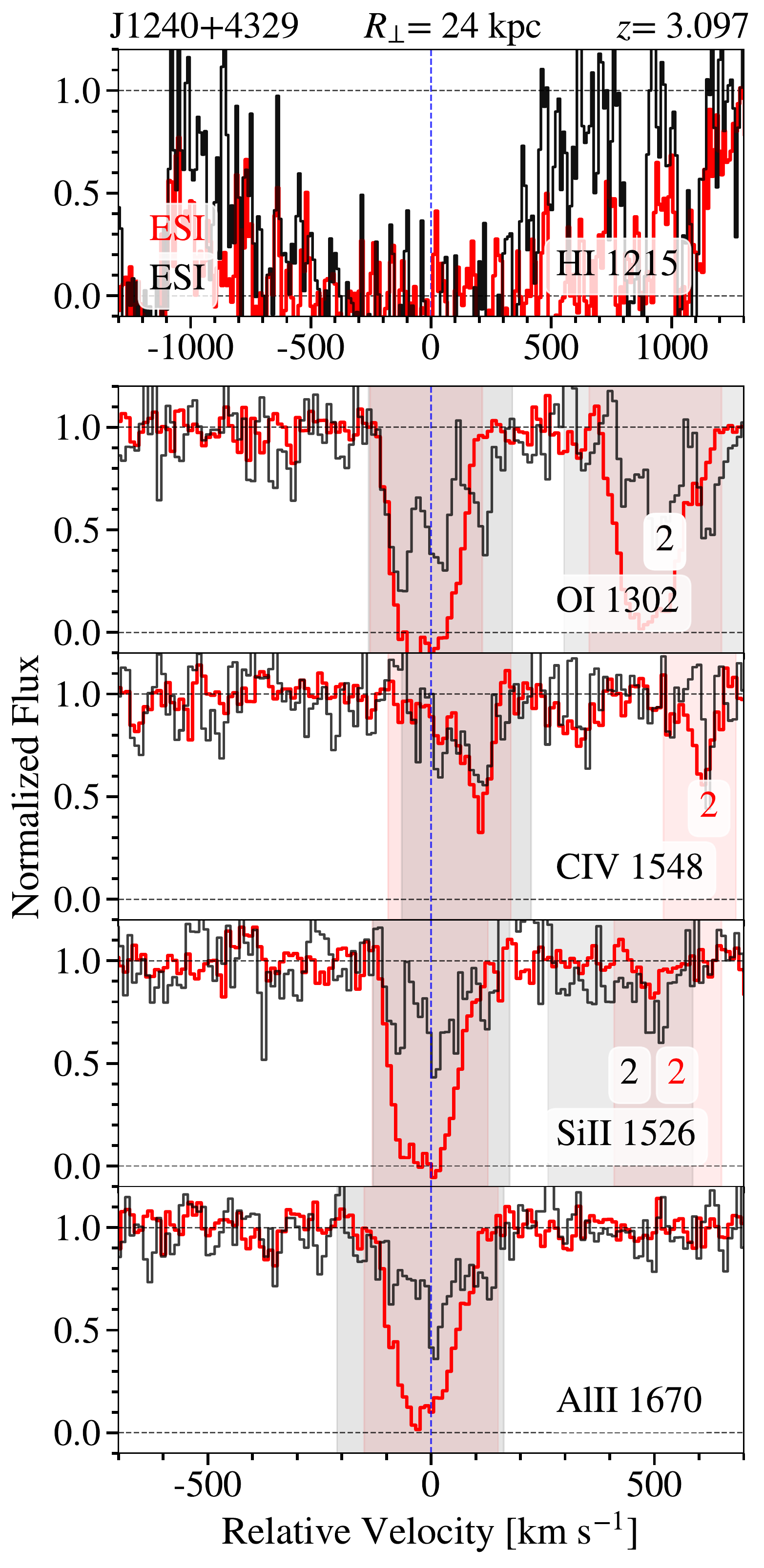}
		\includegraphics[width=0.3175\textwidth]{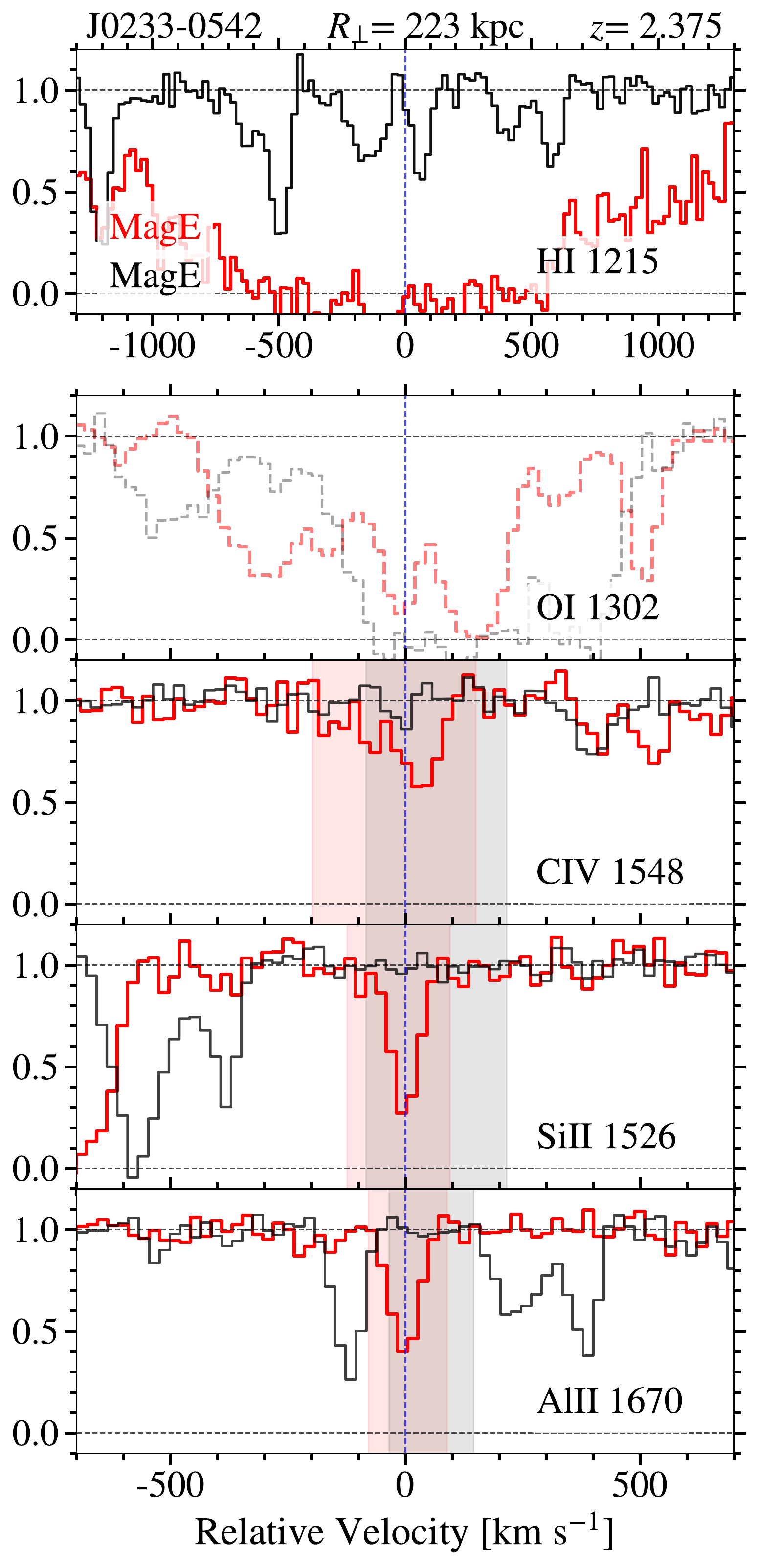}
		\includegraphics[width=0.316\textwidth]{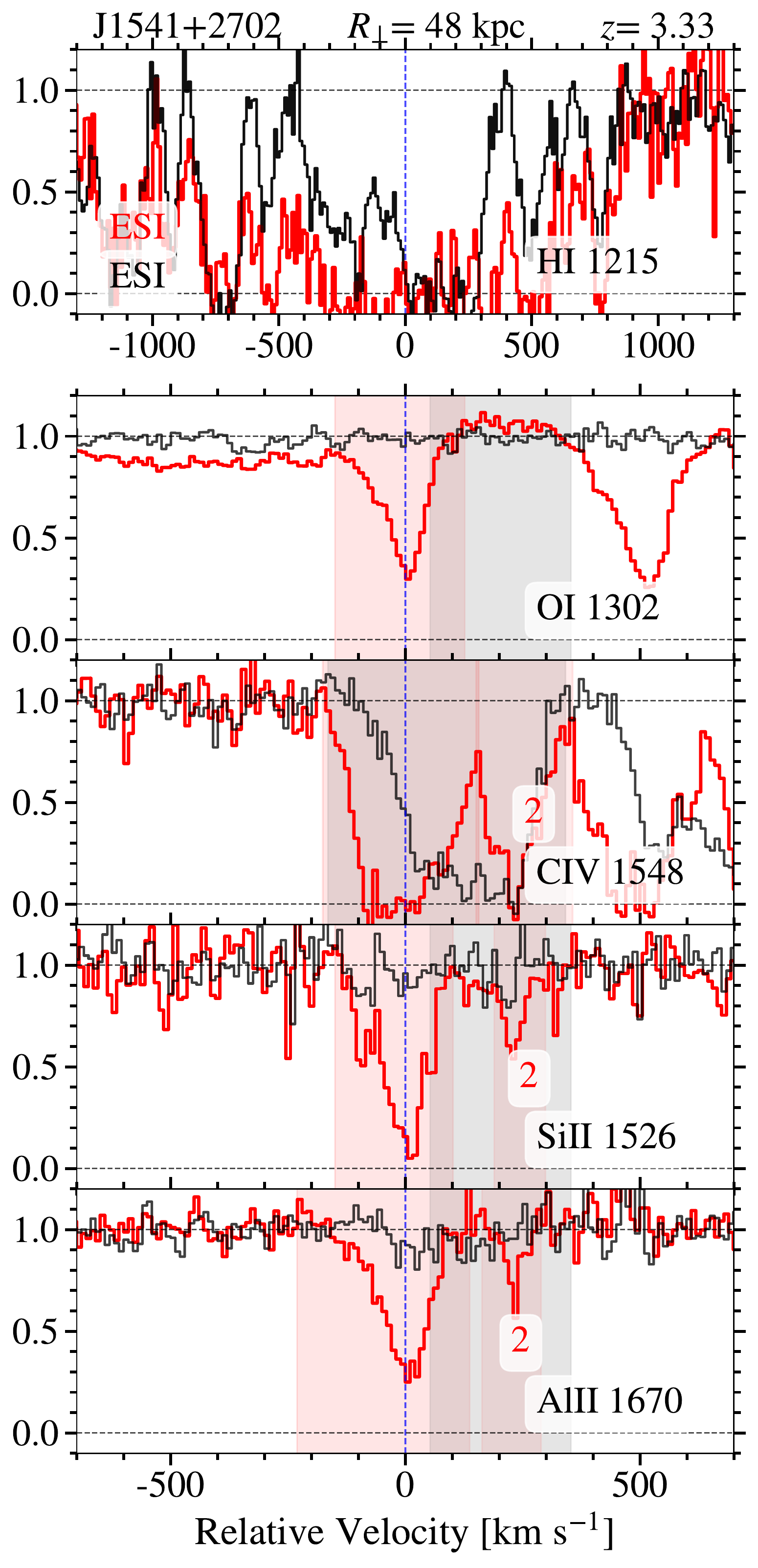}
		\caption{\ion{H}{1} and metal-line absorption profiles for three DLA-CGM sightline pairs.  \ion{H}{1}, \ion{O}{1}, \ion{C}{4}, \ion{Si}{2}, and \ion{Al}{2}  transitions are shown (as indicated at the bottom right of each panel).  The black histograms show CGM sightlines and the red histograms show the corresponding DLA sightlines. The QSO pair ID is shown at the top of each column, along with the projected distance between the sightlines at the redshift of the host DLA. The blue dotted line indicates the systemic velocity of the DLA. The shaded areas indicate the placement of the velocity windows used to measure metal-line absorption strength. In the case of profiles with multiple components, the shaded regions are marked `2' or `3' to indicate a second or third component. The instrument used for each sightline is labeled in the bottom left corner of each \ion{H}{1} panel. Transitions which are not used in this analysis due to extreme blending are shown with dotted histograms.  }
		\label{fig:linecomp}
	\end{minipage}
\end{figure*}

Each metal line in our DLA and CGM sightlines was visually inspected using the interactive GUI \texttt{XAbsSystemGUI} in the package \texttt{linetools}, which displays multiple transitions for a simultaneous comparison. For all strong transitions in each spectrum, we manually set velocity limits over which we measure the associated absorption line by searching within $\pm 1000~\mkms$ of the absorber redshift. This search window was adopted based on the findings of \citet{Rudie2019}, who identified metal-line absorption associated with LBG hosts at relative velocities of up to $\pm 1000~\mkms$. We assigned velocity limits for each velocity component in every sightline. In cases in which no absorption is clearly evident, we adopted velocity limits of $\pm 300\ \mkms$ by default, and adjusted the edges of this window to avoid absorption from unassociated systems. In many cases the blending between components associated with our target system is severe, such that they cannot be separated into two distinct absorption lines. In such cases, we separated components only if the flux rises to $>$ 50$\%$ of the continuum level between the lines. If a given transition is severely blended with an unassociated absorber such that it could not be separated, we excluded the line from our analysis. Representative examples of our chosen absorption line windows, including systems with multiple components, for three quasar pairs are shown in Figure \ref{fig:linecomp}. 

\subsection{Column Densities of Metal Lines}\label{sec:metallines}

Once velocity windows were selected, we used the apparent optical depth method as detailed in \cite{Savage&Sembach_1991_379} to measure column densities. The optical depth per unit velocity ($v$) is defined as: 
\begin{equation}\label{eq:aodm}
\tau(v) = \ln{I(v)/I_c} = \ln \frac{1}{F(v)},
\end{equation}
where $I_c$ is the intensity of the continuum within the set velocity window, $I(v)$ is the absorbed intensity within that window, and $F(v)$ is the continuum-normalized absorbed flux. \cite{Savage&Sembach_1991_379} used the optical depth to find the total column density, $N$, as:
\begin{equation}
    N = \frac{m_e c}{\pi e^2 f \lambda_0} \sum_i {\ln{F(v)_i^{-1}} \ \Delta v},
\end{equation}
where $m_e$ is the mass of an electron, $c$ is the speed of light, $e$ is the elementary charge, $f$ is the oscillator strength of the transition, $\lambda_0$ is the rest wavelength of the transition, and $\Delta v$ is the step in velocity space for each pixel ($i$) within the velocity window. The error ($\sigma_N$) is thus defined as: 
\begin{equation}
    \sigma_N^2 =  \sum_i {\left (\frac{m_e c}{\pi e^2 f \lambda_0} \frac{\sigma_{F(v)_i}}{F(v)_i} \ \Delta v \right )^{2}},
\end{equation}
where $\sigma_{F(v)}$ is the error in the spectral flux.

For our high-resolution spectroscopy (obtained with MIKE), if the absorbed line profile drops below 20$\%$ of the flux uncertainty, or if the relevant spectral pixels reach a normalized flux level of $<$ 0.05, the line was flagged as saturated and our column density estimate was treated as a lower limit. Line saturation is a larger concern for medium-resolution spectroscopy (e.g., from ESI or MagE), and for these sightlines we conservatively defined a line as saturated if the absorbed line profile drops below 50$\%$ of the continuum flux (see \citealt{Prochaska_2003_147}). Three-$\sigma_N$ upper limits are used for non-detections (defined as $N <$ 3$\sigma_N$).

We also investigated the systematic error associated with this measurement due to uncertainty in the placement of our velocity windows. We measured the column density of all single-component systems after broadening the velocity limits by $50 ~\mkms$ on both sides of the line profile. In the case of \ion{Si}{2}, we find this increases our measured column densities by an average of 0.07 dex with a scatter of 0.05 dex. Thus, the choice of a broader velocity window would systematically increase our column density measurements.  However, given that the FWHM velocity resolution of our dataset is $\lesssim 66~\mkms$, our uncertainty in the velocity limit of our absorption profiles is not likely to exceed $\pm 50~\mkms$.  The implied systematic error in our column densities is therefore $\lesssim 0.1$ dex.

We combined multiple column density constraints for each species as follows:
(1) if there is one transition that has yielded a direct measurement of the column density, that measurement is adopted; (2) if there is more than one detection, we adopted the mean $N$ value; (3) if there are no direct measurements and one or more transitions are saturated, the adopted column density is the highest value flagged as a lower limit; and (4)  if \textit{all} the transitions have yielded upper limits on the column density, we adopted the lowest upper limit. 

Finally, we summed the column densities measured from each separate velocity component associated with a given ion. While we include separate components in this final summation, the velocity components which are kinematically consistent with the primary \ion{H}{1} absorber have the largest columns along the line of sight and therefore dominate these measurements.    

\subsection{Metal Line Kinematics}\label{sec:methodskin}

We also assessed the kinematic properties of the metal lines in the DLA and CGM sightlines. Our spectral coverage includes singly-, doubly-, and triply-ionized transitions. We focused on the kinematics of singly- and triply-ionized transitions only. We made two kinematic measurements: the flux-weighted velocity centroid ($\delta v_\text{weight}$) and the $\Delta v_{90}$ velocity width. 
To estimate the former, we first calculated the flux-weighted wavelength centroid, defined as follows:
\begin{equation}\label{eq:vweight}
\lambda_\text{weight} = \frac{\sum_{i} (1-F_i(\lambda_i))\lambda_i }{ \sum_{i} (1-F_i(\lambda_i))}
\end{equation}
where $F_i(\lambda_i)$ is the continuum-normalized flux and $\lambda_i$ is the wavelength at each pixel $i$ within the velocity window for the line. The final $\delta v_\text{weight}$ was then calculated using this wavelength relative to the redshift of the associated DLA. 

We rely upon $\delta v_\text{weight}$ as opposed to the velocity at the peak optical depth ($\delta v_\text{peak}$) for several reasons. First, using $\delta v_\text{peak}$ would bias the low-ion kinematics towards $\sim 0~\mkms$, as they were used to estimate the redshift of the individual absorption systems (see Section \ref{subsec:pairs}). Secondly, many lines are not symmetric about the peak optical depth, such that $\delta v_\text{peak}$ probes the velocity of the strongest absorption rather than the average velocity of the absorbing gas. Furthermore, for saturated lines, the velocity at which the optical depth peaks is ambiguous. For most sightlines, the difference in the \ion{Si}{2} $\lambda 1526$ $\delta v_\text{peak}$ vs.\ $\delta v_\text{weight}$ is $\lesssim 50~\mkms$, and we find that sightlines for which there is a greater than $50~\mkms$ difference in these measures have large widths and strongly asymmetric profiles.

We conducted our $\delta v_\text{weight}$ measurements for an unsaturated high-S/N low-ion transition, as well as for \ion{C}{4} $\lambda$1548. \ion{C}{4} $\lambda$1548 was selected as representative of the velocity profile of high-ionization material due to its high oscillator strength. We do not report high-ion kinematics for sightlines in which either transition in the \ion{C}{4} doublet is not securely detected or is heavily blended. For sightlines with multiple velocity components, we measured each component's $\lambda_\text{weight}$ separately using Equation \ref{eq:vweight}, and computed the corresponding $\delta v_\text{weight}$ relative to the redshift of the corresponding DLA. We were able to assess the $\delta v_\text{weight}$ of low-ionization material (and \ion{C}{4}-absorbing material) in 31 (30) of our DLA sightlines, 7 (6) of which have resolved secondary velocity components. The CGM sightlines have fewer securely-detected metal lines, reducing the number of sightlines we could use for kinematic analysis. We measured the $\delta v_\text{weight}$ of low-ionization material in 8 sightlines, 3 of which have secondary velocity components. We measured the $\delta v_\text{weight}$ of \ion{C}{4} in 20 CGM sightlines, 6 of which have secondary components.

Our second kinematic measurement is the $\Delta v_{90}$ velocity width, which was introduced in \cite{Prochaska_1997_487} as a tracer for kinematics of the  neutral gas content of DLAs. In that work, the authors analyzed the full absorption profiles of unsaturated, low-ionization transitions to assess the bulk neutral gas velocity dispersion, and to ensure that the velocity width is not overestimated due to weak, outlying velocity components.  We take the same approach for each of our DLA and CGM sightlines.  In addition, we assess $\Delta v_{90}$ on a component-by-component basis for both the low-ionization material in each system, and for each \ion{C}{4} $\lambda 1548$ profile (chosen for its high oscillator strength).  
We make use of these latter (component-by-component) measurements when comparing the kinematics of our sightline pairs in Section~\ref{sec:kinematics}, and report these values in Table~\ref{table: kinmeasurements}.  We make use of the former $\Delta v_{90}$ values (measured without component separation) when comparing our sample to global relations in the literature in Section \ref{Disc:DLAZ} and Figure \ref{fig:Zwidth_relation}. 

\cite{Prochaska_2008_672} investigated the artificial broadening associated with $\Delta v_{90}$ measured from medium-resolution spectra. 
In that work, they reduced their measured ESI $\Delta v_{90}$ widths by $20~\mkms$ and adopted an uncertainty of $20~\mkms$. 
We assume that the artificial broadening of $\Delta v_{90}$ in our medium-resolution spectra is proportional to what is measured in \cite{Prochaska_2008_672}; e.g., a FWHM resolution of $45~\mkms$ would broaden $\Delta v_{90}$ by $\sim20~\mkms$, a factor of $0.44$ times the FWHM resolution. Using this factor ($0.44 \times$FWHM resolution), we estimated the artificial broadening of $\Delta v_{90}$ in all of the spectra used herein. The measured $\Delta v_{90}$ widths were then reduced by that estimate to produce the final, reported $\Delta v_{90}$ widths used in the following analysis.

\subsubsection{Uncertainties in Kinematic Measurements}\label{subsubsec:uncertainties_kinematics}

The precision of our kinematic measurements depends on the FWHM resolution and S/N of our spectra. In order to assess the level of uncertainty in our measurements of $\delta v_{\rm weight}$ and $\Delta v_{90}$, we performed a Monte Carlo analysis on mock \ion{C}{4} lines. To be conservative, we use the lowest spectral resolution and S/N among all of our observed sightlines for this analysis, and adopt the resulting uncertainties across our sample. 

We first created a mock spectrum with a velocity resolution consistent with that of our ESI data (FWHM $\sim 66 ~\rm \mkms$).
We then added a single, fake \ion{C}{4} line with a column density equal to the minimum column density detection ($\log N_\text{CIV} = 13.1 ~\rm cm^{-2}$) in our ESI dataset. We adopted the mean Doppler width measured for \ion{C}{4} by \citet[][12.4 km s$^{-1}$]{Rudie2019}. Because our absorption features likely include unresolved velocity components, we also created mock spectra with 2-3 of these absorbers at a maximum velocity separation of $\sim 100 ~\mkms$. 
Finally, we added Gaussian random noise to the mock spectra. 
We generated 100 realizations of each mock spectrum with a S/N equal to the lowest S/N measured in our observed spectra (S/N $\sim 20$ pixel$^{-1}$).
The standard deviation of the $\delta v_\text{weight}$ measurements for our one-component, two-component, and three-component profiles are $8~\mkms$, $10~\mkms$, and $15~\mkms$, respectively. The corresponding values of the dispersion in our $\Delta v_{90}$ measurements are $10~\mkms$, $20~\mkms$, and $35~\mkms$. We adopted the largest of these values as our $1\sigma$ measurement uncertainty for $\delta v_\text{weight}$ and  $\Delta v_{90}$ for all sightlines, regardless of their S/N or spectral resolution. 

\section{Results} \label{sec:results}

\subsection{Our DLA Sample as a Representative DLA Population}

To better understand whether our DLA sample is representative of random populations of DLAs in this redshift range, we compare its properties to those of a larger DLA sample from literature. \cite{Neeleman_2013_769} analyzed 100 DLAs observed at high resolution ($\mathcal{R} \sim 40,000$) with $z_\text{abs} \sim 1.5-4$. With these high-quality data, \cite{Neeleman_2013_769} were able to measure precise metal column densities. We restrict our comparison to a subset of the \cite{Neeleman_2013_769} sample including 72 DLAs with $z_\text{abs}$ $<$ 3.6 (i.e., the highest redshift in our DLA sample). Our sample ranges from $z_\text{abs} \sim 1.6-3.5$, with an average absorber redshift of $\langle z_\text{abs} \rangle \approx 2.5$. The \cite{Neeleman_2013_769} subset has relatively more systems above $z_\text{abs} > 2.9$, yielding an average $\langle z_\text{abs} \rangle \approx 2.6$. We perform a two-sample Kolmogorov-Smirnov (K-S) test on the two distributions to test the null hypothesis that the two samples are drawn from the same parent distribution. The maximum absolute difference between the distributions calculated from the two-sample K-S statistic is low ($D_\text{K-S} =$ 0.27) and has a P-value of 0.06, suggesting we cannot reject the null hypothesis at a $>95\%$ confidence level. The standard deviation of redshifts for our DLA sample is 0.47, similar to the standard deviation of the \cite{Neeleman_2013_769} subset (0.48). These comparisons suggest that the redshift distributions of these two samples are similar.  


\begin{figure}[ht]
	\centering
	\begin{minipage}{\linewidth}
		\centering
		\includegraphics[width=\linewidth]{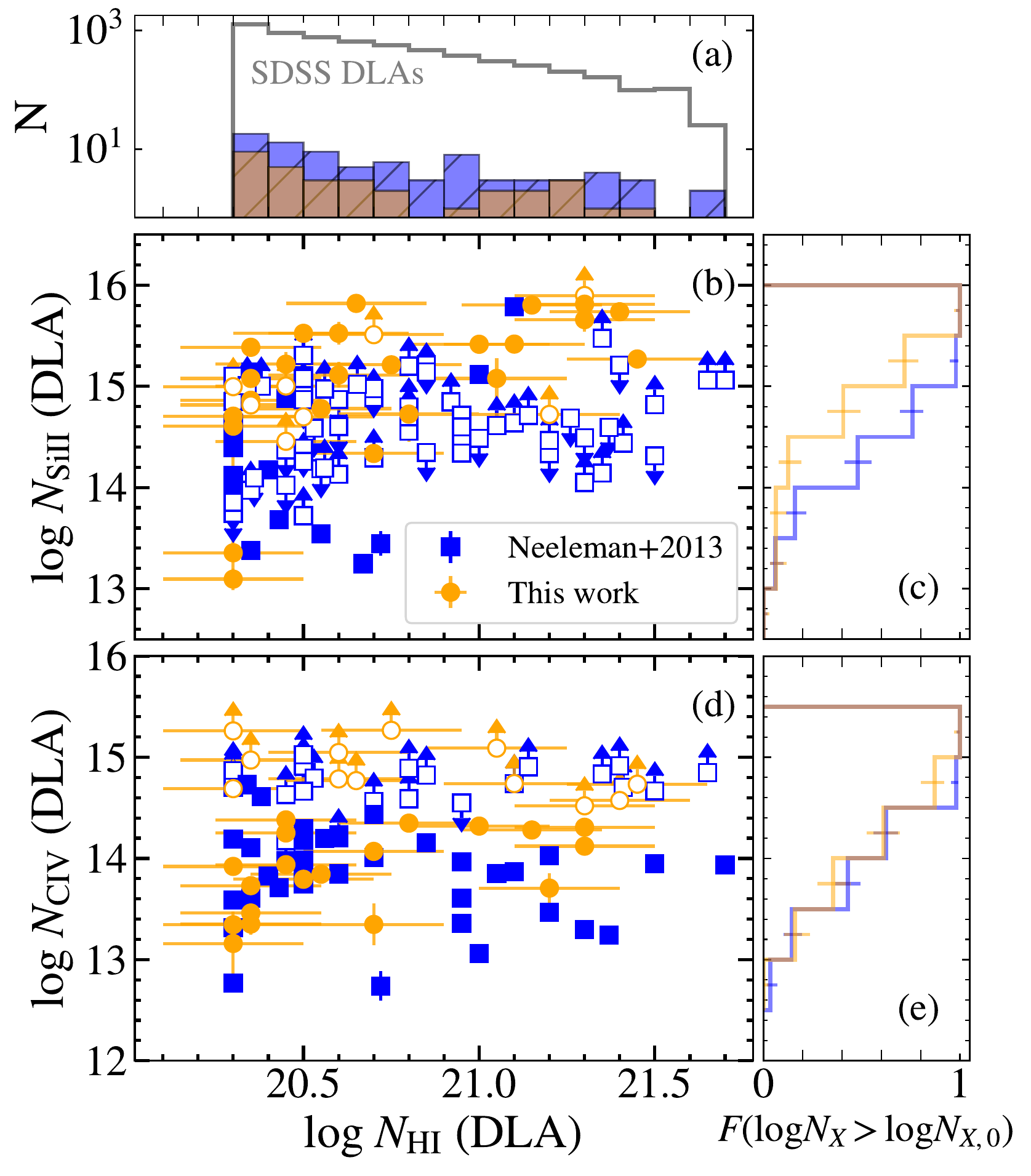}
		\caption{(a) Distribution of $N_\text{H\textsc{i}}$ for our DLA sample (orange) and the subset of the \cite{Neeleman_2013_769} sample having $z < 3.6$ (blue). The distribution of $N_\text{H\textsc{i}}$ for DLAs with redshifts $1.6 < z < 3.5$ discovered in SDSS-III DR9 QSO spectra \citep{Noterdaeme2012} is shown in gray. Bins have a width of $\Delta \log N_\text{H\textsc{i}}=0.1$. All three samples have an average $\log N_\text{H\textsc{i}}/\rm cm^{-2}$ of $\approx 20.7$.  (b) $N_\text{Si\textsc{ii}}$ vs.\ $N_\text{H\textsc{i}}$ for our sample and that of \citet{Neeleman_2013_769}. Open symbols represent limits. (d) $N_\text{C\textsc{iv}}$ vs.\ $N_\text{H\textsc{i}}$ for our sample and that of \citet{Neeleman_2013_769}. Panels (c) and (e) show normalized cumulative distributions of $N_\text{Si\textsc{ii}}$ and $N_\text{C\textsc{iv}}$, respectively, with a bin width of $\Delta \log N_\text{Si\textsc{ii},C\textsc{iv}} =0.5$. The 1$\sigma$ Wilson score intervals for these bins are shown as errorbars. We include only detections and saturations in these distributions and exclude upper limits. The distributions are similar, suggesting our sample is representative of typical DLAs around $z \sim 2.5$. \label{fig: neelcomp}}
	\end{minipage}
\end{figure}

We now consider how the physical properties of our sample DLAs relate to those of the parent DLA population during this epoch by comparing their distributions of $N_\text{H\textsc{i}}$, $N_\text{Si\textsc{ii}}$, and $N_\text{C\textsc{iv}}$. The latter two ions were selected to be representative of low-ionization and high-ionization metal absorption. We first compare the  $N_\text{H\textsc{i}}$ distribution of our sample to that of a much larger sample of 6132 DLAs with redshifts $1.6 < z < 3.5$ discovered in QSO spectroscopy from the SDSS-III DR9 \citep{Noterdaeme2012}, as well as to that of \citet{Neeleman_2013_769}. These comparisons are shown in Figure \ref{fig: neelcomp}a. A two-sample K-S test comparing the $N_\text{H\textsc{i}}$ distribution of our sample and that of \citet{Noterdaeme2012} yields a maximum difference value of $D_\text{K-S} =$ 0.16 with a P-value of 0.33.  The same test comparing our sample and that of \citet{Neeleman_2013_769} yields a maximum difference value of $D_\text{K-S} =$ 0.1 with a P-value of 0.96.
Thus we find no evidence that either of the two $N_\text{H\textsc{i}}$ distribution pairs are drawn from different parent populations.

We also compare our sample distributions of $N_\text{Si\textsc{ii}}$ and $N_\text{C\textsc{iv}}$ with those of \citet{Neeleman_2013_769}. 
$N_\text{H\textsc{i}}$ and $N_\text{Si\textsc{ii}}$ are similarly correlated in both samples, though we note that the DLAs with the highest $N_\text{H\textsc{i}}$ values in our sample appear to have higher values of $N_\text{Si\textsc{ii}}$ (shown in Figure \ref{fig: neelcomp}b). These values are however consistent with the lower limits on $N_\text{Si\textsc{ii}}$ in the \cite{Neeleman_2013_769} data. Our measurements of \ion{Si}{2} column density are somewhat more sensitive than those of \citet{Neeleman_2013_769}, as measurements from the latter study relied on \ion{Si}{2} $\lambda$1546 (which is typically saturated in DLA sightlines), whereas we make use of the weaker \ion{Si}{2} $\lambda$1808 when calculating \ion{Si}{2} column density. The more highly ionized material, traced by \ion{C}{4}, does not have column densities that are strongly correlated with $N_\text{H\textsc{i}}$. However, the distributions of $N_\text{C\textsc{iv}}$ are similar between both samples (Figure \ref{fig: neelcomp}d). We conclude that the DLAs in our QSO pair sample are representative of typical DLAs at redshifts $1.6 \lesssim z \lesssim 3.5$ from the point of view of column density distributions.

\subsection{$N_\text{H\textsc{i}}$ in DLA Environments}\label{sec:dlahI}
\begin{figure}[ht]
	\centering
	\begin{minipage}{\linewidth}
		\includegraphics[width=\linewidth]{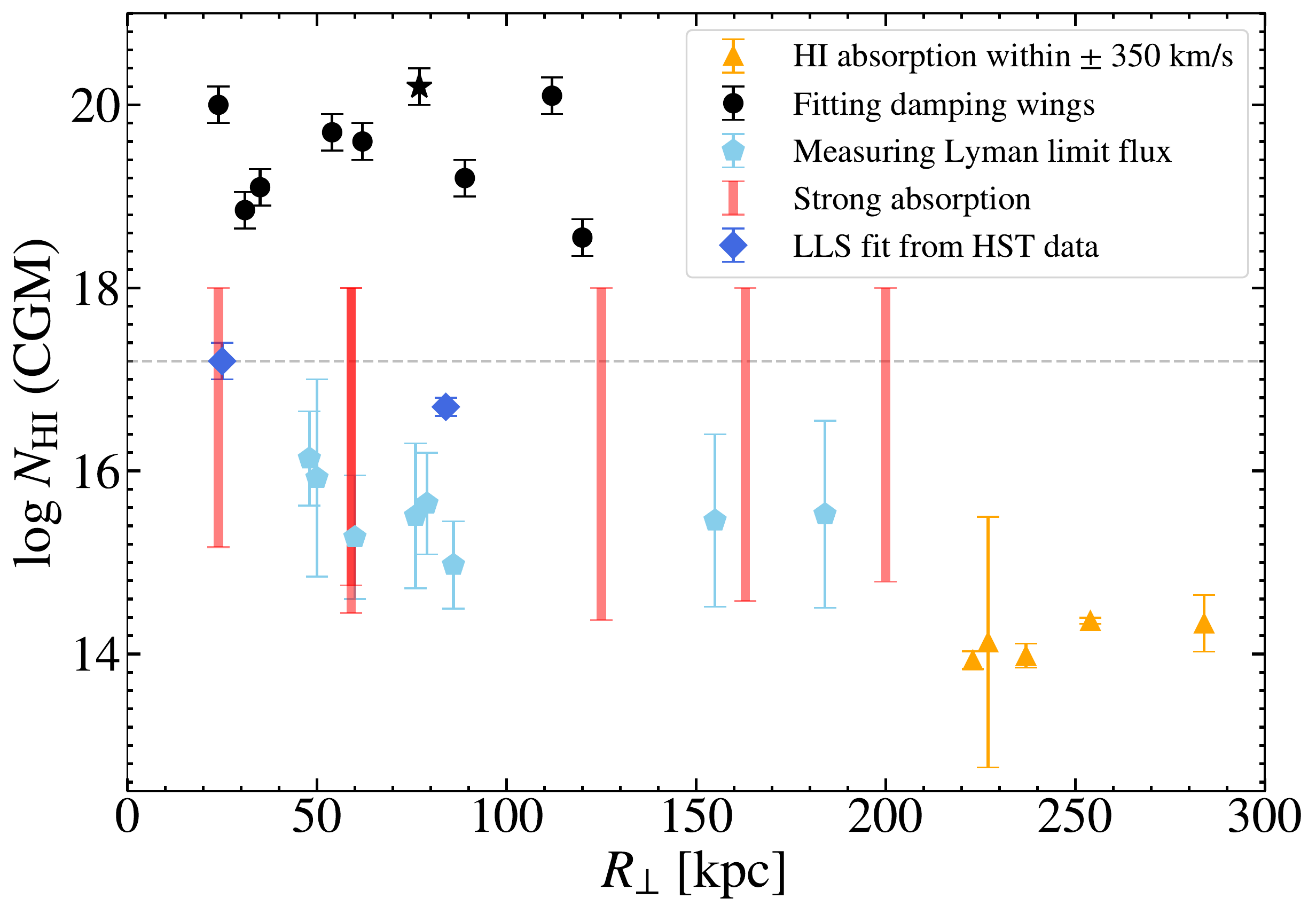}
		\caption{$N_\text{H\textsc{i}} $ measurements in our CGM sightlines vs.\ $R_\bot$. Black circles represent sightlines for which we constrain $N_\text{H\textsc{i}}$ by fitting the damping wings of the Ly$\alpha$ absorption profile. The black star indicates a double-DLA system. Dark blue diamonds indicate $N_\text{H\textsc{i}}$ measurements obtained from analysis of the Lyman limit observed with \emph{HST}/WFC3 grism spectroscopy. Light blue pentagons represent sightlines for which we place an upper limit on $N_\text{H\textsc{i}}$ by fitting the flux blueward of the system's Lyman limit. Red bars indicate systems with strong, undamped Ly$\alpha$ absorption for which we lack coverage of the Lyman limit. Orange triangles represent systems with no single, strong Ly$\alpha$ absorption line. The dashed line shows the limit for optically thick gas, $N_\text{H\textsc{i}}$ $\geq$ $10^{17.2} \ \text{cm}^{-2}$. We find that $\log N_\text{H\textsc{i}}$ is anti-correlated with $R_\bot$ overall, but exhibits significant scatter at $R_\bot < 120$ kpc. \label{fig: cgmHI}}
	\end{minipage}
\end{figure}

As DLAs are the dominant reservoirs of neutral gas at $z < 5$, the environments of DLAs can elucidate how \ion{H}{1} gas is distributed in the Universe. We first investigate the absorption strength of  \ion{H}{1} as a function of distance from our host DLAs. Figure \ref{fig: cgmHI} shows $\log N_\text{H\textsc{i}}$ measured in each CGM sightline vs.\ transverse distance ($R_\bot$) from the corresponding DLA. All optically thick CGM systems (including those with damped Ly$\alpha$ absorption, indicated in black, as well as one sightline shown as the blue diamond at $R_\bot = 25$ kpc) are located within $R_\bot < 120$ kpc. However, Figure \ref{fig: cgmHI} also includes numerous sightlines within 120 kpc of DLAs that are optically thin, suggesting that neutral gas near DLAs exhibits a wide range of densities. Weak absorption (with $N_\text{H\textsc{i}} \sim 10^{14} \ \text{cm}^{-2}$, indicated in orange) is only found further than 200 kpc from DLAs, indicating that \ion{H}{1} column densities may decrease with increasing $R_\bot$. To evaluate the significance of an anti-correlation between $R_\bot$ and \ion{H}{1} column density, we calculate the Kendall rank correlation coefficient ($\tau_{K}$). We caution that six \ion{H}{1} absorbers (indicated by the red bars) are on the flat part of the curve of growth, with large errorbars that are not accounted for in this calculation. Nevertheless, we find $\tau_{K} = - 0.4$ with a two-sided probability of no correlation of P $=$ 0.002, indicative of an anti-correlation. Such anti-correlations between \ion{H}{1} column density and projected distance are ubiquitous features of CGM sightline samples, including those probing LBG environments \citep{Rudie_2012_750, Rakic2012}, QSO host environments \citep{Prochaska_2013_776_136P}, and the environments around Ly$\alpha$ emitters \citep{Liang_2020_arXiv} at $z\sim 2-3$.

\begin{figure}[ht]
	\centering
	\begin{minipage}{\linewidth}
		\includegraphics[width=\linewidth]{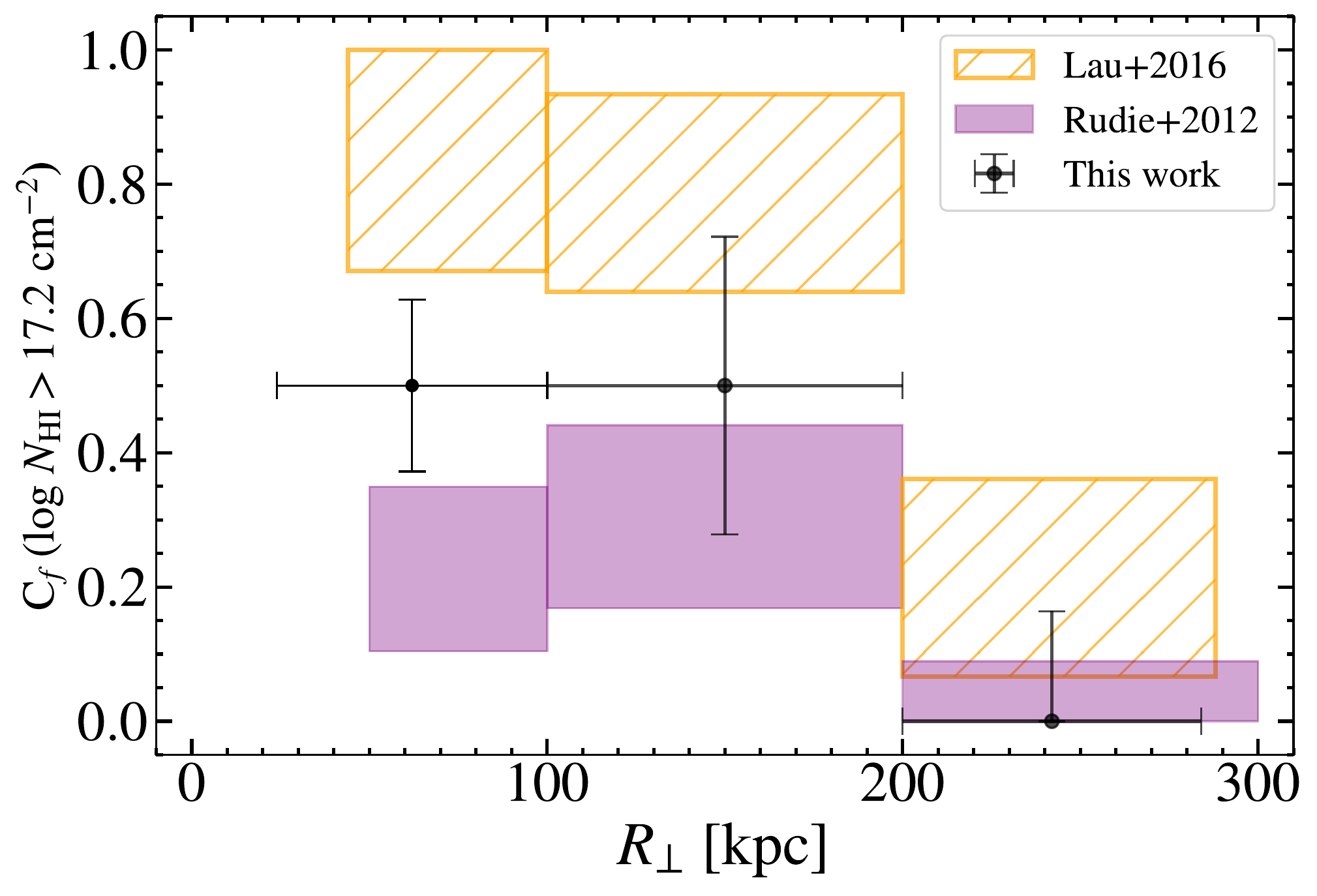}
		\caption{Covering fractions of optically thick \ion{H}{1} ($N_\text{H\textsc{i}} > 10^{17.2}$ cm$^{-2}$) measured in bins of $\Delta  R_\bot \approx 100 $ kpc. The black points represent covering fractions in the CGM of DLAs (this work). The orange boxes represent the covering fraction of optically thick \ion{H}{1} measured in the CGM of quasar hosts (\citealt{Lau_2016_226}). Purple boxes represent the covering fraction of optically thick \ion{H}{1} in the CGM of LBGs (\citealt{Rudie_2012_750}). The vertical errorbars represent the 1$\sigma$ Wilson score interval for each bin. The $x$-axis errorbars indicate the span of $R_\bot$ within each bin. We find DLA halos exhibit more than twice the covering fraction of optically thick \ion{H}{1} relative to that measured in LBG halos within $R_\bot < 100$ kpc.
		\label{fig: cfHI}}
	\end{minipage}
\end{figure}

We also explore the spatial extent of optically thick ($N_\text{H\textsc{i}}$ $\geq$ $10^{17.2} \ \text{cm}^{-2}$) gas in DLA halos, calculating covering fractions within $R_\bot < 100$ kpc, at $100~\mathrm{kpc} < R_\bot < 200~\mathrm{kpc}$, and at $R_\bot > 200$ kpc. The error associated with these covering fractions is estimated by calculating the 1$\sigma$ Wilson score interval for each bin. The systems marked by red bars in Figure \ref{fig: cgmHI}, with strong but undamped absorption, 
have column density constraints that are ambiguous with respect to the optically thick threshold, and therefore are excluded from these calculations. Similarly, we exclude the one blue point which lies at the limit of optically thick gas. The resulting covering fractions ($C_f$) are shown in Figure \ref{fig: cfHI}, with the $x$-axis errorbars indicating the $R_\bot$ span of each bin. The halos of DLAs have an incidence of optically thick \ion{H}{1} of $50\pm13\%$ and $50\pm22\%$ for bins with  24 kpc $< R_\bot \leq 100$ kpc and  100 kpc $< R_\bot \leq 200$ kpc, respectively. At 200 kpc $ < R_\bot \leq 284$ kpc, we place an upper limit on the covering fraction of $<16\%$ for optically thick \ion{H}{1}. Combining measurements within the first two bins, we find the incidence of optically thick \ion{H}{1} to be $50\pm11\%$ within $R_\bot < 200 $ kpc of DLAs. All together, these findings suggest that the extent of optically thick gas around DLAs is  $\lesssim 200 $ kpc.  Moreover, our finding that DLAs are very rarely detected along both sightlines in our QSO pairs suggests that the total extent of DLAs themselves is likely $\lesssim 50 $ kpc (see Urbano Stawinski et al. in prep.).

We compare these measurements to the results of two other surveys at similar redshifts: one focused on massive quasar host galaxy halos (\citealt{Lau_2016_226}), and the other assessed halos of Lyman Break Galaxies (LBGs; \citealt{Rudie_2012_750}). Within 100 kpc, the latter survey implies covering fractions of optically thick material around LBGs of $20^{+15}_{-13}\%$.  The $C_f$ we measure around DLAs, $50\pm13\%$, is $\approx 1.5\sigma$ larger than this value. While this offset is not statistically significant, it is nevertheless suggestive that DLA halos may have more uniformly distributed optically thick \ion{H}{1} than LBG halos (i.e., it is more likely to find optically thick gas near a DLA than near an LBG). Such a finding may moreover be a natural result of our selection criteria for CGM sightlines -- i.e., they must arise close to a region that is already known to have a high neutral column density. On the other hand, quasar halos have nearly 100$\%$ optically thick covering fractions within 100 kpc, $>1\sigma$ larger than DLA halos.  It is therefore even more likely that optically thick material will be found close to quasar host galaxies. Beyond 100 kpc, the errorbars on these $C_f$ constraints overlap, such that the covering fractions measured around these three samples are statistically consistent. Previous work has demonstrated that DLAs are clustered to LBGs, with the DLA-LBG correlation length being statistically consistent with that of the LBG-LBG autocorrelation length ($r_0 = 2.81^{+1.4}_{-2.0} ~h^{-1}$Mpc; \citealt{Cooke2006}). While this implies that LBGs and DLAs occupy similar environments, these clustering studies do not sample scales $<400$ kpc as we do here.    

\subsection{Column Densities and Covering Fractions of Metal Lines}\label{sec:colmden}

\begin{figure*}[ht]
	\centering
	\begin{minipage}{\textwidth}
		\includegraphics[width=0.5\linewidth]{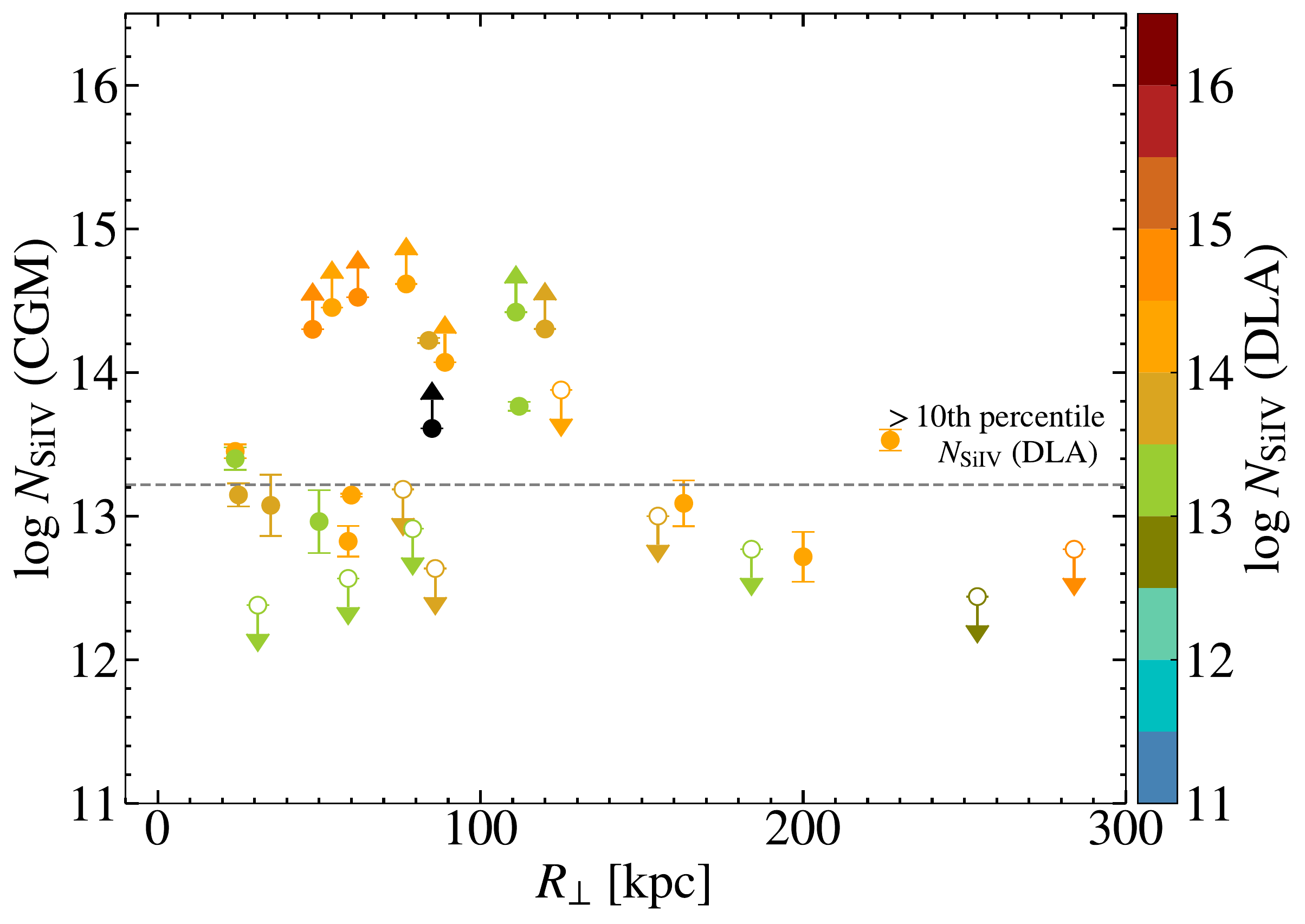}
		\includegraphics[width=0.5\linewidth]{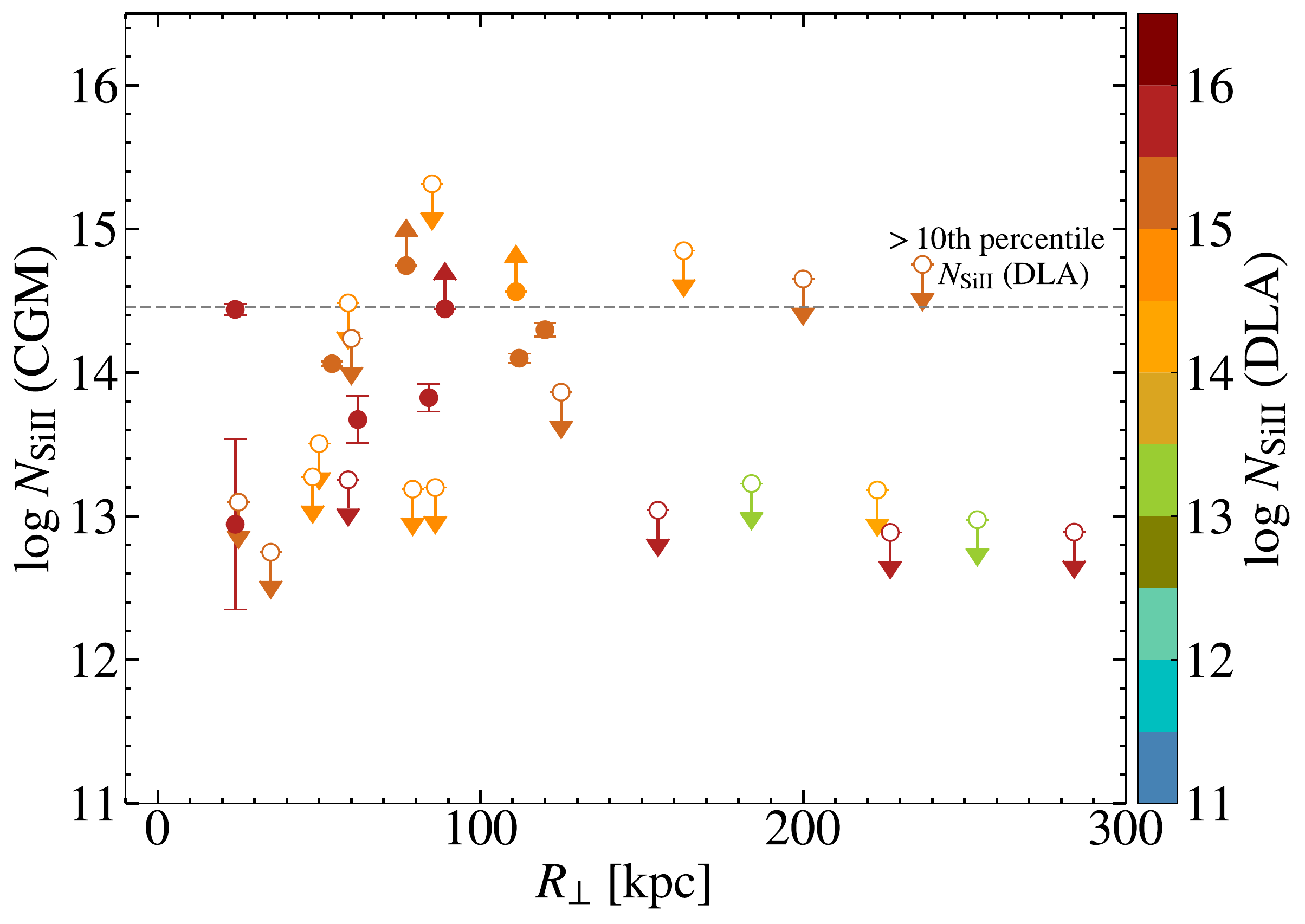}
		\caption{Column densities of \ion{Si}{4} (left panel) and \ion{Si}{2} (right panel) measured in our CGM sightlines. Measurements are color-coded according to the corresponding DLA column density for that ion. The black point represents a system for which the DLA sightline has an ambiguous column density. Open symbols indicate that our constraint on the CGM sightline column is an upper limit. The horizontal dashed lines represent the threshold above which 90$\%$ of metal line column densities for the DLA sightlines fall. The column densities of \ion{Si}{2} in the CGM drop below what we measure for the vast majority of DLA sightlines at small $R_\bot$. In contrast, the column densities of \ion{Si}{4} remain similar to those measured in our DLAs up to projected distances of $\sim 150$ kpc. \label{fig:Rpropsi} }
	\end{minipage}
\end{figure*}

Our medium- and high-resolution spectroscopy uniquely allows us to investigate the properties of metal-enriched halo gas and compare them directly to those observed in the DLA hosts. First, we compare the column densities for different ions measured in the CGM sightlines to column densities of the same ion in the associated DLA sightlines. An illustration of this comparison is shown in Figure \ref{fig:Rpropsi}, which presents the column densities of \ion{Si}{4} and \ion{Si}{2} in our CGM sightlines vs.\ $R_\bot$. The colors represent the corresponding DLA column densities for these ions. The horizontal dashed lines on Figure \ref{fig:Rpropsi} represent the threshold above which 90$\%$ of metal line column densities for the DLA sightlines fall and can be used to compare individual CGM sightlines to the column densities of the majority of our DLA sample. The vast majority of DLA column density upper limits are below these thresholds, and most lower limits are above them.

First, we note that we measure overall higher DLA column densities of \ion{Si}{2} relative to \ion{Si}{4}, consistent with previous metal-line absorption studies for DLAs (e.g.,  \citealt{Vladilo_2001_557}, \citealt{Fox_2007_473}, \citealt{Mas-Ribas_2017_846}), and indicative that DLAs probe overall neutral environments. Second, we find that the CGM column densities of \ion{Si}{2} are significantly lower (below the dashed line) than those measured in the DLAs across the full range in $R_\bot$ of our sample. By contrast, the CGM column densities of \ion{Si}{4} are similar to those of the DLAs within $R_\bot$ $\leq$ 150 kpc. Similar patterns are apparent in all elements analyzed in this work for which we have access to both low- and high/intermediate-ionization species, including \ion{C}{2}, \ion{C}{4}, \ion{Al}{2} and \ion{Al}{3} (see Figure \ref{fig:Rpropall} in Appendix \ref{chAppend: column}). These results imply that high-ionization species observed in DLA sightlines trace halo gas out to distances of $\approx 150$ kpc. This finding verifies the results of studies such as \citet{Wolfe_2000_545}, who showed that \ion{C}{4} and \ion{Si}{4} velocity profiles in DLAs are consistent with those arising from halo gas in semianalytic cold dark matter models.

\begin{figure}[h]
	\centering
	\vskip 0.2in
	\begin{minipage}{\linewidth}
		\includegraphics[width=\linewidth]{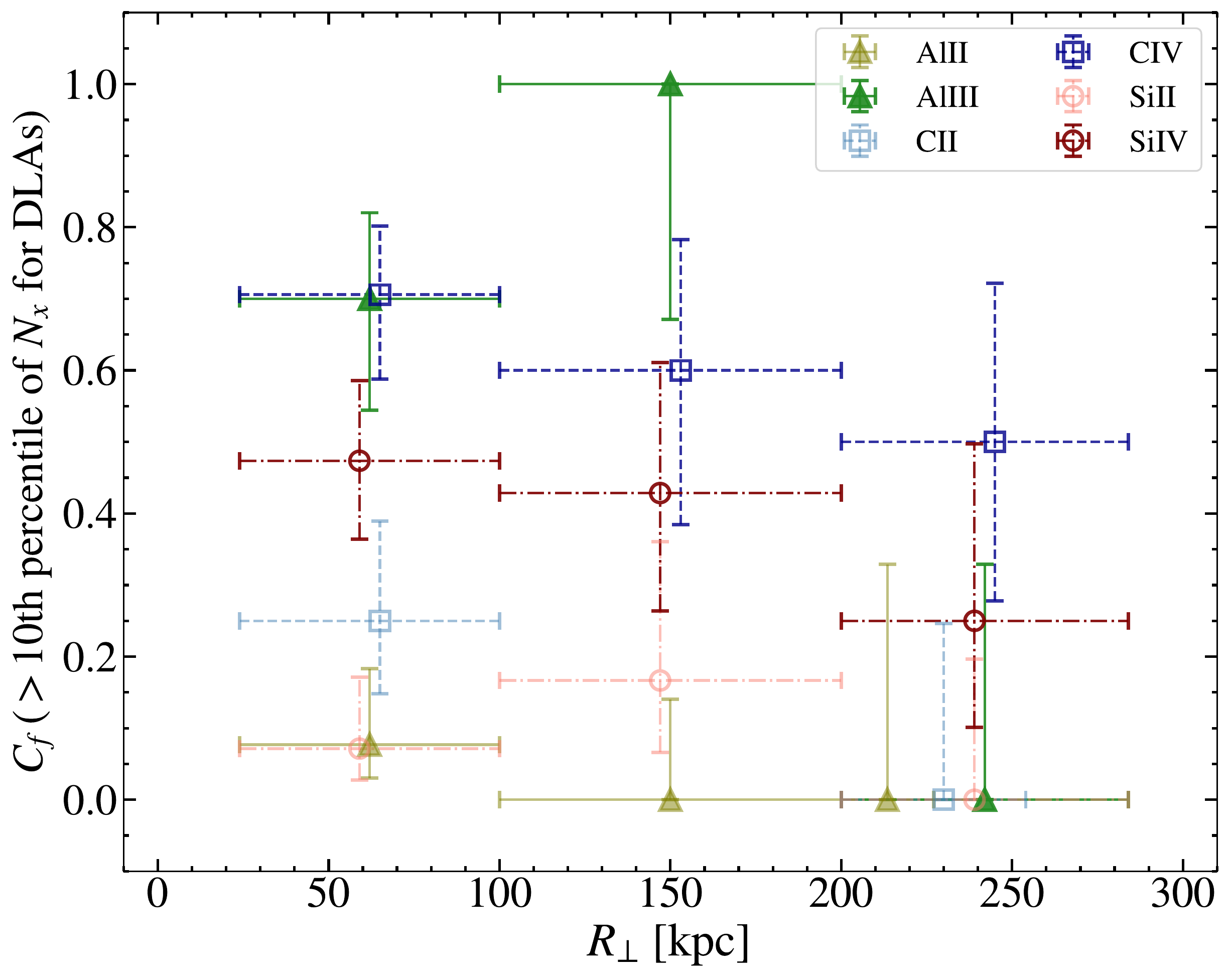}
		\caption{Covering fractions for \ion{Al}{2}, \ion{Al}{3}, \ion{C}{2},  \ion{C}{4}, \ion{Si}{2}, and \ion{Si}{4} with the threshold set to encompass the upper 90$\%$ of column densities from the DLA sightlines for each species. The $x$-axis errorbars represent the bins used to compute each covering fraction. We reduce the ranges of the first and last bins to indicate the span of $R_\bot$ for our sample. In some cases, the bins are sparsely populated and therefore may not span the full range as represented by the $x$-axis errorbars. The $y$-axis errorbars represent the 1$\sigma$ Wilson score interval for each covering fraction estimate. We find that (1) the halos of DLAs exhibit a higher incidence of high- and intermediate-ionization gas across all distance bins, and (2) covering fractions decrease as a function of projected distance for most species.  \label{fig:covall}}
	\end{minipage}
\end{figure}

To further investigate the extent of high- (represented by  \ion{Si}{4} and  \ion{C}{4}), intermediate- (represented by \ion{Al}{3}), and low- (represented by \ion{Si}{2}, \ion{Al}{2}, and \ion{C}{2}) ionization gas in DLA halos, we calculate covering fractions for each of these ionic transitions (shown in Figure \ref{fig:covall}). We require that at least two column density measurements be available in a given bin to compute the corresponding covering fraction. To assess how the column densities in the CGM compare to those of the DLA sightlines and account for the relative abundances of each individual ion, we use a column density threshold set at the 10th percentile value of the column densities measured in the DLA sightlines for each species (i.e., 10$\%$ of all DLA metal columns for that species lie below the chosen threshold). Thus, these covering fractions trace the incidence of absorption similar in strength to that observed in DLAs. 

We find in general that all high- and intermediate-ionization species have large covering fractions compared to low-ionization species. Within 200 kpc of DLAs, high- and intermediate-ionization species have covering fractions above 40$\%$, while the incidence of low-ionization species never exceeds 30$\%$ even at 24 kpc $< R_\bot < 100$ kpc. This indicates that the warm, ionized material associated with DLAs frequently extends over ${\gtrsim}100$ kpc scales, whereas cool, photoionized or neutral material seldom exhibits DLA-level absorption strengths across more than ${\gtrsim}25~\rm kpc$. We place upper limits on the covering fractions of all intermediate and low ions beyond $R_\bot$ $>$ 200 kpc (yielding an incidence of $<$32$\%$ for \ion{Al}{3}, $<$32$\%$ for \ion{Al}{2}, $<$25$\%$ for \ion{C}{2}, and $<$20$\%$ for \ion{Si}{2}); however, these species do exhibit some absorption that is weaker than the corresponding 10th-percentile column density threshold.

\begin{figure*}[h]
		\centering
		\includegraphics[width=0.47\linewidth]{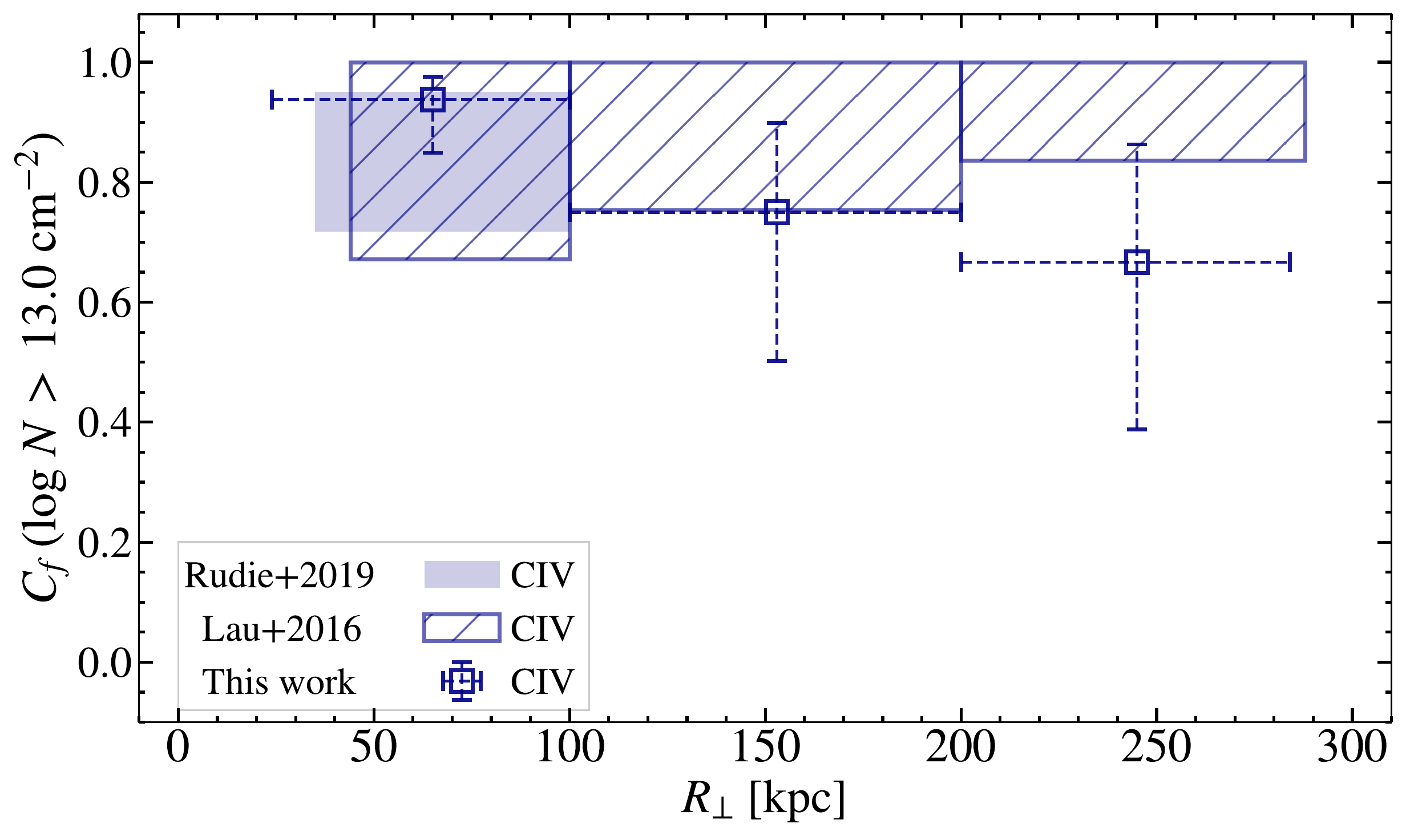}
  		\includegraphics[width=0.47\linewidth]{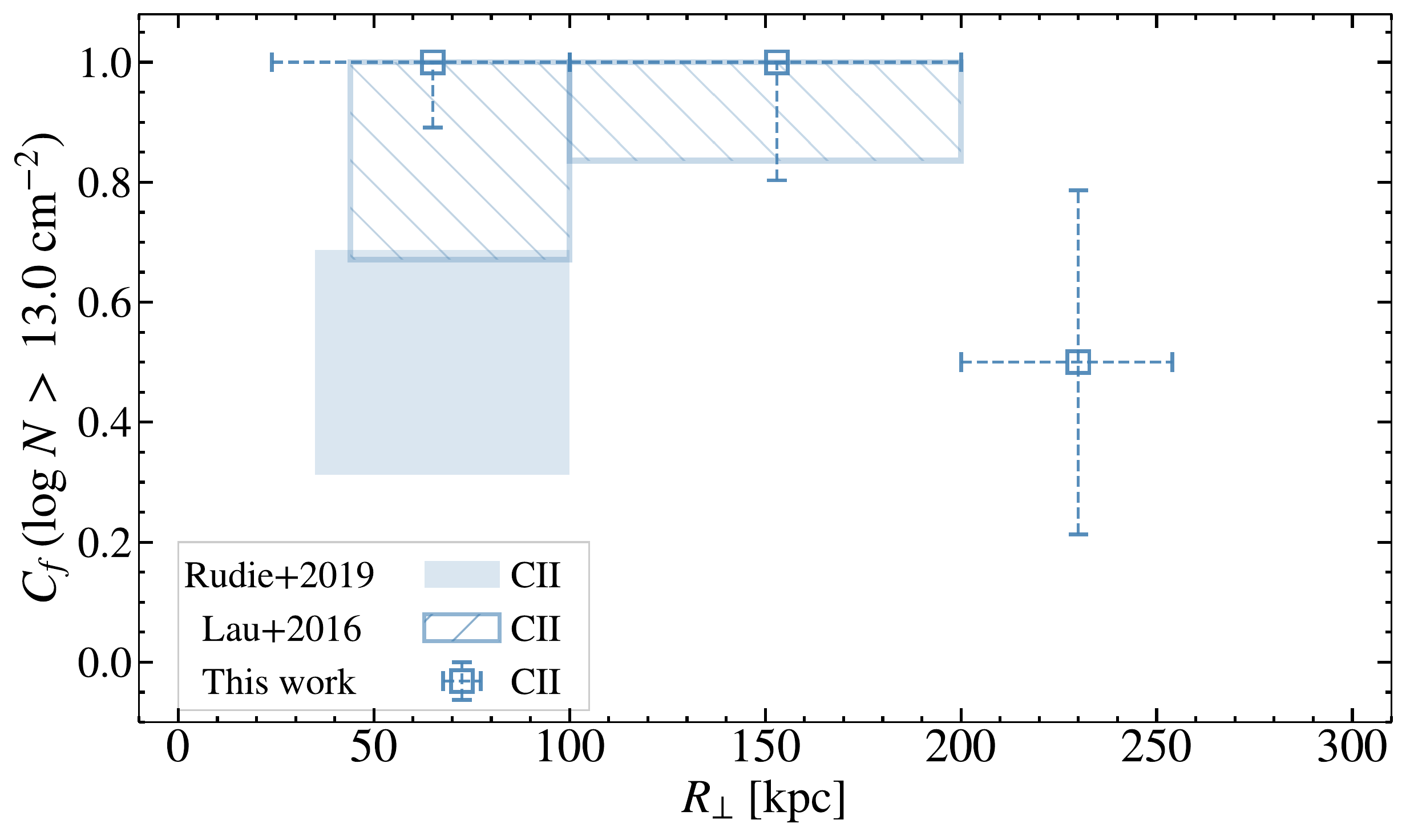}
		\includegraphics[width=0.47\linewidth]{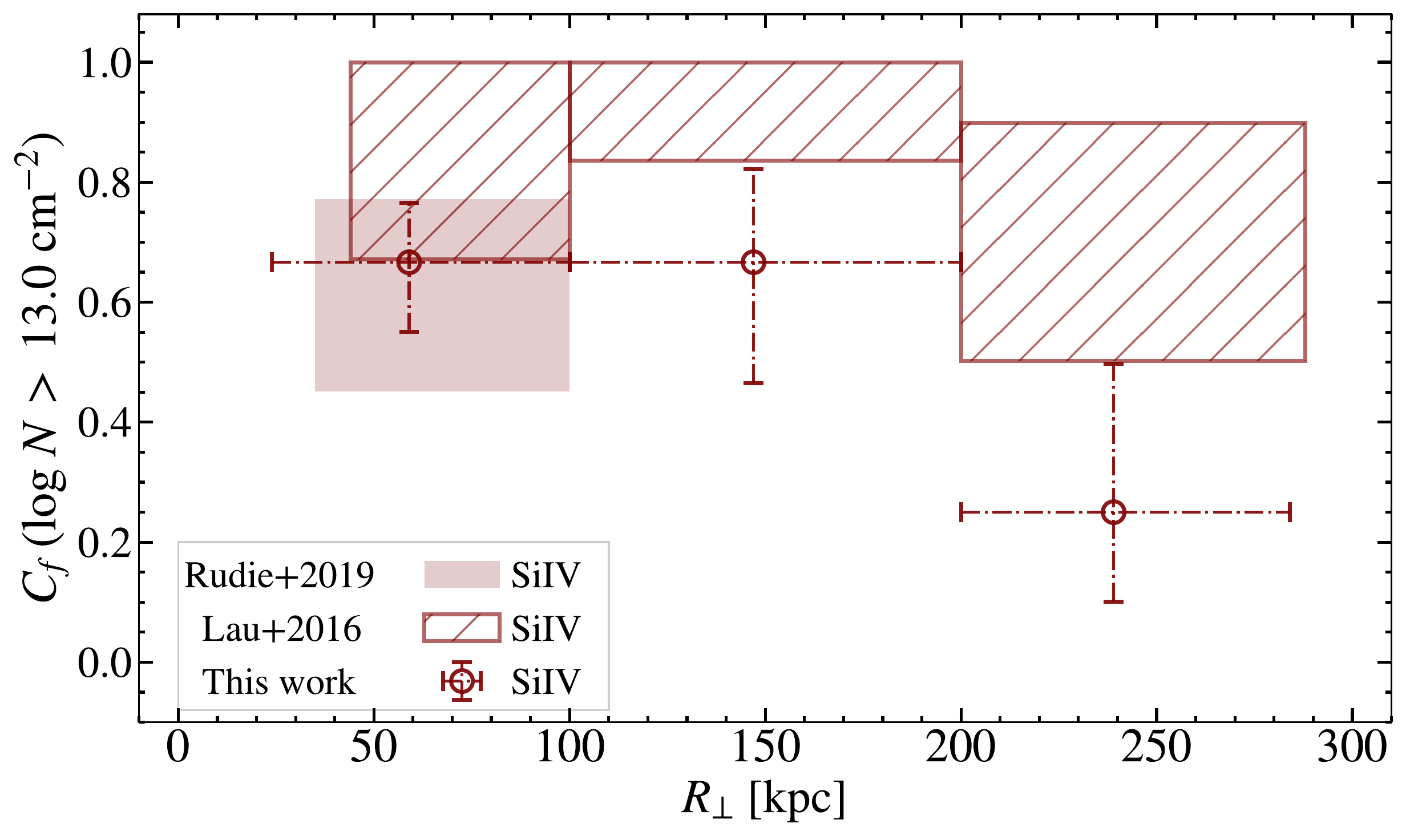}
		\includegraphics[width=0.47\linewidth]{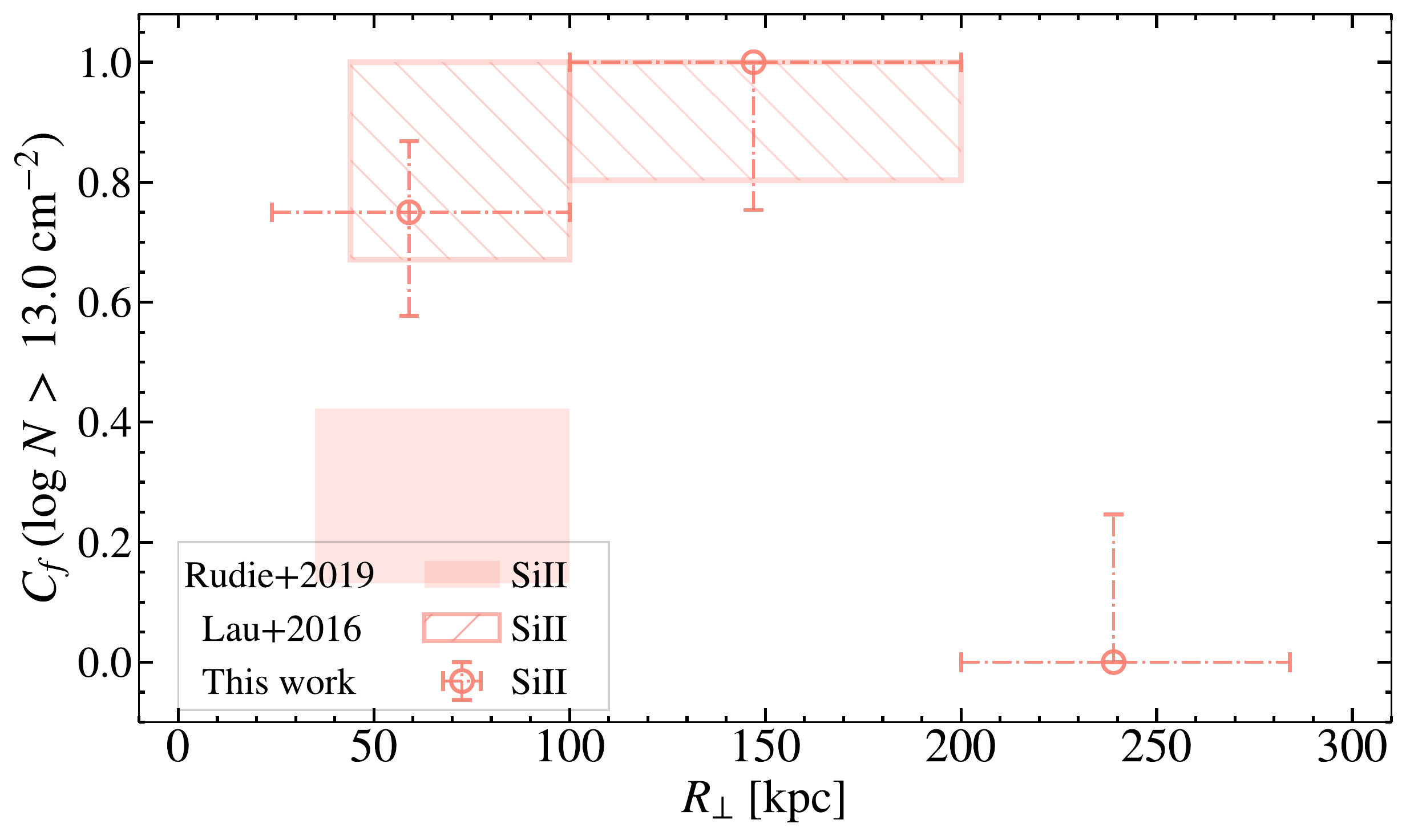}
		\includegraphics[width=0.47\linewidth]{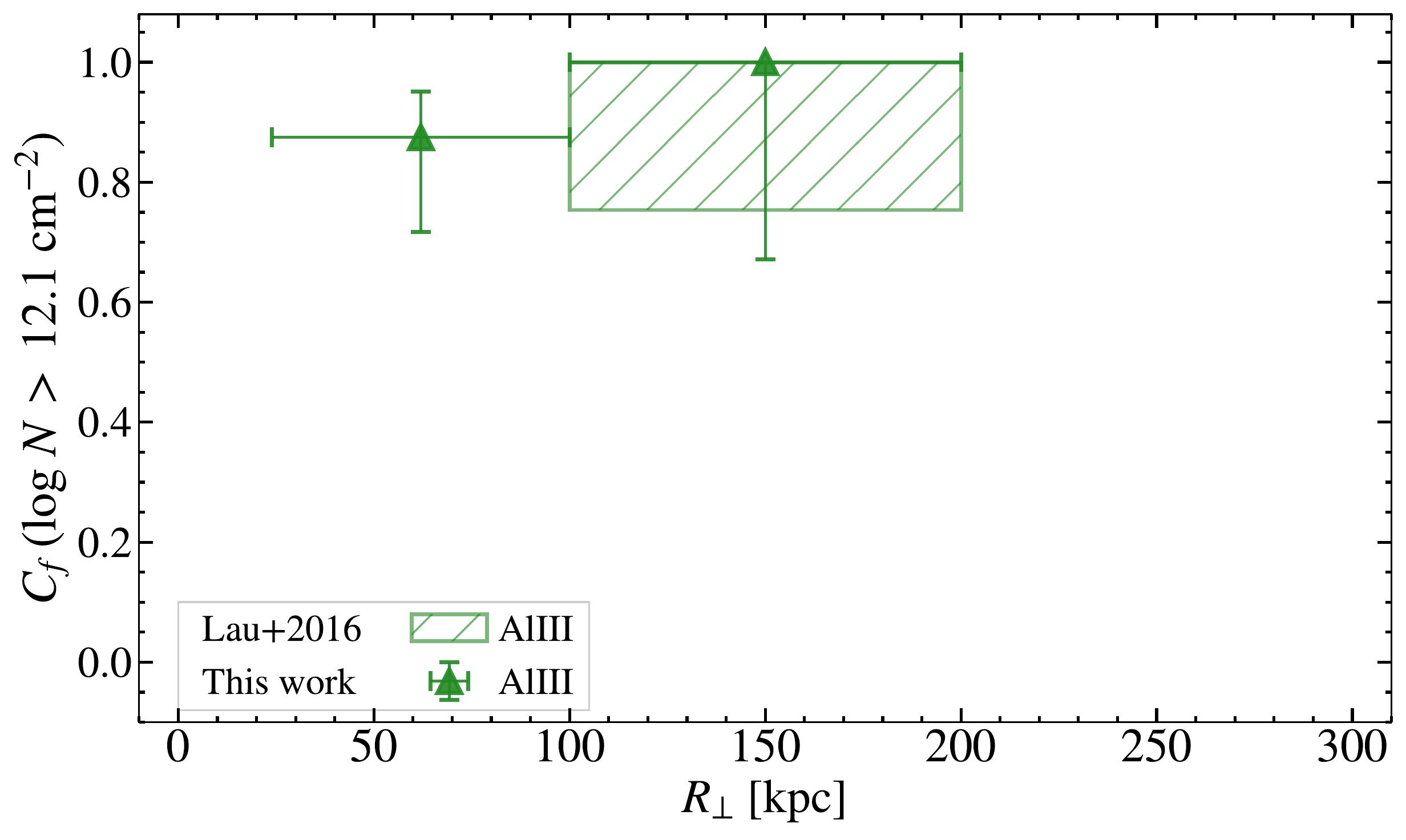}
		\includegraphics[width=0.47\linewidth]{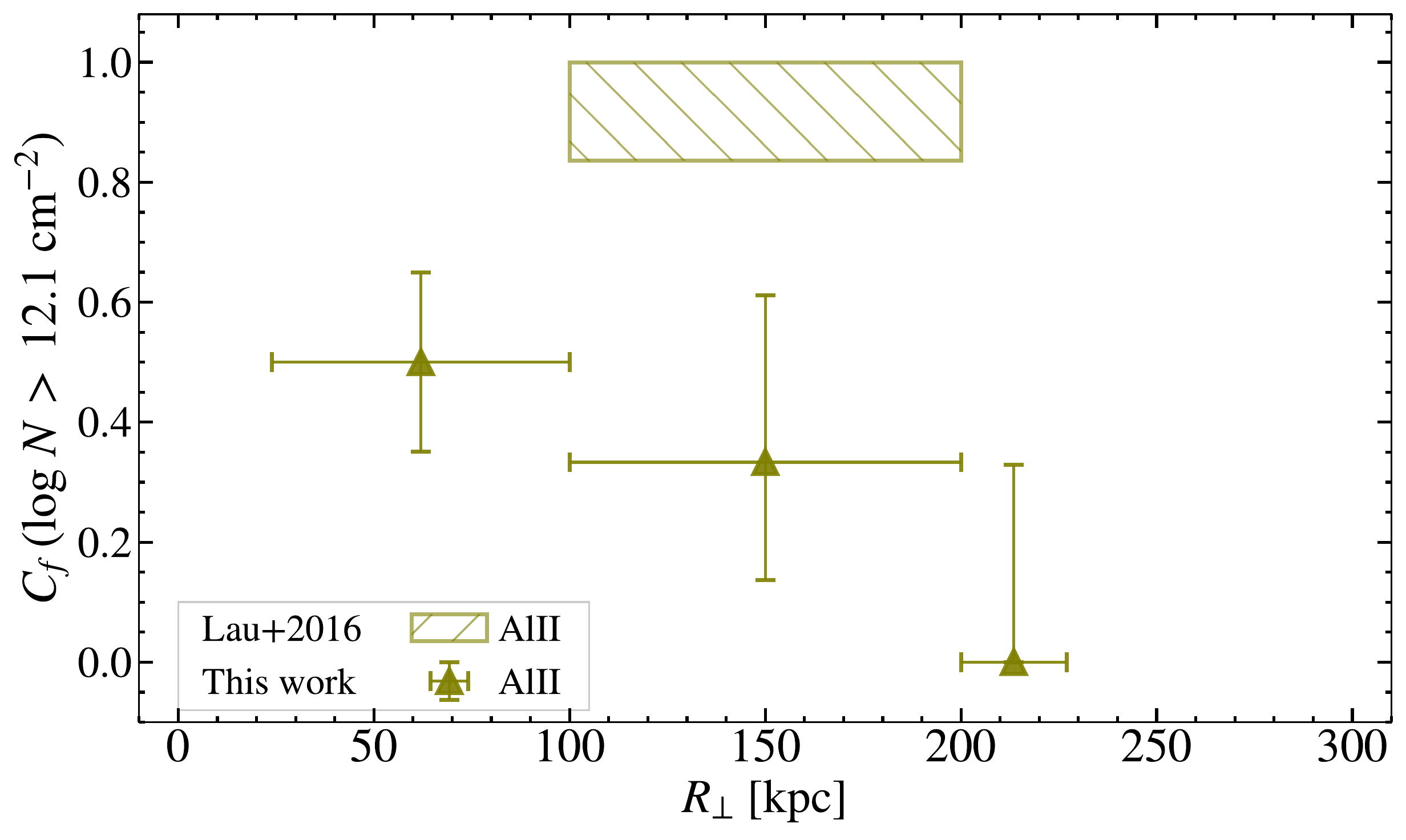}
		\caption{Covering fractions of carbon species (top row), silicon species (middle row), and aluminum species (bottom row) in DLA halos (points with error bars), quasar halos (\citealt{Lau_2016_226}; hatched boxes), and LBG halos (\citealt{Rudie_2019_arXiv}; filled boxes). High-ionization species are shown in the left column and low-ionization species are shown in the right column. Covering fractions are calculated with a threshold of $N > 10^{13} \text{cm}^{-2}$ for carbon and silicon species and $N > 10^{12.1} \text{cm}^{-2}$ for aluminum species. Vertical errorbars and box heights represent 1$\sigma$ Wilson score intervals. The $x$-axis errorbars represent the bins used to compute each covering fraction. These errorbars also show the span of $R_\bot$ of our sample, however in some cases the bins are sparsely populated and may not span the full range. The covering fractions in DLA halos are similar to those in LBG halos for high-ionization species and larger by $\sim 2\sigma$ for low-ionization species. DLA halo covering fractions are similar to those measured in QSO halos within 200 kpc (with the exception of covering fractions of \ion{Si}{4} and \ion{Al}{2}), but decline relative to the latter beyond this distance.
\label{fig: comp_cfmetals}}
\end{figure*}

Lastly, we investigate the extent of metals in the CGM of DLAs in comparison to other CGM environments at similar redshifts: in quasar halos (\citealt{Lau_2016_226}) and the halos of LBGs (\citealt{Rudie_2019_arXiv}). The results are shown in Figure \ref{fig: comp_cfmetals}. Here we measure covering fractions with a threshold of $N > 10^{13} \ \text{cm}^{-2}$ for species of carbon and silicon and $N > 10^{12.1} \ \text{cm}^{-2}$ for species of aluminum to account for the expected differences in abundance of each element. We only report covering fractions calculated with at least two measurements within each $R_\bot$ bin. The  y-axis errorbars indicate the 1$\sigma$ Wilson score interval, which accounts for the sample size in each bin. For reference, the DLA-CGM sample has the largest number of measurements (between eight and 18) within 100 kpc, between three and eight measurements at 100 kpc $< R_\bot <$ 200 kpc, and between two and four measurements at 200 kpc $< R_\bot <$ 300 kpc. The \cite{Rudie_2019_arXiv} sample includes between six and eight measurements within 100 kpc, and the \cite{Lau_2016_226} sample includes two measurements within 100 kpc, between three and five measurements at 100 kpc $< R_\bot <$ 200 kpc, and either four or five measurements at 200 kpc $< R_\bot <$ 300 kpc.

We find the covering fractions of high-ionization species (\ion{Si}{4} and \ion{C}{4}) around DLAs and LBGs to be similarly high, suggestive of a volume-filling medium that extends to comparable projected distances in both environments. In contrast, our sample of DLA halos exhibits higher covering fractions for \ion{C}{2} and \ion{Si}{2} than the LBG sample by $\sim2\sigma$, though we caution that the statistical uncertainties are significant. These differences may indicate, e.g., that we are preferentially selecting regions with a higher incidence of neutral (and hence also low-ionization) gas by targeting the CGM of DLAs, and/or that the metallicity of halo material around LBGs is generally lower than that in DLA halos.

Metal covering fractions in quasar halos are either larger than or consistent with those measured around DLAs in all ions. Within $R_\bot <100$ kpc, all ionic covering fractions for DLA and QSO halos are consistent within $<1\sigma$. Within $100~\mathrm{kpc} < R_\bot < 200$ kpc, the covering fractions of \ion{Si}{2}, \ion{C}{2}, and \ion{Al}{3} are consistent within $<1\sigma$. Beyond 200 kpc, all high- and low-ion covering fractions in DLA halos are $\gtrsim$ 1$\sigma$ lower than those measured in  QSO halos, suggesting that metal-enriched gas pervades these latter massive halos to larger impact parameters.

\subsection{Metallicity of DLAs and their Halos}\label{sec:metallicity}

With column densities in hand, we can provide new constraints on the metallicities of high-redshift DLAs and their associated CGM. In an optimal scenario, the metallicities of absorption-line systems are estimated via photoionization modeling, which can simultaneously constrain the ionization parameter of the system along with its metallicity \citep[e.g.,][]{Crighton_2015_446,Fumagalli_2016_455,Prochaska_2017_837}. However, given that our column density measurements include numerous upper limits with relatively high values (e.g., $N_\text{Si\textsc{ii}}\lesssim 10^{13-14} \text{cm}^{-2}$), it is unlikely that photoionization modeling will yield robust metallicity constraints for our dataset. Instead, we adopt a simpler approach using ionic ratios to assess the ionization state of each system as described in \cite{Prochaska_2015_221}. We then explore the ratios of low-ionization metal column densities to those of neutral hydrogen, which may be used as a proxy for metallicity in systems with negligible ionization corrections \citep[e.g.,][]{Prochaska_2015_221}.

\subsubsection{Constraining Metallicities}\label{secsub:metallicitiesmethods}

To estimate the metallicities of our DLA and CGM absorption systems, we make use of a quantity introduced by \citet{Prochaska_2015_221}:
\begin{equation}\label{eq:solarmetals}
	\{\text{X$i$}/\text{H$j$}\} = \log(N_\text{X$i$}/N_\text{H$j$}) - \epsilon_\text{X} + \epsilon_\text{H}.
\end{equation}
Here, $\epsilon$ is the logarithmic solar abundance for the element X, while $i$ and $j$ represent ionization levels. The bracket notation indicates an ionic ratio of two different elements that ignores ionization corrections. In cases in which the ionization correction is small, we will assume that $\{\mathrm{X}i/\mathrm{H}j\} =$ [X/H]. We adopt solar elemental abundances from \citet{Asplund_2009_47}.

Previous studies have assessed approximate ionization corrections via the ratio $\nSiIV / \nSiII$.  We discuss these ratios in detail for our sample, along with other ionic ratios sensitive to ionization state, in Appendix \ref{app:metallicities}. Our measurements of these ratios imply negligible ionization corrections for only a small subset of our CGM sightlines. We therefore 
rely primarily on the ratio \{O\textsc{i}/H\textsc{i}\} as a direct indicator of metallicity.  For CGM sightlines with $\nHI \gtrsim 10^{19}~\rm cm^{-2}$, this ionic ratio is insensitive to ionization state due to the similar ionization potentials of \ion{H}{1} and \ion{O}{1}, the possibility of charge exchange between them \citep[e.g.,][]{FieldSteigman_1971_166,Prochaska_2015_221}, and because oxygen is only weakly depleted by dust \citep{Jenkins2009}. It is commonly assumed that \{O\textsc{i}/H\textsc{i}\} $\approx$ [O/H] for systems with $\nHI \gtrsim 10^{19}~\rm cm^{-2}$ \citep[e.g.,][]{Crighton_2013_776L}. There are six CGM sightlines which have both $\nHI \gtrsim 10^{19}~\rm cm^{-2}$ and a constraint on \{O\textsc{i}/H\textsc{i}\}. 
We also include in the following analysis three more CGM systems with both $\nHI > 10^{18.5}~\rm cm^{-2}$ and a constraint on \{Si\textsc{ii}/H\textsc{i}\}. Similar to \{O\textsc{i}/H\textsc{i}\}, \{Si\textsc{ii}/H\textsc{i}\} can be used to trace [Si/H] for mostly neutral systems, although it is somewhat more sensitive to ionization state than \{O\textsc{i}/H\textsc{i}\} and overestimates the metallicity as the ionized fraction increases. Since these three systems may have non-negligible ionization corrections, we report these \{Si\textsc{ii}/H\textsc{i}\} constraints as upper limits on [Si/H].  For completeness, we also report the ionic ratios \{Si\textsc{ii}/H\textsc{i}\}, \{C\textsc{ii}/H\textsc{i}\}, \{Fe\textsc{ii}/H\textsc{i}\}, and \{O\textsc{i}/H\textsc{i}\} for all DLAs in our sample in Appendix~\ref{app:metallicities}.



\subsubsection{Metallicity of DLA Halos}\label{secsub:metallicitycompare}

\begin{figure*}[hb]
	\centering
	\begin{minipage}{\linewidth}
		\centering
		\includegraphics[width=0.98\linewidth]{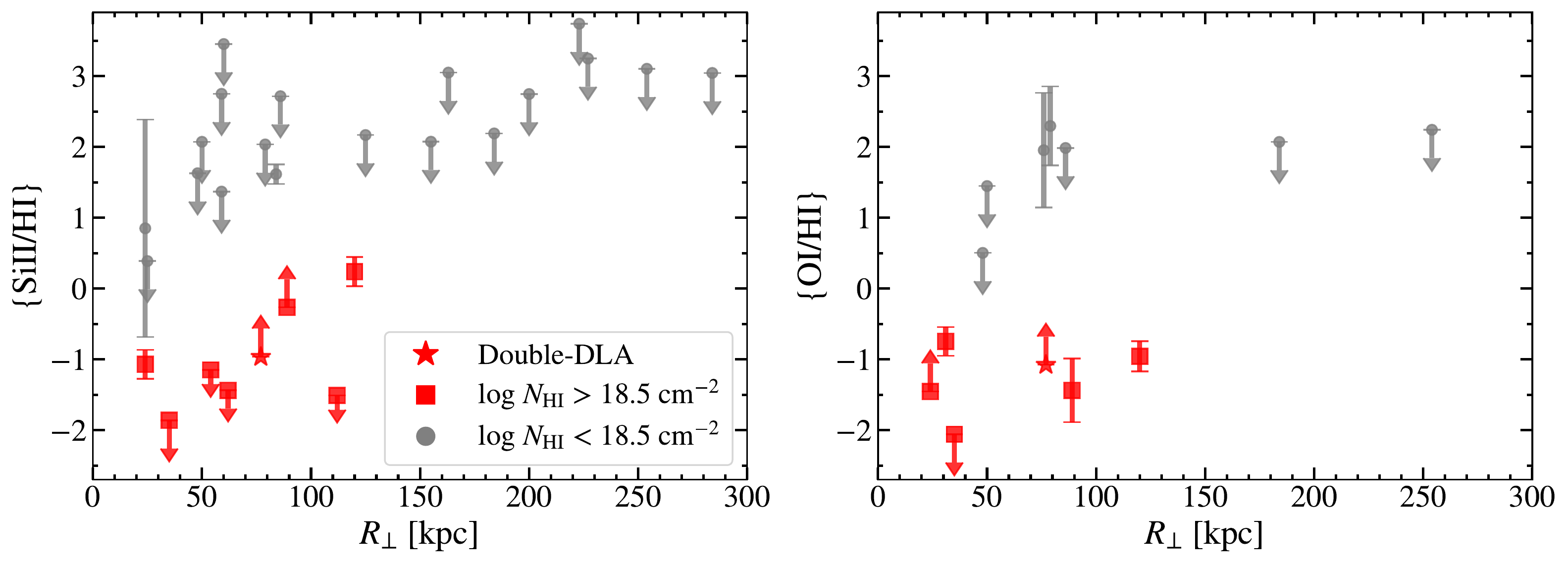}
		\caption{Logarithmic ionic ratios \{Si\textsc{ii}/H\textsc{i}\} (left) and \{O\textsc{i}/H\textsc{i}\} (right) vs.\ $R_{\perp}$ for our CGM sightlines. Metallicity constraints
  for systems with $\nHI > 10^{18.5}~\rm cm^{-2}$ are represented by red symbols. Measurements shown with gray circles are likely overestimates of [X/H]. The double-DLA system is shown with a red star.
\label{fig:cgmmetalabund}}
	\end{minipage}
\end{figure*}

Our estimates of the metallicity of individual CGM sightlines, assessed via \{Si\textsc{ii}/H\textsc{i}\} and \{O\textsc{i}/H\textsc{i}\},
are shown in Figure \ref{fig:cgmmetalabund}. 
The six \{O\textsc{i}/H\textsc{i}\} measurements with small ionization corrections and the three \{Si\textsc{ii}/H\textsc{i}\} metallicity limits are shown with red symbols. 
Two of these measurements, represented by the red square and the red star, are lower limits with \{O\textsc{i}/H\textsc{i}\} $>$ $-1.08$ and $-1.45$ dex, respectively. The four other sightlines with \{O\textsc{i}/H\textsc{i}\} constraints have metallicities ranging from $-0.75$ dex to at least as low as $-2.09$ dex. A comparison of these  measurements with CGM metallicities reported in the literature will be discussed in more detail in Section \ref{sec:discuss}.

\begin{figure*}[ht]
	\centering
	\begin{minipage}{\linewidth}
		\centering
		\includegraphics[width=0.47\linewidth]{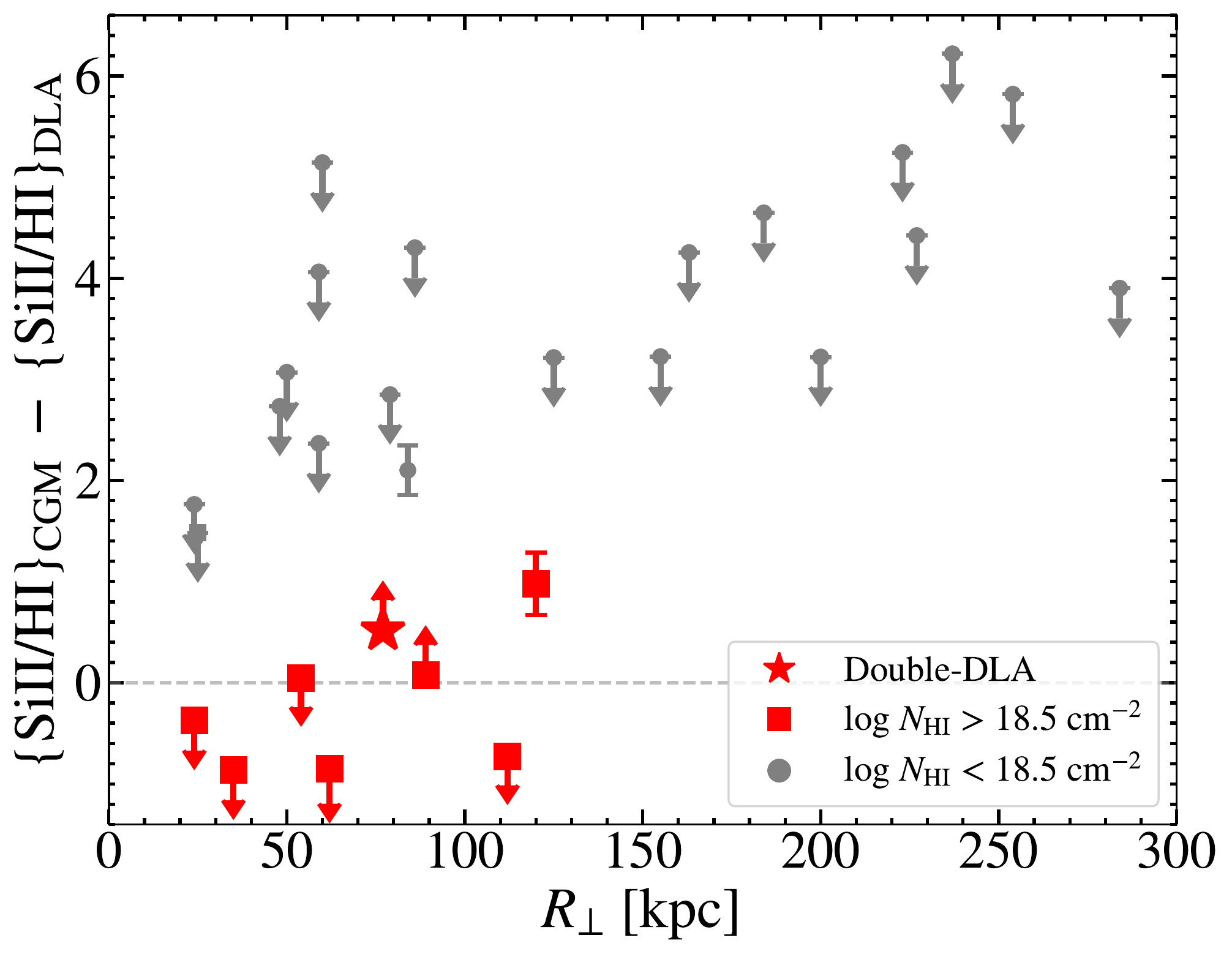} \hspace{8mm}
		\includegraphics[width=0.47\linewidth]{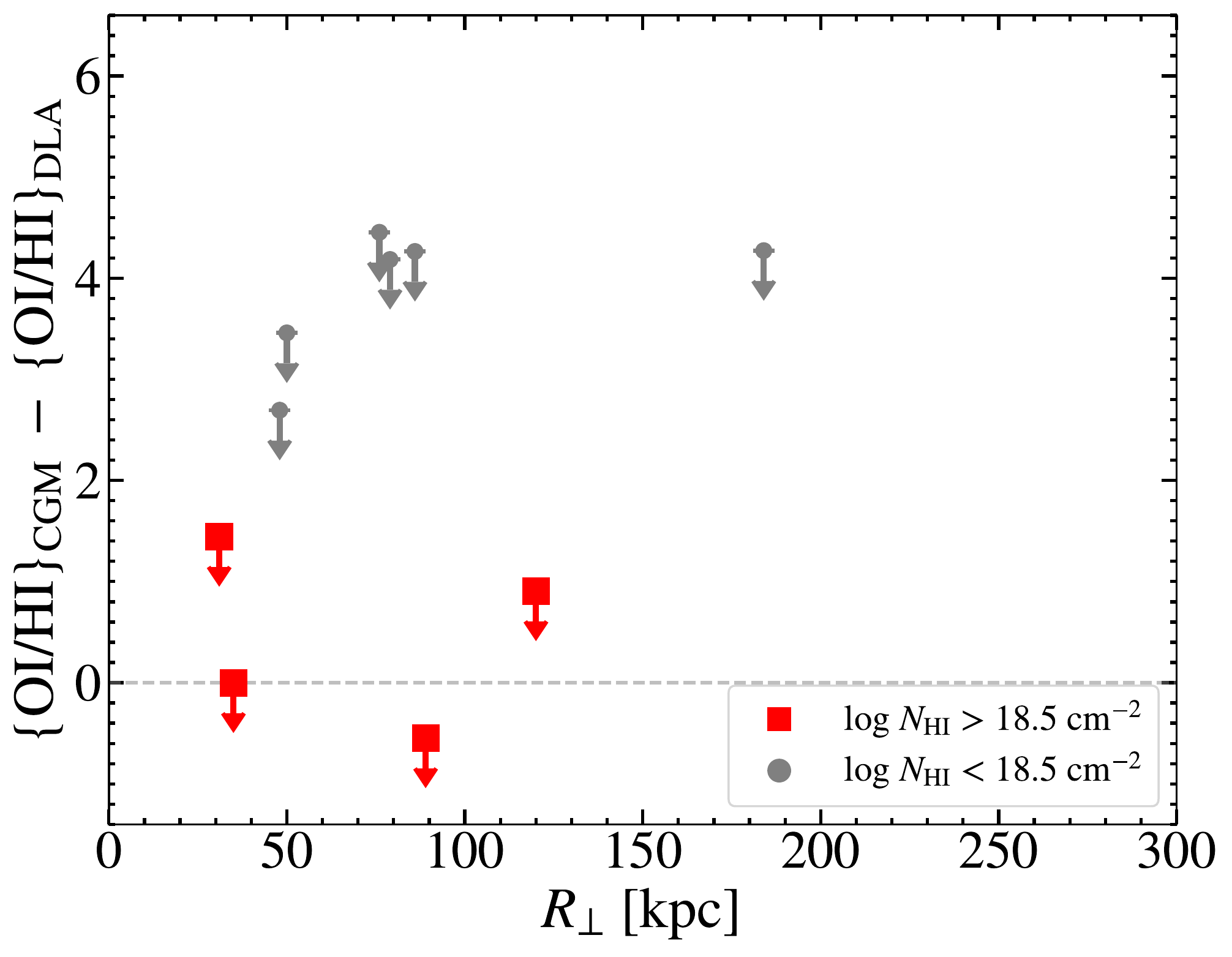}
		\caption{Comparison of metallicity constraints in our DLAs and the associated CGM as a function of $R_\bot$. Offsets that rely on 
  ionic ratio measurements for CGM systems with $\nHI > 10^{18.5}~\rm cm^{-2}$ are represented by red symbols. Metallicity offsets shown with gray circles are overestimates, as our ionic ratios likely overestimate [X/H] for the CGM in these systems. The double-DLA system is shown with a red star. All systems for which the CGM metallicity estimate is lower than that of the DLA have optically thick CGM gas. There are four systems for which we can say with confidence that \{Si\textsc{ii}/H\textsc{i}\}$_{\rm CGM} - $\{Si\textsc{ii}/H\textsc{i}\}$_{\rm DLA} < 0$ dex and two systems for which \{O\textsc{i}/H\textsc{i}\}$_{\rm CGM} - $\{O\textsc{i}/H\textsc{i}\}$_{\rm DLA}$ is likely $ < 0$ dex. 
		\label{fig:metalabundcomp}}
	\end{minipage}
\end{figure*}

We also compare our CGM metallicities  to the metallicities measured in the corresponding DLA sightlines as a function of $R_{\perp}$. For this analysis, we subtract the \{X$i$/H\textsc{i}\} measured in each DLA sightline from the same ionic ratio measured in its CGM (shown in Figure \ref{fig:metalabundcomp}). We find the majority of the points are upper limits, due to the preponderance of upper limits on $\nSiII$ and $\nOI$ in our CGM sightline sample, and are likely overestimates given the unknown CGM ionization correction. 

We comment here on a small subset of these sightline pairs that yield interesting constraints on the relative levels of enrichment in the DLA vs.\ CGM sightlines. There are five systems with \{X$i$/H\textsc{i}\}$_{\rm CGM} - $\{X$i$/H\textsc{i}\}$_{\rm DLA}$ values less than $0$ dex (four with \{Si\textsc{ii}/H\textsc{i}\}$_{\rm CGM}$ $-$ \{Si\textsc{ii}/H\textsc{i}\}$_{\rm DLA} < 0$, one of which has both \{Si\textsc{ii}/H\textsc{i}\}$_{\rm CGM}$ $-$ \{Si\textsc{ii}/H\textsc{i}\}$_{\rm DLA}$ and \{O\textsc{i}/H\textsc{i}\}$_{\rm CGM} - $\{O\textsc{i}/H\textsc{i}\}$_{\rm DLA} < 0$, and an additional sightline with \{O\textsc{i}/H\textsc{i}\}$_{\rm CGM} - $\{O\textsc{i}/H\textsc{i}\}$_{\rm DLA} < 0$).  Each of these systems has optically thick gas in the CGM sightline ($ N_\text{H\textsc{i},\text{CGM}} > 10^{18.55} \ \text{cm}^{-2}$), and all occur within $R_{\perp} < 120$ kpc. In these five cases, we may state with at least $\sim 1\sigma$ confidence that the metallicities in the CGM sightlines are lower than those in the corresponding DLAs by at least $-0.2$ dex. 

There is one sightline with a robust measurement of \{Si\textsc{ii}/H\textsc{i}\}$_{\rm CGM} - $\{Si\textsc{ii}/H\textsc{i}\}$_{\rm DLA} > 0.5$ dex. This sightline is the double-DLA, and therefore may probe a different environment than is typical of the other CGM sightlines in our sample. Nonetheless, it is the only system where we are certain the metallicity in the sightline with the lower $N_\text{H\textsc{i}}$ is higher than in the so-called host DLA.

Taken together, these results point to a significant degree of scatter in the level of enrichment in the CGM at $24~\mathrm{kpc} < R_\bot < 120$ kpc relative to that in the DLA gas in the corresponding galaxy host.
 
\subsubsection{Investigation of $\alpha$/Fe Ratios in DLAs and their Halos}\label{secsub:alpharatios}
In the above section we demonstrated that the CGM around DLAs has a wide range of metal enrichment, with robustly-estimated metallicities ranging from as high as $-0.75$ dex to at least as low as $-2.09$ dex. We expect this metal content was originally formed in the interiors of stars and ejected via supernovae (SNe). The comparison of the abundance of $\alpha$ elements to that of Fe is useful in dissecting the specific nucleosynthetic processes that ultimately produced this enriched gas. $\alpha$ elements are produced in massive stars and are ejected by Type II SNe, a process which happens on relatively short timescales (10$^{6-7}$ years). On the other hand, Fe is produced in both Type II and Type Ia SNe. Type Ia SNe occur on longer timescales, on the order of 10$^{8-9}$ years (\citealt{Kobayashi_2009}). Once Type Ia SNe begin within a stellar population, the overall $\alpha$/Fe of the surrounding gas will decrease \citep{Tinsley_1979,Matteucci_2001}. An intermediate-redshift ($0.1 < z < 1.24$) study from \citet{Zahedy_2016} measured the $\alpha$/Fe ratio in halo gas close to galaxies ($R_\bot < 60$ kpc), and showed increased $\alpha$-enrichment in star-forming galaxy halos ($\alpha$/Fe $= 0.25\pm0.21$ dex) compared to that of quiescent galaxy halos ($\alpha$/Fe $= 0.06\pm0.15$ dex). They also found that the $\alpha$-enrichment increased at larger distances ($R_\bot > 60$ kpc), measuring $\alpha$/Fe $= 0.9\pm0.4$ dex around star-forming galaxies and $\alpha$/Fe $>0.3$ dex around quiescent galaxies at these distances. They concluded the higher $\alpha$-enrichment around star-forming galaxies is a consequence of the presence of young star-forming regions, whereas the higher $\alpha$-enrichment in the outer halos of quiescent galaxies is suggestive of core-collapse dominated enrichment histories. This work thus successfully uses the $\alpha$/Fe ratio measured in CGM material to trace differences in the stellar populations dominating its enrichment. 

\begin{figure}[ht]
	\centering
	\begin{minipage}{\linewidth}
		\centering
		\includegraphics[width=\linewidth]{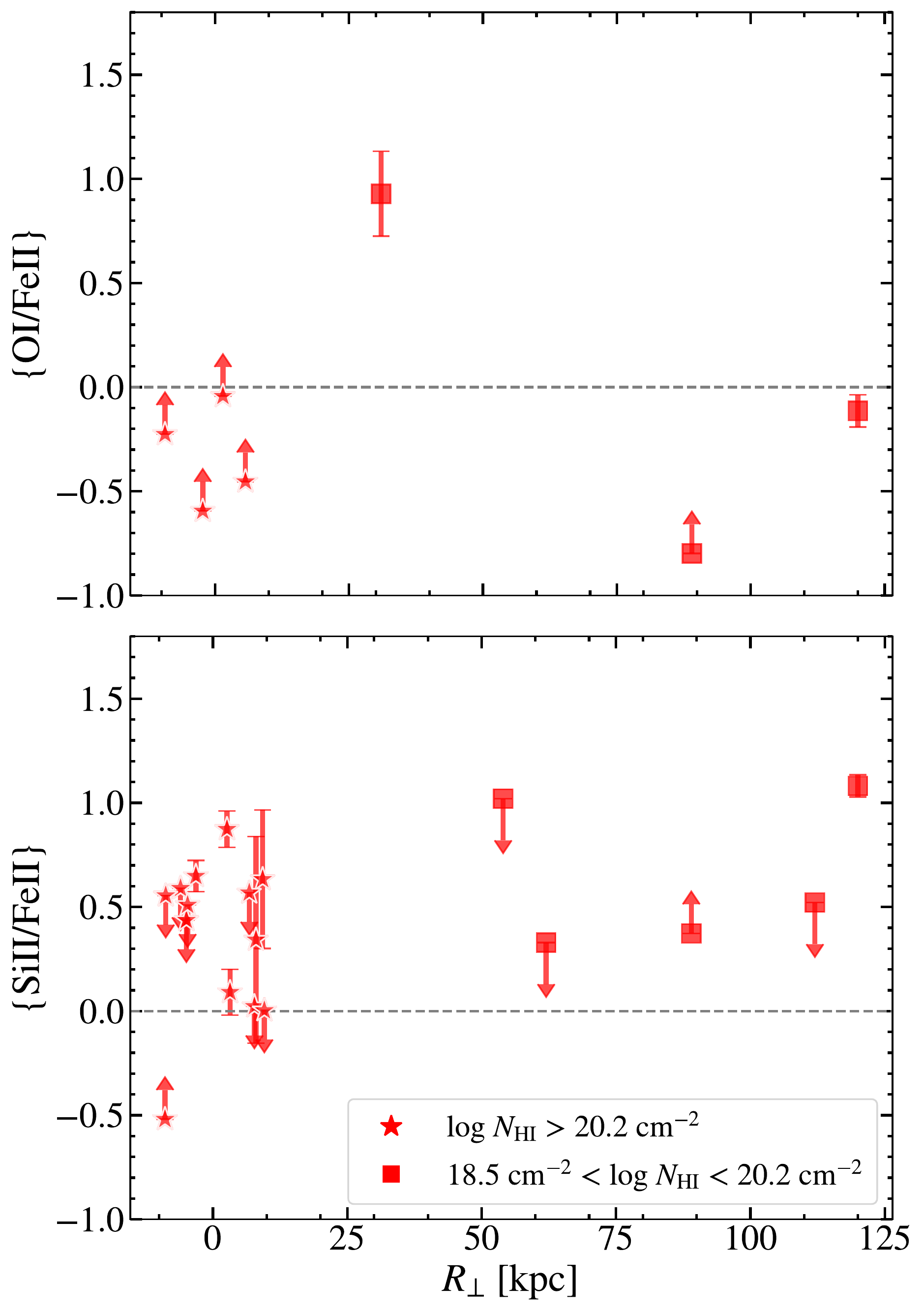}
		\caption{\{O\textsc{i}/Fe\textsc{ii}\} (top) and \{Si\textsc{ii}/Fe\textsc{ii}\} (bottom) vs. $R_\bot$ for sightlines with \{X/H\textsc{i}\} $< -1$ dex. DLAs are represented by stars and LLSs are shown as squares.
		\label{fig:alpharatios}}
	\end{minipage}
\end{figure}

When measuring the $\alpha/$Fe ratios in our sample, we must consider the depletion of Fe due to dust. This depletion scales with metallicity and therefore has a larger impact on the measured $\alpha/$Fe ratio in higher metallicity systems. A study of $\alpha$-enrichment in a larger sample of DLAs found a positive correlation between $\alpha/$Fe and metallicity for higher metallicity systems ([X/H] $> -1$ dex), suggesting the depletion of Fe due to dust makes the $\alpha/$Fe measurement unreliable \citep{Rafelski_2012_755}. We therefore limit this analysis to only include sightlines in which we robustly measure a metallicity $< -1$ dex. 

Figure \ref{fig:alpharatios} shows two ionic ratios that trace $\alpha/$Fe: \{O\textsc{i}/Fe\textsc{ii}\} (top panel) and  \{Si\textsc{ii}/Fe\textsc{ii}\} (bottom panel). 
Higher levels of ionization tend to elevate \{Si\textsc{ii}/Fe\textsc{ii}\} ratios, so our values may be overestimates of [Si/Fe] for our CGM sightlines.
Conversely, \{O\textsc{i}/Fe\textsc{ii}\} is not very sensitive to ionization corrections; however, most sightlines yield lower limits on this quantity because \ion{O}{1} 1302 is saturated in high-$N_\text{HI}$ sightlines, or because \ion{Fe}{2} is typically not securely detected for lower-$N_\text{HI}$ sightlines. 

From our constraints on \{O\textsc{i}/Fe\textsc{ii}\}, we find that at least one CGM sightline is $\alpha$-enriched. There is also one CGM sightline at $R_{\perp} = 120$ kpc for which our constraints imply that the system is not $\alpha$-enriched, with \{O\textsc{i}/Fe\textsc{ii}\} = $-0.11 \pm 0.08$ dex. This sightline probes high-metallicity gas at $ Z \approx 0.1 Z_\odot$. The kinematics of this gas (discussed in more detail in Section \ref{sec:kinematics}) are relatively quiescent, with a $\Delta v_{90}$ width = 87 $\mkms$ for low-ionization material. Overall from \{O\textsc{i}/Fe\textsc{ii}\}, we find the $\alpha$-enrichment of this population of DLAs and their halos is ambiguous due to the small sample size and preponderance of limits. 

\{Si\textsc{ii}/Fe\textsc{ii}\} ratios yield more detections than \{O\textsc{i}/Fe\textsc{ii}\}, but are overestimates for sightlines with significant ionization. The majority of our CGM sightlines are ionized (implying large ionization corrections to \{Si\textsc{ii}/Fe\textsc{ii}\}), and therefore we focus here on the DLA population (red points in the bottom panel of Figure \ref{fig:alpharatios}). For the low-metallicity DLA population, we find that at least six sightlines are $\alpha$-enriched, with a median value of \{Si\textsc{ii}/Fe\textsc{ii}\} $= 0.52$ dex among the six detections. This is larger than what was measured in \cite{Rafelski_2012_755}. In their examination of DLA abundances at $z > 1.5$, they likewise found low-metallicity DLAs to be mostly $\alpha$-enriched, however they reported a mean value of [$\alpha$/Fe] in their low-metallicity ([X/H] $< - 1$ dex) DLA sample of 0.27$\pm 0.02$ dex. 

Together, these results confirm that (1) our low-metallicity DLA population is mostly $\alpha$-enriched with a median value of \{Si\textsc{ii}/Fe\textsc{ii}\} = 0.52 dex; and (2) for CGM sightlines in which we robustly measure $\alpha/$Fe, we find one sightline is $\alpha$-enriched and one has an abundance ratio near solar.  However, due to uncertainties in both ionization state and the degree of dust depletion across our sample, we cannot comment more generally on the $\alpha$-enrichment of the CGM of DLAs. 

\subsection{Kinematics}\label{sec:kinematics}

\begin{figure*}[ht]
	\centering
	\begin{minipage}{\linewidth}
		\centering
		\includegraphics[width=0.48\linewidth]{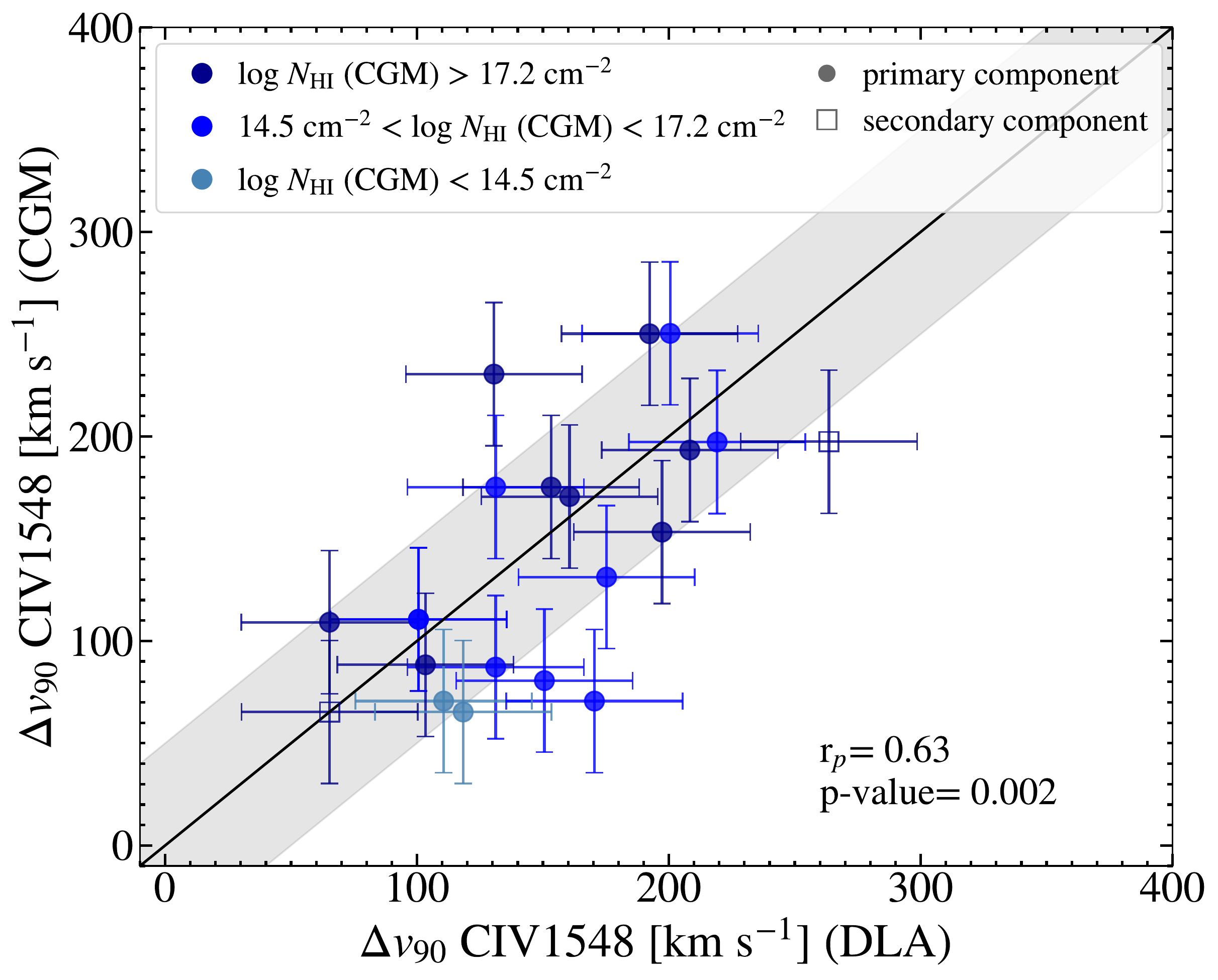} \includegraphics[width=0.48\linewidth]{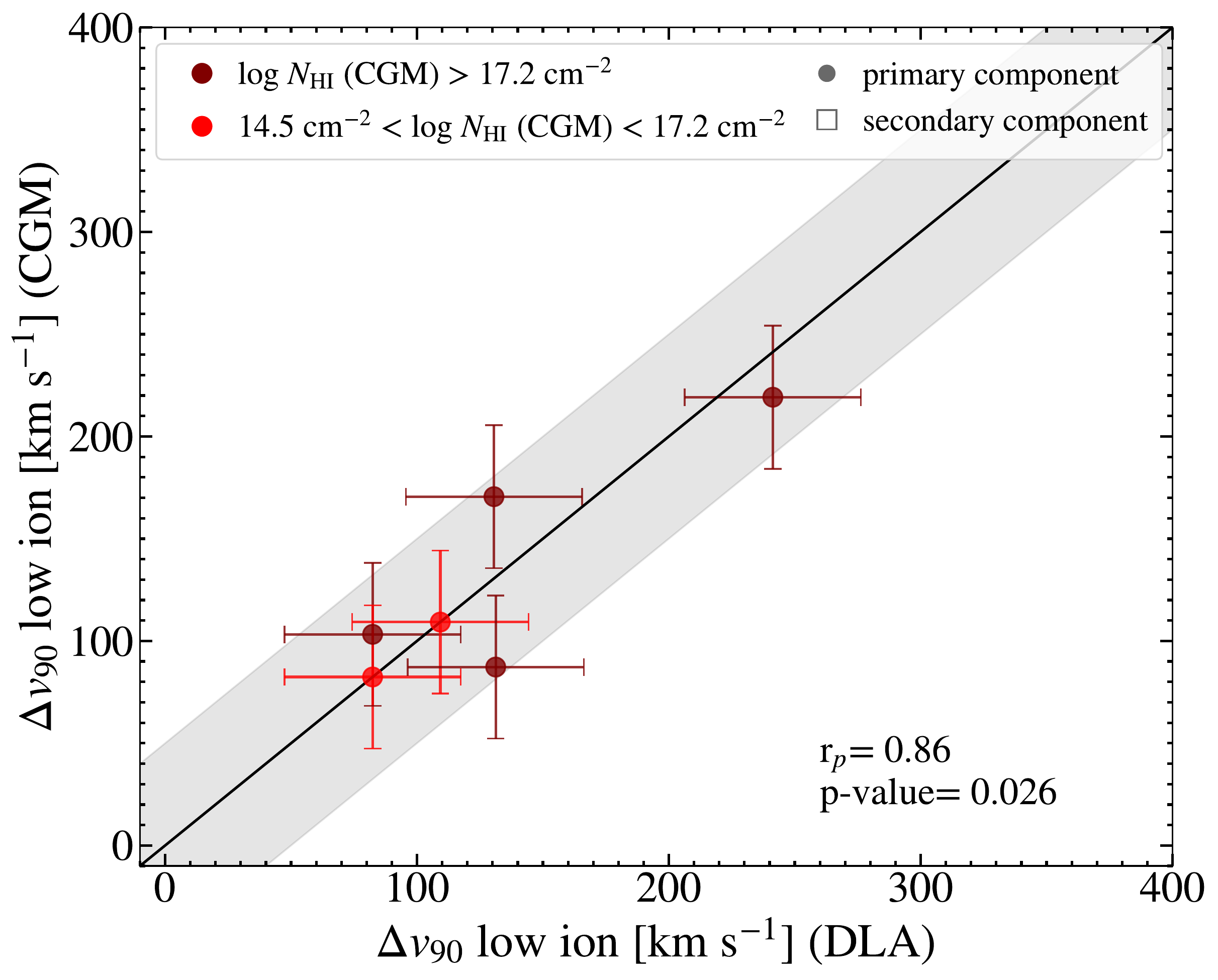} 
		\caption{Comparison of $\Delta v_{90}$  velocity widths for the DLA and CGM sightlines measured from \ion{C}{4} 1548 (left) and from an unsaturated low-ionization transition (right). Open squares represent secondary velocity components. The black lines show a 1:1 relation, and the gray bars show $\pm$50 $\mkms$ offsets from this relation in the $x$- and $y$- directions. The color of each point is indicative of the \ion{H}{1} column density in the corresponding CGM sightline, as shown in the legend. The $\Delta v_{90}$  velocity widths measured from both \ion{C}{4} and low-ionization transitions closely follow a 1:1 relation.  The statistical significance of the correlations between these quantities is indicated with the correlation coefficient (r$_p$) and the P-value in the bottom right of each plot.
		\label{fig:dv_compare}}
	\end{minipage}
\end{figure*}

In this section, we investigate the kinematics of high- and low-ionization gas surrounding DLAs. By necessity, this analysis is limited to sightlines with significantly-detected metal-line absorption profiles. As a result, we caution that our conclusions will be biased toward those systems with significant metal content. In many CGM sightlines, there are few detected metal absorption lines, so we choose the strongest transitions as follows: to trace high-ionization gas, we choose \ion{C}{4} $\lambda1548$ for its high oscillator strength; and to trace low-ionization gas, we choose a low-ion transition with the highest S/N at the peak of the optical depth profile (i.e., the profile with the highest value of $\tau_{\rm peak}/\langle \sigma_\tau \rangle$). We make use of two quantities (fully described in Section \ref{sec:methodskin}): the flux-weighted velocity centroids ($\delta v_\text{weight}$) measured relative to the DLA redshifts, and the velocity widths measured between the locations where the cumulative optical depth profile reaches 5$\%$ of the total integrated optical depth on either side ($\Delta v_{90}$). For some sightlines, the \ion{C}{4} $\lambda 1548$ transition is affected by saturation, and in these cases the associated $\Delta v_{90}$ will likely overestimate the width of 90$\%$ of the total line optical depth to some degree. However, because the column densities of \ion{C}{4} in our DLA vs.\ CGM sightlines have overall similar values at $R_\bot < 200$ kpc (as shown in Appendix Figure~\ref{fig:Rpropall}), we posit that saturation effects should not systematically impact our measured DLA velocity widths more than our CGM velocity widths (or vice versa). We adopt uncertainties on these quantities as described in Section~\ref{subsubsec:uncertainties_kinematics}.

First, we investigate the difference in the velocity widths for high- and low-ionization gas between DLAs and the corresponding CGM sightlines. We show these results in Figure \ref{fig:dv_compare}. The colors in Figure \ref{fig:dv_compare} represent different bins of CGM \ion{H}{1} column density. We include secondary components as squares where there is a detection in both the CGM and DLA sightline. We also include gray bars to show an offset of $\pm 50 \ \mkms$, representing the uncertainty in the $x$- and $y$- directions (35 $\mkms$ uncertainty for our $\Delta v_{90}$ values) added in quadrature. For both high- and low-ionization material, this comparison reveals clear correlations between the DLA and CGM line widths. We perform a Pearson rank correlation test on the two datasets to quantify the strength and significance of a linear correlation of these quantities. For \ion{C}{4} $\lambda 1548$, we find a Pearson correlation coefficient (r$_p$) of 0.63 with a P-value of 0.2$\%$, indicative of a very low probability that these quantities are uncorrelated. For our low-ion $\Delta v_{90}$ measurements, we find r$_p = 0.86$, indicative of a close to 1:1 relation, with a P-value of 2.6$\%$. While these results are suggestive of strong correlations in both cases, we also note a larger degree of scatter in the $\Delta v_{90}$ measurements of \ion{C}{4} $\lambda 1548$: 35$\%$ of our sightline pairs have $\Delta v_{90}$ values that differ by more than 50 $\mkms$, larger than the uncertainty associated with our $\Delta v_{90}$ measurements ($35 \ \mkms$). For low-ionization gas, none of the $\Delta v_{90}$ values in our sightline pairs differ by more than 50 $\mkms$. This distinction may reflect a larger degree of variation in the kinematics of high-ionization gas in nearby sightlines, or may be driven by saturation effects in our $\Delta v_{90}$(\ion{C}{4}) values. We note that systems with differing values of CGM \ion{H}{1} column density appear to yield a consistent level of scatter in these quantities. Overall, these results reveal a close correspondence in velocity widths over $24-237$ kpc scales. This in turn suggests that these $\Delta v_{90}$ widths are a consistent tracer of the potential well of the host halo, regardless of the impact parameter of the sightline. 

Both \citet{Christensen_2019} and \citet{Moller_2020} performed a close examination of the $\Delta v_{90}$ values measured for DLAs as a function of the projected distance from their host galaxies (identified in emission).  In particular, \citet{Moller_2020} measured a gradient of $-0.017~\rm dex~kpc^{-1}$ in the quantity $\log \Delta v_{90}/\sigma_{\rm em}$ over an impact parameter range $0~\mathrm{kpc} < R_{\perp} < 60~\mathrm{kpc}$, with $\sigma_{\rm em}$ equal to the velocity width of strong emission lines.  These authors demonstrated that this trend is consistent with the projected velocity dispersion profile predicted for a \citet{Dehnen1993} dark matter halo potential model.  Moreover, they pointed out that this latter profile flattens at impact parameters $R_{\perp} > 60$ kpc.  Our finding of a close correspondence between $\Delta v_{90}$ values over scales of $\gtrsim 100$ kpc is fully consistent with this prediction, and may be viewed as further confirmation of the interpretation of $\Delta v_{90}$ as an effective measure of halo dynamics.

\begin{figure*}[hb]
	\centering
	\begin{minipage}{\linewidth}
		\centering
		\includegraphics[width=0.35\linewidth]{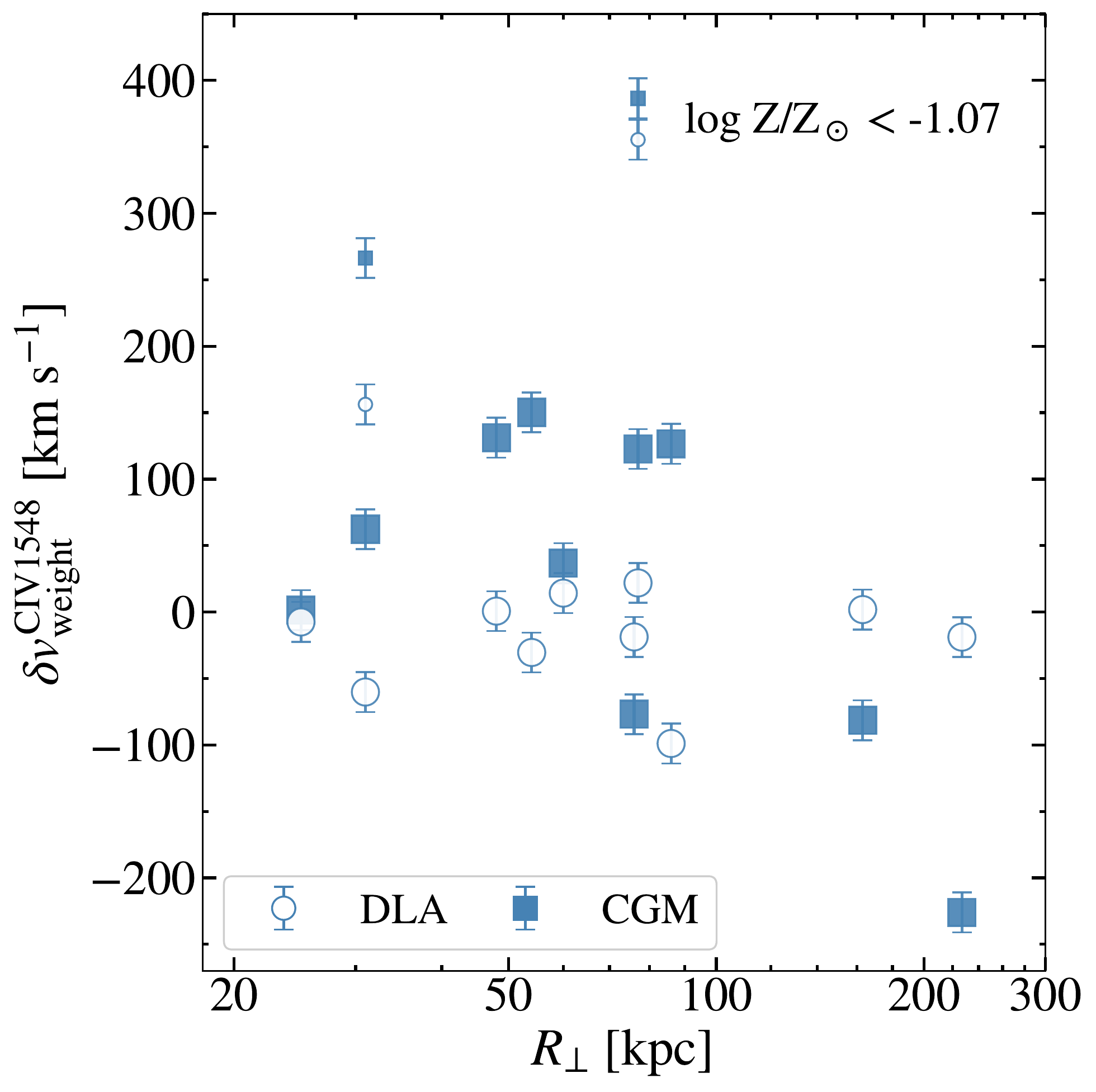}
		\includegraphics[width=0.35\linewidth]{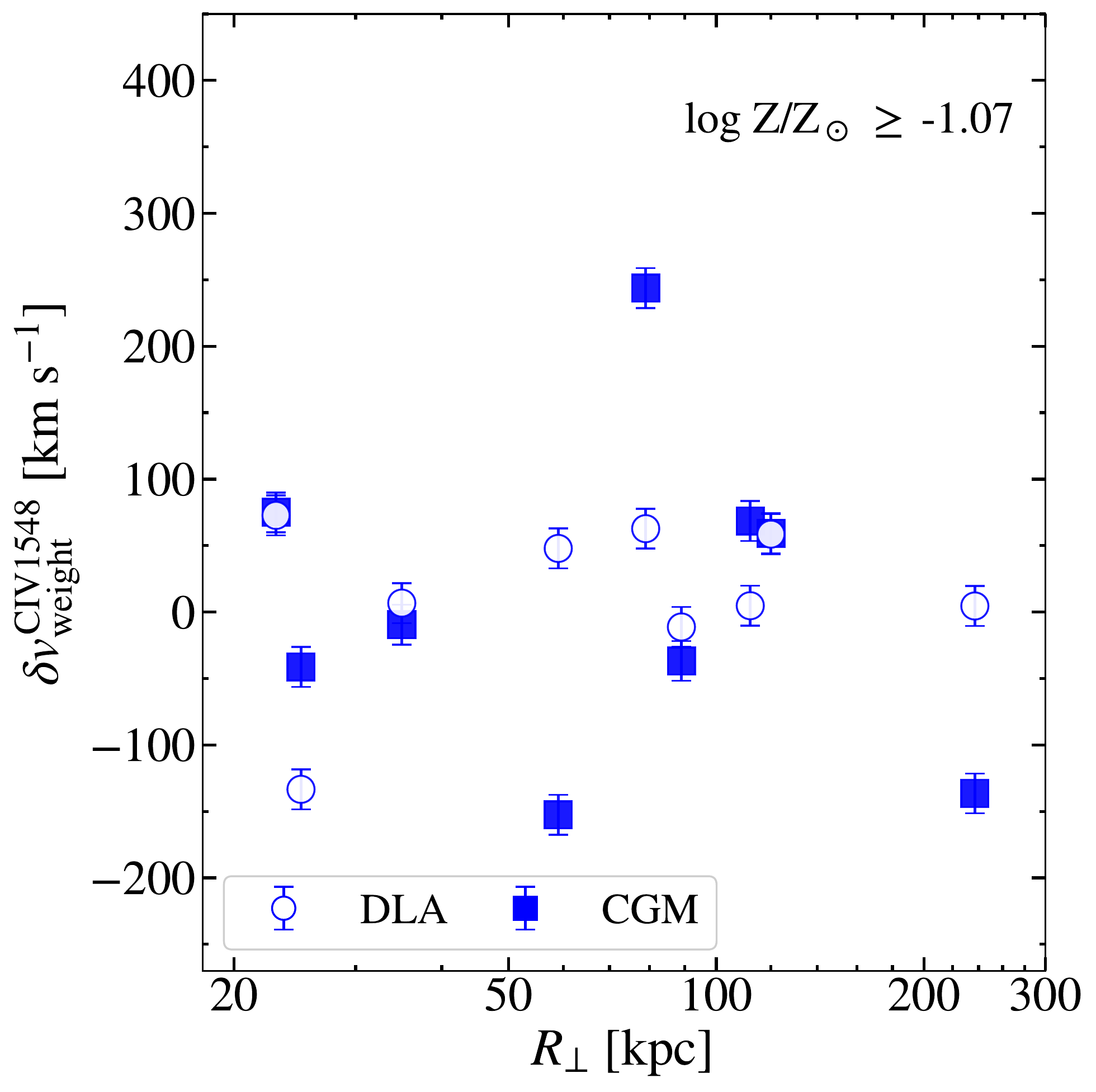}
         \includegraphics[width=0.262\linewidth]{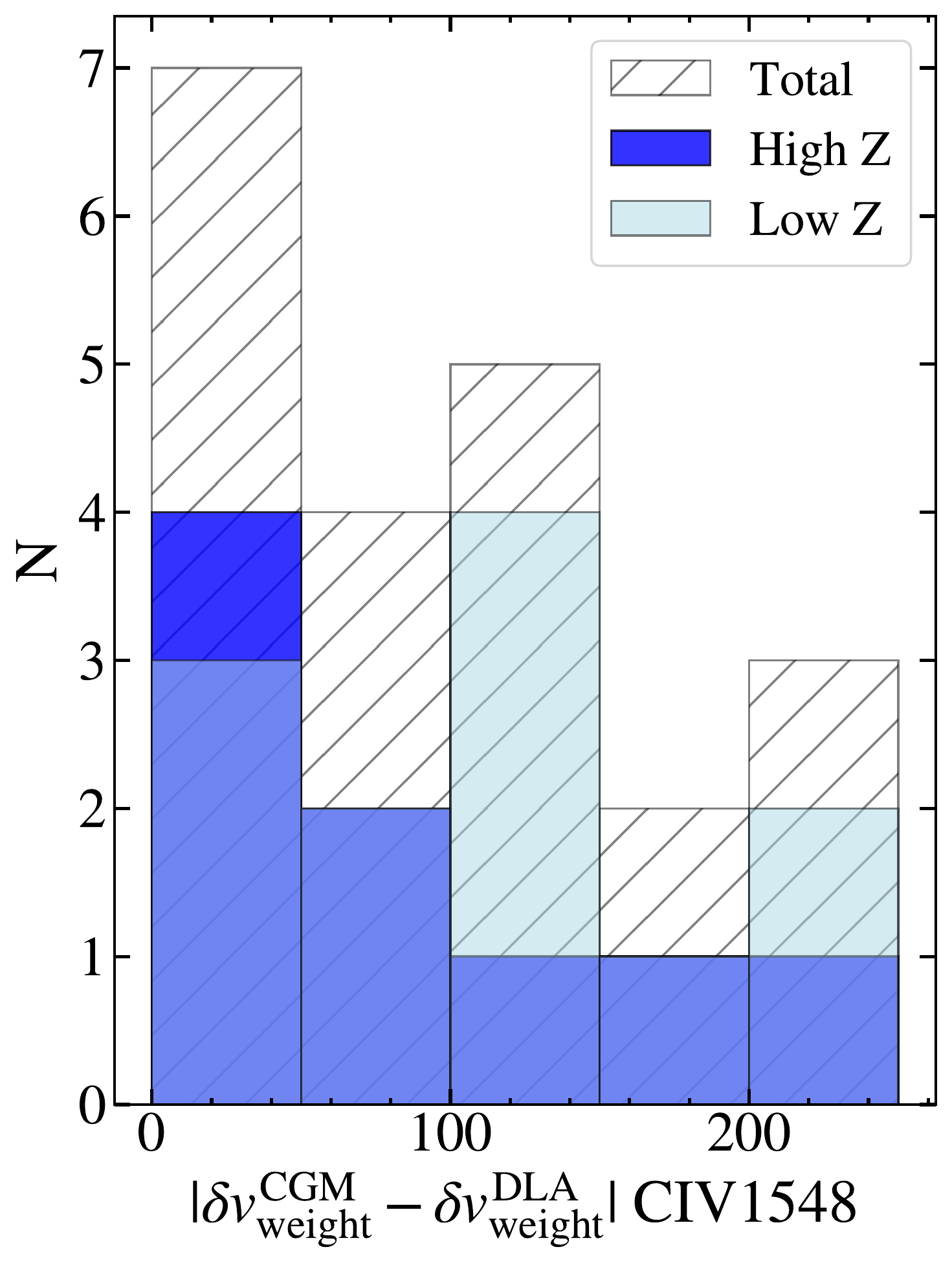} 
		\caption{Flux-weighted velocity centroids for \ion{C}{4} ($\delta v_\text{weight}^{\text{CIV} \lambda1548}$) for our DLA (open circles) and CGM (filled squares) sightlines plotted vs.\ $R_{\perp}$ for each sightline pair. For direct comparison between the CGM and DLA $\delta v_\text{weight}^{\text{CIV} \lambda1548}$, the DLA measurements are placed at the same $R_{\perp}$ as the corresponding CGM sightline. Smaller symbols indicate secondary components. The left-hand panel shows systems with DLA metallicities $\log Z/Z_\odot < -1.07$, while the middle panel includes systems with higher DLA metallicities. The right-most panel shows a histogram of the difference between $\delta v_\text{weight}^{\text{CIV} \lambda1548}$ measured for each DLA and CGM sightline pair with a bin width of 40 $\mkms$. The hatched histogram shows the full sample; the dark blue shading shows the high-metallicity sample; and the light blue shading shows the low-metallicity sample. We find all systems have a $\delta v_\text{weight}^{\text{CIV} \lambda1548}$ difference of ${<}225 ~\rm \mkms$, and 52$\%$ have a difference of ${<}100 ~\rm \mkms$.
			\label{fig:vweight_compare}}
	\end{minipage}
\end{figure*}

\begin{figure*}[hb]
	\centering
	\begin{minipage}{\linewidth}
		\centering
		\includegraphics[width=0.48\linewidth]{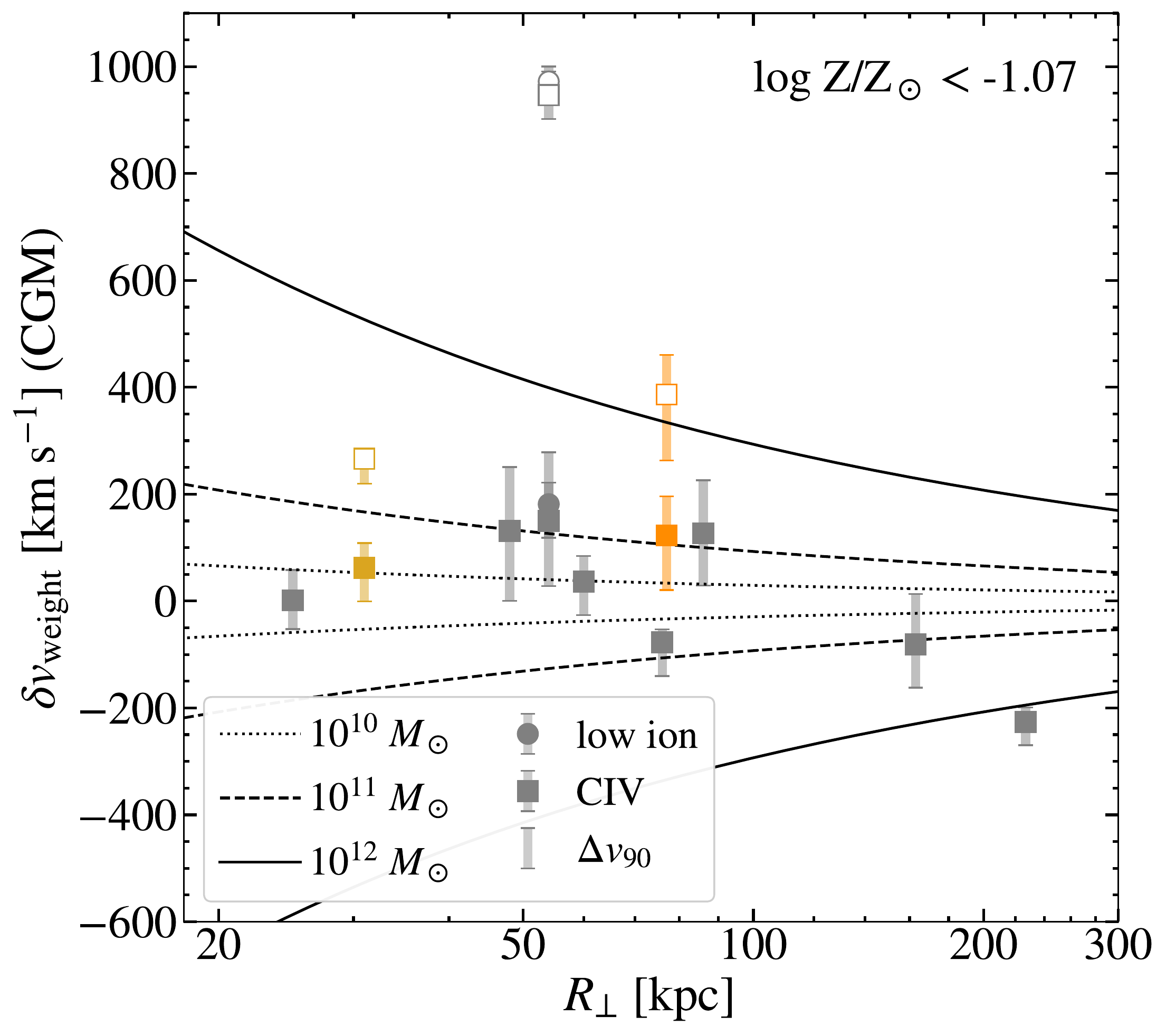} \includegraphics[width=0.48\linewidth]{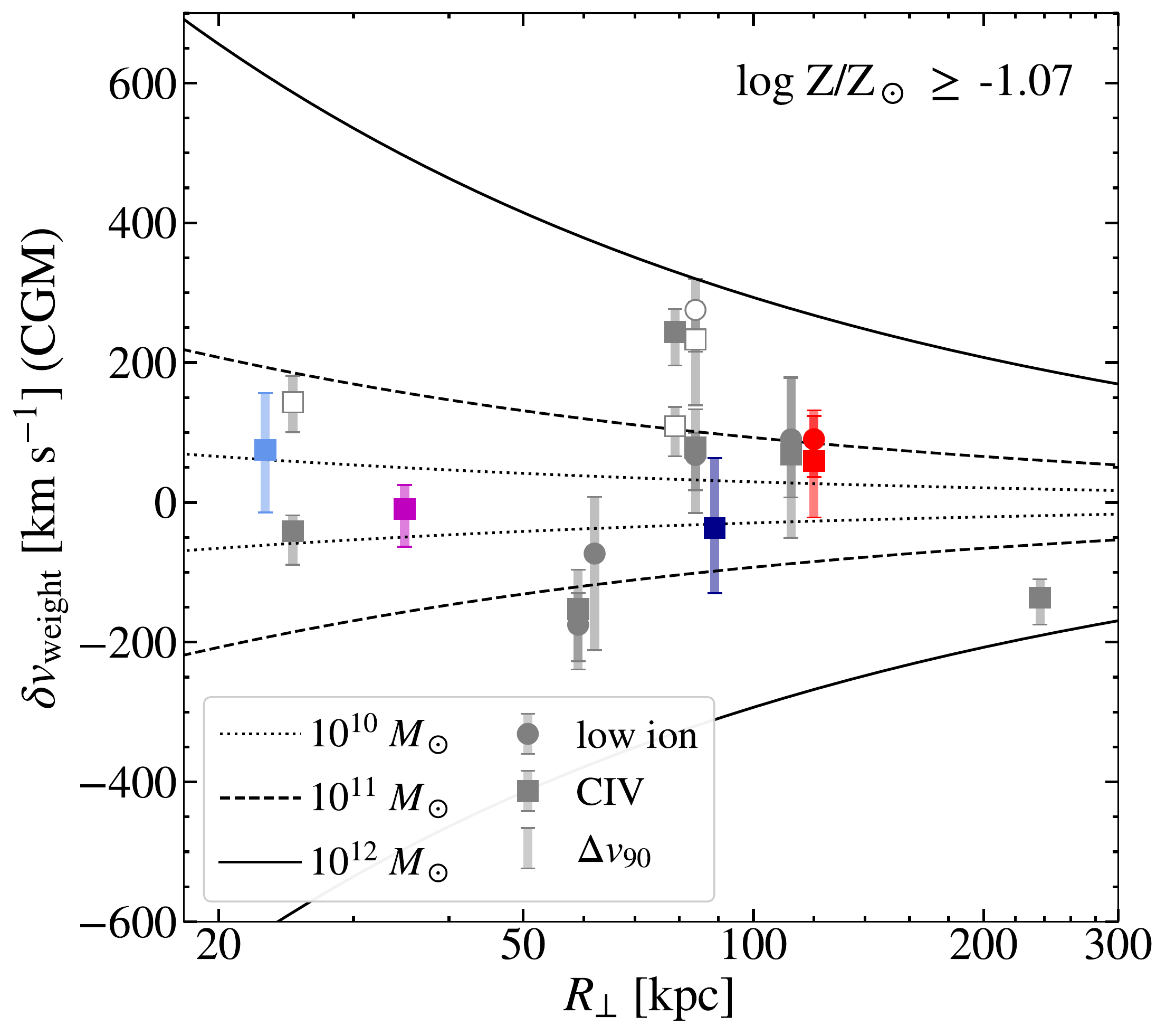} 
		\caption{Flux-weighted velocity centroids for low-ions (circles) and \ion{C}{4} (squares) in our CGM sightlines vs. $R_\bot$ for systems with low-metallicity DLAs (left) and high-metallicity DLAs (right). The error bars here represent the velocity range encompassed by the $\Delta v_{90}$ width relative to the component line center, and therefore show the approximate velocity range exhibited by each absorption component. Primary velocity components are shown with filled markers, and secondary velocity components are shown with open markers. Colors highlight systems in which we have robust metallicity constraints mentioned in Section \ref{sec:metallicity}, which will be used in discussion later in Section \ref{sec:discuss}. The double-DLA system is represented by the orange points in the left panel. Curves represent radial escape velocities for halos with masses 10$^{10} M_\odot$ (dotted lines), 10$^{11}  M_\odot$ (dashed lines), and 10$^{12} M_\odot$ (solid lines). Most points lie within these bounds for halos of 10$^{11-12}$ $M_\odot$, suggesting that if DLAs reside in such massive halos, the 
		bulk of the absorbing gas traced by both \ion{C}{4} and low-ionization lines will remain bound.}
		\label{fig:vesc}
	\end{minipage}
\end{figure*}

In our previous work (\citealt{Rubin_2015_808}), we compared the $\delta v_\text{weight}$ values for \ion{C}{4} $\lambda1548$ for 12 DLA-CGM sightline pairs, 8 with medium resolution spectroscopy ($\mathcal{R} \gtrsim 4000$) and 4 with low resolution spectroscopy ($\mathcal{R} \lesssim 2400$). We found that the differences in $\delta v_\text{weight}^{\text{CIV} \lambda1548}$ did not exceed 105 $\mkms$ across the full sample, which included sightlines with separations up to $R_\bot = 176$ kpc. We interpreted this finding as suggestive of strong coherence in \ion{C}{4} absorption over scales of ${>}100 ~\rm kpc$. Here we expand on this analysis with a larger sample of medium- and high-resolution spectroscopy (drawing on 21 absorber pairs, including 19 primary and 2 secondary components). The results are shown in Figure \ref{fig:vweight_compare}. We split the sample into two bins of DLA metallicity at the median metallicity of the DLA sightlines ($\log Z/Z_\odot = -1.07$). The high-metallicity DLAs are more likely to trace more massive halos (with $10^{11.5} M_\odot \lesssim M_h \lesssim 10^{12} M_\odot$) with SFR $\gtrsim 1~M_\odot ~\rm yr^{-1}$ (\citealt{Krogager2017}), while lower-metallicity DLAs are associated with lower SFRs and halo masses ($10^{10} M_\odot \lesssim M_h \lesssim 10^{11.5} M_\odot$; e.g., \citealt{Bird_2014_445}). We highlight the differences in $\delta v_\text{weight}^{\text{CIV} \lambda1548}$ in our CGM vs.\ DLA sightlines as a function of DLA metallicity in the right-most panel in Figure \ref{fig:vweight_compare}.

Overall, we find no evidence for a correlation between the difference in $\delta v_\text{weight}^{\text{CIV} \lambda 1548}$ for the DLA and CGM sightlines and metallicity, separation between the sightlines ($R_\bot$), \ion{C}{4} column density, or \ion{H}{1} column density. In addition, this sample yields  larger differences in $\delta v_\text{weight}^{\text{CIV} \lambda1548}$ than that analyzed in  \cite{Rubin_2015_808}. We find that 52$\%$ of these sightline pairs have $\delta v_\text{weight}^{\text{CIV} \lambda1548}$ values that differ by ${\leq} 100~\mkms$. Moreover, 86\% of our pairs yield differences in $\delta v_\text{weight}^{\text{CIV} \lambda1548}$ of ${\leq} 200 \ \mkms$. There are three pairs that have differences larger than 200 $\mkms$. This expanded sample shows clear evidence that the velocities of \ion{C}{4} in the outer halos of DLA hosts are frequently more than $100\ \mkms$ different from that measured in the inner CGM. We use the relations given in \citet{Maller&Bullock_2004_355} to estimate the virial velocities of the DLA host halos (assuming they have halo masses $M_h \sim 10^{11-12} M_\odot$ at $z_\text{abs} \sim 2.45$), finding that they span the range $100-216 \ \mkms$. This suggests that our measured  differences in  $\delta v_\text{weight}^{\text{CIV} \lambda1548}$ are less than or approaching the virial velocity of the host halos for ${\sim} 52\%$ of sightlines. Overall, with access to a larger sample, we show there is not a strong coherence (${\lesssim}100\ \mkms$) in \ion{C}{4} velocity centroids over large scales as seen in our previous work.  

Finally, to investigate the possibility of CGM gas escaping the DLA host halos, we compare our measurements of $\delta v_\text{weight}$ and $\Delta v_{90}$ to the radial escape velocities ($v_{\rm esc} = \sqrt{G M_h /R}$, with $R=R_{\perp}$) of halos with three different values of total mass ($M_h =$ 10$^{10} M_\odot$, 10$^{11} M_\odot$, and 10$^{12} M_\odot$). The results are shown in Figure~\ref{fig:vesc}. The error bars above and below $\delta v_\text{weight}$ represent the velocities encompassed by the $\Delta v_{90}$ interval relative to the line center. Low-ionization gas is represented by circles, and high-ionization gas traced by \ion{C}{4} $\lambda1548$ is represented by squares. Primary and secondary velocity components are shown with filled and open markers, respectively. As we did in the above analysis, we divide our sample by the median DLA metallicity, to differentiate between sightlines that likely trace lower-mass halos ($\lesssim 10^{11.5} M_\odot$) and those that are more likely to probe higher-mass halos ($\gtrsim 10^{11.5} M_\odot$).

Before interpreting these results, we consider two caveats. First, we note that for any given sightline pair, the $R_\bot$ we measure does not necessarily reflect the true projected distance of the CGM sightline from the center of its host halo, as DLAs do not always lie at the centers of their hosts. However, the difference between our measured $R_{\perp}$ and the true $R_{\perp}$ is likely small, as observational and theoretical studies typically measure DLA-galaxy separations to be $< 25$ kpc (\citealt{Krogager2017}). Second, we caution that our analysis assesses velocities along the line of sight, rather than the total radial velocity of gas measured with respect to each halo's center. Our velocities should therefore be interpreted as lower limits on this latter quantity. 

Within 100 kpc of the DLAs, we frequently detect velocity components in both high- and low-ionization absorption profiles that have a $\delta v_\text{weight}$ which exceeds the escape velocity for halos with $M_h \le$ 10$^{11} M_\odot$. Moreover, there are two secondary velocity components detected that have a $\delta v_\text{weight}$ exceeding the escape velocity of a halo with $M_h =$ 10$^{12} M_\odot$. One of these components is detected in the double-DLA sightline, shown in orange in Figure \ref{fig:vesc}, and therefore may trace a different CGM environment, possibly a different halo, from that of the typical isolated DLA. The other high-velocity component, detected in both low-ions and \ion{C}{4}, is at the edge of our search window at $946~\mkms$. This gas is likely unbound from the central DLA host halo. Beyond 150 kpc, we detect \ion{C}{4} absorption from a single system at a $\delta v_\text{weight}$ that exceeds escape for a halo with $M_h =$ 10$^{12} M_\odot$. All three of these velocity components that are detected in excess of the escape velocity for a  $M_h =$ 10$^{12} M_\odot$ halo are in systems which have low-metallicity DLAs ($\log Z/Z_\odot < -1.07$).

Overall, we find no significant correlation between the measured distribution of CGM gas velocities and $R_\bot$ or DLA metallicity. We find that 32 of these 35 components are likely bound under the assumption that they reside in halos with masses of $\sim 10^{12} M_\odot$.  The remaining three components are in turn very likely to escape their host halos regardless of their precise dark matter mass (given that they are almost certainly $\lesssim 10^{12}M_{\odot}$).  In the case that these DLAs predominately reside in halos with $M_h \sim 10^{11}M_{\odot}$, approximately half (17) of the 35 components in our sample have velocities that exceed that required for escape.

\section{Discussion} \label{sec:discuss}

\subsection{Implications for the Metallicity of DLA Halos}\label{Disc:CGMZ}

\begin{figure}[ht]
	\centering
	\begin{minipage}{\linewidth}
		\includegraphics[width=\linewidth]{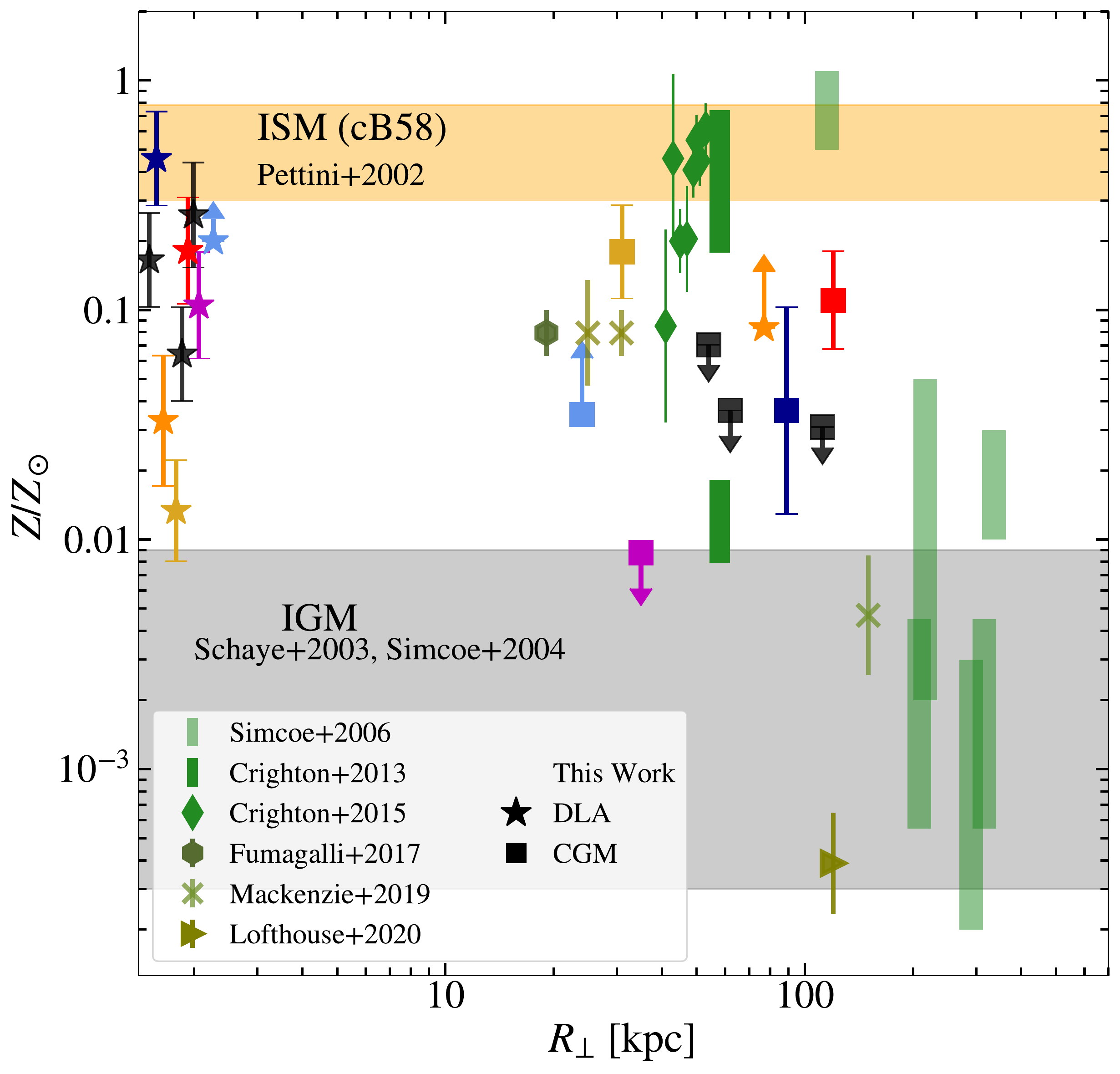}
		\caption{DLA and CGM metallicities vs.\ $R_\perp$. Metallicities measured along CGM sightlines having $10^{18.5}~\mathrm{cm}^{-2} < \nHI < 10^{20.3}~\rm cm^{-2}$ are shown as light blue, dark blue, purple, red, gold, and black squares. The double DLA sightlines are represented by orange stars. Black squares show metallicities constrained by \{Si\textsc{ii}/H\textsc{i}\}. The DLA sightlines are shown with stars near $R_\perp \approx 2\,\rm kpc$, and are colored to indicate the corresponding CGM sightline. Green diamonds show metallicities of distinct absorption components detected in the halo of a ${\sim}0.2 L^*$ galaxy at $z=2.5$ \citep{Crighton_2015_446}. The green rectangles indicate metallicity constraints on the CGM of LBGs from \citet[][dark green]{Crighton_2013_776L} and \citet[][light green]{Simcoe_2006_637}. The dark green hexagon is a DLA located 19.1 kpc from a compact galaxy at $z = 3.25$ \citep{Fumagalli_2017_471}. The green cross-hairs represent the metallicities of DLAs at $3.2 < z < 3.5$ detected close to confirmed LAEs \citep{Mackenzie2019}. The olive green triangle represents a LLS located 120 kpc from a LAE at $z= 3.53$ \citep{Lofthouse_2020_491}. The yellow bar indicates the metallicity measured from interstellar absorption lines in the spectrum of cB58 \citep{Pettini2002}, and the gray bar shows the range in abundances observed in the Ly$\alpha$ forest \citep{Schaye_2003_596,Simcoe_2004_606}.
	 \label{fig:ismigm}}
	\end{minipage}
\end{figure}

We may now place constraints on the metallicities of DLAs and their CGM in the context of other circumgalactic environments at $z\sim2$. We focus our discussion on the subset of our constraints that we consider to be most robust. As described in Section \ref{sec:metallicity}, we include six CGM systems for which we can measure \{O\textsc{i}/H\textsc{i}\} and which are highly optically thick (i.e., with $\nHI > 10^{18.5}~\rm cm^{-2}$), such that we may assume \{O\textsc{i}/H\textsc{i}\} $\approx$ [O/H] \citep[e.g.,][]{Crighton_2013_776L,Prochaska_2015_221}. 
We also include three more CGM systems with $\nHI > 10^{18.5}~\rm cm^{-2}$ for which we constrain metallicity using \{Si\textsc{ii}/H\textsc{i}\},
and report these measurements as upper limits (shown in black in Figure~\ref{fig:ismigm}). For the corresponding DLA sightlines, we assume \{Si\textsc{ii}/H\textsc{i}\} $\approx$ [Si/H] (as \ion{O}{1} $\lambda1302$ is typically saturated in these systems and ionization corrections are likely small). Metallicities for the nine sightline pairs in this subsample are indicated in Figure~\ref{fig:ismigm} with colored/black stars for the DLAs and colored/black squares for the CGM (with the point colors pairing CGM systems to the associated DLA). 

We also compare our metallicity measurements to CGM metallicities from the literature. Following Figure 4 of \citet{Crighton_2013_776L}, we indicate the metallicity of ISM absorption measured in a lensed LBG spectrum (cB58; \citealt{Pettini2002}) with a yellow horizontal bar.  We note that oxygen abundances measured from \ion{H}{2} region emission from LBGs also fall within this range ($Z/Z_{\odot} \sim 0.4-0.7$; \citealt{Strom2018}). We indicate the range in abundances measured in the Ly$\alpha$ forest in gray \citep{Schaye_2003_596,Simcoe_2004_606}. Finally, we include measurements of the metallicity in CGM material  detected around sub-$L^*$ systems and LBGs at $z\sim2.1-3.6$ reported in the literature \citep{Simcoe_2006_637,Crighton_2013_776L,Crighton_2015_446,Fumagalli_2017_471,Mackenzie2019,Lofthouse_2020_491}.

In interpreting these results, we first emphasize that our analysis approach cannot reveal order-of-magnitude variations in metallicities along individual sightlines as observed by \citet{Crighton_2013_776L}, \citet{Crighton_2015_446}, and \citet{Simcoe_2006_637}. Instead, the bulk column densities we use to compute ionic ratios are dominated by the absorption components with the largest columns along the line-of-sight -- and these dominant components need not arise at the same velocity across all ions. Our metallicities assess the overall level of enrichment integrated along each sightline. Nevertheless, these measurements exhibit a large range of values consistent with that observed at much higher spectral resolution. 

Looking at these results in detail, we find three of our CGM sightlines (indicated in red, gold, and orange) exhibit the high metallicities ($\gtrsim0.05 \ Z_{\odot}$) that are observed within $R_{\perp} \le 100$ kpc of emission-selected galaxies at $z\sim2.0-2.5$. At the same time, the host DLAs of these systems have metallicities well below that typical of the ISM of LBGs at this epoch; moreover, two of these DLAs have metallicities lower than that measured in their respective CGM sightline. On the other hand, five of our CGM sightlines have metallicities lower than those of their respective DLA, including some of the highest metallicity DLAs included in this analysis. One of these CGM sightlines, shown as the purple square at $R_{\perp}=35$ kpc, has a metallicity consistent with that typical of the IGM \citep{Simcoe_2004_606,Schaye_2003_596} and of the CGM of LBGs measured at $R_{\perp}> 100$ kpc \citep[][$\log Z/Z_{\odot}<-2.09$]{Simcoe_2006_637}. The low metallicity of this sightline is likely inconsistent with enriched galactic outflows, and instead suggests the origin of this gas is from the surrounding IGM. We further discuss the implications of our metallicity measurements for the origins of the CGM material on a system-by-system basis in Appendix \ref{sec:origin/fate}. Overall, under the assumption that high metallicity DLAs trace higher-mass halos than low metallicity DLAs, our sample of DLA-CGM metallicities is not indicative of any strong dependence of CGM metallicity on halo mass. 

Our findings are consistent with the picture that the CGM is inhomogeneous, containing both enriched (nearing the metallicity of the typical ISM of a LBG) and low-metallicity (near or within the enrichment level of the surrounding IGM) gas. The incidence of higher metallicity (${>} 0.01 ~\rm dex$) versus low metallicity (${<} 0.01 ~\rm dex$) gas along sightlines with \{O\textsc{i}/H\textsc{i}\} constraints is high (5:1) and could suggest lower covering fractions for low-metallicity material. Recent cosmological zoom simulations \citep[e.g.,][]{Hafen_2019_488,Stern2021} also predict a qualitatively inhomogeneous CGM at $z=2$, and find that its mean metallicity similarly decreases significantly with distance from the host galaxy. \citet{Hafen_2019_488} found that the fractions of the total CGM mass arising from wind material vs.\ from accreted IGM gas are approximately equal within ${\lesssim} 0.5R_{\rm vir}$ for halos with masses $10^{10} M_{\odot} \lesssim M_h \lesssim 10^{12} M_{\odot}$, whereas wind material contributes only ${\sim}30-40\%$ of the mass at ${\sim} R_{\rm vir}$. This in turn yields a broad distribution of predicted metallicities throughout these environments, which systematically shift to lower enrichment levels at larger radii. Further analysis is required to enable detailed comparisons between these predictions and the metallicities implied by our pencil-beam probes (which are sensitive to gas over a broad range of physical radii, and which may be dominated by the highest-metallicity material along the line of sight).  The combined datasets shown in Figure~\ref{fig:ismigm} represent  substantive observational constraints to motivate such a comparison.

\subsection{A Global Velocity Width-Metallicity Relation for Absorption-Line Systems}\label{Disc:DLAZ}

\begin{figure}[hb]
	\centering
	\begin{minipage}{\linewidth}
		\centering
		\includegraphics[width=0.97\linewidth]{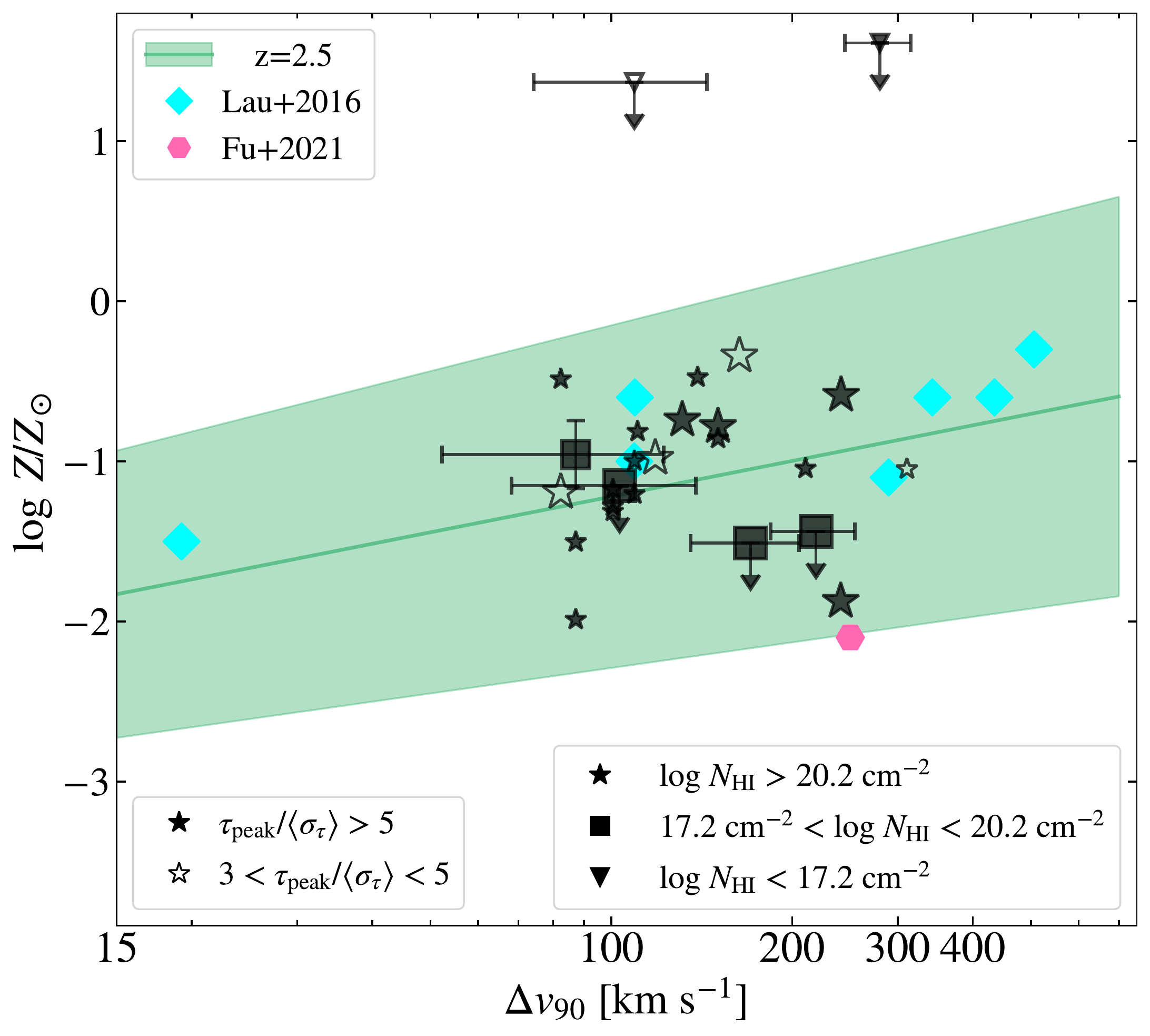} 
		\caption{Metallicity vs.\ velocity width for our DLA (black stars) and CGM (black squares and triangles) sightlines. All systems marked with large squares and stars are also included in Figure~\ref{fig:ismigm}.  Additional DLA and CGM systems from our sample are marked with smaller stars and triangles, respectively. The linear relationship measured by \cite{Neeleman_2013_769} for DLAs at the average redshift of our sample ($z=2.5$) is shown with a light green solid line. The width of the colored contour around this line indicates the measured $\pm1\sigma$ scatter. Filled symbols indicate systems with high-S/N measurements of $\Delta v_{90}$ (i.e., they have $\tau_{\rm peak}/\langle \sigma_\tau \rangle > 5$), whereas open symbols indicate lower-S/N assessments ($3 \le \tau_{\rm peak}/\langle \sigma_\tau \rangle \le 5$).
		We also include measurements of absorption from the CGM of a SMG at $z=2.674$ (\citealt{Fu_2021}, shown with a pink hexagon) and the CGM of QSOs at $z=2-3$ (\citealt{Lau_2016_226}, shown with cyan diamonds).
		\label{fig:Zwidth_relation}}
	\end{minipage}
\end{figure}

Our unique data set allows us to investigate the $\Delta v_{90}$ width-metallicity relation in our CGM sightlines and compare them to both the corresponding $\Delta v_{90}$ for the DLA sightlines and impact parameter. We find the $\Delta v_{90}$ width in the same way for both sets of sightlines, using an unsaturated low-ion transition with the highest ratio $\tau_{\rm peak}/\langle \sigma_\tau \rangle$, and here do so without first separating the profiles into distinct velocity components.
The majority of low-ion transitions in our CGM sightlines are not significantly detected, so this analysis is limited to six CGM sightlines. Two of these six systems have $\nHI < 10^{17.2} ~\rm cm^{-2}$, so the majority of this subset represent LLSs around DLAs (and one of these CGM sightlines is also a DLA). The $\Delta v_{90}$-metallicity distribution of these systems is shown in Figure~\ref{fig:Zwidth_relation}, along with the same measurements for 20 of the DLAs in our sample. We show those systems with robust $\Delta v_{90}$ measurements (having $\tau_{\rm peak}/\langle \sigma_\tau \rangle > 5$) with filled symbols, and show those with lower-S/N constraints on $\Delta v_{90}$ (with $3 \le \tau_{\rm peak}/\langle \sigma_\tau \rangle \le 5$) with open symbols.  Figure~\ref{fig:Zwidth_relation} also indicates the mean relation between $\Delta v_{90}$ and metallicity for the DLA population as reported by \citet{Neeleman_2013_769} at $z=2.5$ (the average redshift of our sample). The shaded region indicates the measured $\pm1\sigma$ scatter in this relation. We find that the $\Delta v_{90}$-metallicity distribution of the DLAs in our sample are overall consistent with the relation fit by \citet{Neeleman_2013_769}.

To increase our sample size, we have included other CGM measurements from the literature. We consider measurements from \cite{Fu_2021}, who analyzed CGM absorption at a transverse separation of 93 kpc from a confirmed submillimeter galaxy (SMG) at $z = 2.674$, 
as well as measurements from \cite{Lau_2016_226}, who studied the CGM within $R_{\perp}<300$ kpc of several QSO host galaxies at $z \sim 2-3$. For the \cite{Lau_2016_226} sample, we used the metallicities reported in that work, and used their publicly-available spectra to calculate the $\Delta v_{90}$ width for these systems in the same manner as for our sample. Seven of these QSO-CGM sightlines have low-ionization transitions that are sufficiently strong to yield a robust $\Delta v_{90}$ measurement. 
Overall, the DLA-CGM and other CGM systems included here occupy a similar region of this parameter space as the DLA population. 

The $\Delta v_{90}$-metallicity relation for DLAs is thought to be driven by the well-known relationship between mass and metallicity among star-forming galaxies \citep{Tremonti2004,Erb2006,Moller_2013_430}, under the assumption that $\Delta v_{90}$ is an effective tracer of the dynamical mass of the system \citep{Ledoux_2006_457,Prochaska_2008_672}.  \citet{Neeleman_2013_769} bolstered evidence for this assumption by demonstrating that DLAs exhibit a $\Delta v_{90}$-metallicity relation that evolves with cosmic time in a manner similar to the stellar mass-metallicity relation.  The sample of $\Delta v_{90}$ and metallicity measurements for CGM systems we have assembled permits two novel tests of this picture.

First, we may compare the values of $\Delta v_{90}$ measured for the QSO-CGM vs.\ DLA-CGM samples.  Because the dark matter halos hosting high-redshift QSOs have masses $M_h > 10^{12.5}$ \citep{Wild2008,White2012,Font-Ribera2013}, we posit that the associated $\Delta v_{90}$ values must be overall larger than those exhibited by DLAs and their CGM, if this quantity is indeed an effective tracer of halo mass. We find that the average values of $\Delta v_{90}$ for the DLA sightlines, DLA-CGM sightlines, and QSO-CGM sightlines are $141 ~\rm \mkms$, $161 ~\rm \mkms$, and $258 ~\rm \mkms$, respectively. This latter value is $97-117 ~\rm \mkms$ larger than the DLA and DLA-CGM sample means. At the same time, we see from Figure~\ref{fig:Zwidth_relation} that there is some overlap among the $\Delta v_{90}$ values of these three subsamples.  
Six of the seven QSO-CGM sightlines (86\%) have $\Delta v_{90} > 100~\mkms$, and three (${\approx}43\%$) have $\Delta v_{90} > 300~\mkms$. Among our DLA and DLA-CGM sightlines,  $80\%$ have $\Delta v_{90} > 100~\mkms$, and just one (${\approx} 4\%$) has $\Delta v_{90} > 300~\mkms$. These findings are consistent with the idea that $\Delta v_{90}$ is indeed correlated with halo mass, and furthermore bolsters previous indications that DLAs arise in halos across a broad mass range, including those as massive as QSO hosts \citep[e.g.,][]{Font-Ribera_2012_11,Mackenzie2019}.

Second, we may use these samples to test for a dependence of $\Delta v_{90}$ on the location of the background QSO sightline with respect to the halo center. We note first that the absolute differences in the $\Delta v_{90}$ values measured for five of the six available DLA-CGM sightline pairs are small: these offsets have a median value of $21~\mkms$ and a maximum value of $44~\mkms$. Only one sightline pair exhibits a much larger difference of $197~\mkms$. Given that the typical uncertainty associated with our $\Delta v_{90}$ measurements is $35~\mkms$, this suggests that $\Delta v_{90}$ is likely an effective tracer of dynamical mass regardless of the location of the background sightline (with the caveat that this paired sightline comparison sample is small).

The suggestion that the $\Delta v_{90}$-metallicity relation may change with distance from the host center was previously studied in literature. As discussed in Section \ref{sec:kinematics}, the work of \citet{Christensen_2019} and \citet{Moller_2020} investigated this topic, 
finding a negative correlation 
in $\log \Delta v_{90}/\sigma_\text{em}$ versus projected distance out to $60$ kpc.  They concluded that the concomitant decline in DLA metallicity with $R_{\perp}$ implies that the $\Delta v_{90}$-metallicity relation does  not depend on impact parameter.
On the other hand, the analysis by \cite{Fu_2021} found a CGM absorber detected at $R_{\perp} = 93 ~\rm kpc$ from a host SMG that deviates from the $\Delta v_{90}$-metallicity relation for DLAs by $+230 ~\rm \mkms$ (see Figure~\ref{fig:Zwidth_relation}). \cite{Fu_2021} argued that this could be due to the large impact parameter of the sightline relative to typical DLA-host galaxy projected separations.

To investigate the nature of offsets from the $\Delta v_{90}$-metallicity relation within our sample, we focus on those systems with secure metallicity measurements (including one DLA-CGM sightline, the one SMG-CGM sightline, and all seven QSO-CGM sightlines). We use $\Delta v_{90}$ and the system redshifts to calculate the expected DLA metallicities from the relation reported in \citet{Neeleman_2013_769}. We then compare the expected value to the measured metallicities by calculating the difference divided by the $\pm 1\sigma$ scatter in the $\Delta v_{90}$-metallicity relation ($|Z/Z_{\rm \odot, observed} - Z/Z_{\rm \odot,expected}|/\sigma_{Z/Z_\odot}$), and relate that to $R_\bot$ for each system. These latter values fall across the range $40~\mathrm{kpc} < R_{\perp} < 180~\mathrm{kpc}$.

The median value of $|Z/Z_{\rm \odot, observed} - Z/Z_{\rm \odot,expected}|/\sigma_{Z/Z_\odot}$ is 0.02, with the SMG-CGM sightline having the maximum value of 1.08. Excluding this sightline, all eight values of this offset are ${<}0.6$, and therefore consistent with the DLA $\Delta v_{90}$-metallicity relation within $\pm1\sigma$. 
We also find no evidence for a dependence of this offset on $R_\perp$:  the four sightlines within 100 kpc have values in the range $0.05 < |Z/Z_{\rm \odot, observed} - Z/Z_{\rm \odot,expected}|/\sigma_{Z/Z_\odot} < 1.08$, while at large impact parameters (${>}100$ kpc) we measure values $0.02 < |Z/Z_{\rm \odot, observed} - Z/Z_{\rm \odot,expected}|/\sigma_{Z/Z_\odot} < 0.19$. 

While this analysis is limited, it is nevertheless suggestive that there is no clear relation between $|Z/Z_{\rm \odot, observed} - Z/Z_{\rm \odot,expected}|/\sigma_{Z/Z_\odot}$ and $R_\bot$ over the impact parameter range $40~\mathrm{kpc} < R_{\perp} < 180~\mathrm{kpc}$. We interpret this result as being consistent with the conclusions of \citet{Christensen_2019} and \citet{Moller_2020} discussed above.
Their analysis implied that the $\Delta v_{90}$-metallicity  relation for DLAs is local in nature, meaning that the metallicity measured along the line of sight follows the local value of $\Delta v_{90}$ (i.e., the local gravitational potential).
Our results may similarly suggest that on average, CGM gas detected at much larger impact parameters than previously explored follows the same mean $\Delta v_{90}$-metallicity relation. 
However, a larger sample size is needed to confidently rule out a relation between $|Z/Z_{\rm \odot, observed} - Z/Z_{\rm \odot,expected}|/\sigma_{Z/Z_\odot}$ and $R_\bot$ over large scales.

Our comparison of metallicities between DLA-CGM and QSO-CGM sightlines has strengthened existing lines of evidence indicating that $\Delta v_{90}$ is correlated with halo mass. Under this assumption, given that 
(1) in the foregoing analysis, we have used the measured $\Delta v_{\rm 90}$ width of individual sightlines to calculate the expected metallicity, and that 
(2) CGM gas metallicities are likely to span a larger range than the metallicities of DLA material, we suggest that any deviation from the $\Delta v_{\rm 90}$-metallicity relation for CGM sightlines is due to the scatter in metallicities extant in the CGM at fixed halo mass.
While DLAs are likely confined to star-forming regions or the ``inner'' CGM \citep[e.g.,][]{Krogager2017,Stern2021,Theuns_2021_500}, the cool circumgalactic medium at this epoch is fed by both pristine inflow and metal-enriched outflow \citep[e.g.,][]{Crighton_2013_776L,Crighton_2015_446,Fumagalli_2017_471,Lofthouse_2020_491}, and it is likely that many of these gaseous structures are destroyed by hydrodynamical instabilities before they can mix with the surrounding material \citep{Schaye2007,Crighton_2015_446,Gronke2021}. This interpretation could also explain the slight deviation of the \citet{Fu_2021} system from the $\Delta v_{\rm 90}$-metallicity relation. Based on its relatively low metallicity, \cite{Fu_2021} argued that this sightline is probing a cold inflowing stream around the host SMG; therefore, while the $\Delta v_{\rm 90}$ width may fall in the range expected for a massive halo typically associated with SMGs, the metallicity in this particular line of sight may scatter away from the $\Delta v_{\rm 90}$-metallicity relation due to stochasticity in the contents of the CGM.

In sum, we have found that optically thick CGM absorbers around both DLAs and QSO hosts occupy a similar region of $\Delta v_{\rm 90}$-metallicity parameter space as DLAs themselves. In addition, we find no evidence for a relationship between the offset from the $\Delta v_{\rm 90}$-metallicity relation and $R_\bot$ (though our sample is small). Instead, we propose that circumgalactic systems that are outliers from this relation arise due to poorly-mixed halo material.

\section{Conclusion} \label{sec:concl}

In previous work, we introduced a technique to study the CGM of high-redshift DLAs (which act as signposts for high-redshift galaxies) in absorption using close quasar pairs. This technique probes gas in the extended CGM in one QSO sightline and permits a direct comparison to the ISM/inner CGM material traced by the DLA in the second sightline. In this work, we have analyzed medium and high resolution ($\mathcal{R} \geq 4000$) spectroscopy for 32 such quasar pairs, each of which has an intervening DLA in the redshift range $1.6 < z_\text{DLA} < 3.5$, and which have sightline separations in the range $24~\mathrm{kpc} < R_\bot < 284$ kpc. We have reported column densities and assessed kinematics for several ionic species in each sightline pair.  We have also performed a novel comparison between the metallicities of DLAs and nearby circumgalactic material.  

Here we summarize the main results of this paper:
\begin{itemize}
    \item We find that the \ion{H}{1} column densities measured in our CGM sightlines are anticorrelated with $R_\bot$ with 99.8$\%$ confidence. We also report a high incidence of optically thick \ion{H}{1} ($\nHI > 10^{17.2}~\rm cm^{-2}$) around DLAs, with $C_f = 50 \pm 13$\% for sightlines within $24 ~\mathrm{kpc} \leq R_\bot \leq 100$ kpc, and $C_f = 50 \pm 22$\%  for sightlines at 100 kpc $< R_\bot \leq 200$ kpc.  These results suggest both that DLAs are located close to the centers of their host halos, and that these systems are located in environments in which neutral gas extends over large scales ($> 100$ kpc). The CGM of Lyman Break Galaxies (LBGs) at $z\sim2-3$ exhibits a marginally lower incidence of optically thick \ion{H}{1} at 50 kpc $\leq R_\bot \leq 100$ kpc ($C_f = 20^{+15}_{-13}\%$; \citealt{Rudie_2012_750}).
    
    \item We report covering fractions within 100 kpc-wide $R_\bot$ bins for several metal species. \ion{C}{2} and \ion{C}{4} yield the largest covering fractions within $R_\bot < 100$ kpc, with $C_f (\nCII > 10^{13} \ \rm cm^{-2}) > 89 \%$ and $C_f (\nCIV > 10^{13} \ \rm cm^{-2}) = 94_{-9}^{+4} \%$. We likewise find high incidences of singly- and triply-ionized silicon, with  $C_f (\nSiII > 10^{13} \ \rm cm^{-2}) = 75_{-17}^{+12} \%$ and $C_f (\nSiIV > 10^{13} \ \rm cm^{-2}) = 67_{-12}^{+10} \%$. Comparing to LBG and QSO-host halos at this epoch, we find that the covering fractions of high-ionization species are similar (consistent within $\pm 1\sigma$) within $R_\bot < 100$ kpc. However, the  covering fractions for \ion{C}{2} and \ion{Si}{2} around DLAs are larger than around LBGs at $0~ \mathrm{kpc} < R_\bot < 100$ kpc by ${\sim}2\sigma$. DLA covering fractions for all species drop below those measured in QSO halos by ${\gtrsim}1\sigma$ beyond $R_\bot > 200$ kpc, suggesting that QSO halos have enriched gas that extends to larger impact parameters.    
        
    \item We identify, species by species, the thresholds above which 90\% of metal line column densities in our DLA sample lie (i.e., the 10th percentile column density values).  We then assess the covering fraction of CGM systems with column densities above these thresholds.  At $24~\mathrm{kpc} < R_\bot <200~\mathrm{kpc}$, both intermediate- and high-ionization species exhibit covering fractions ${>}40\%$ relative to the corresponding 10th percentile thresholds; however, even within ${<} 100$ kpc, the incidence of low-ionization species does not exceed $C_f = 30\%$. This suggests that the warm material traced by \ion{Si}{4} and \ion{C}{4} associated with DLAs frequently extends over $100-200$ kpc scales, whereas cool, photoionized or neutral material seldom exhibits DLA-level absorption strengths across length scales ${\gtrsim} 30$ kpc. 
    
    \item We identify nine DLA-CGM systems, all having impact parameters $24~\mathrm{kpc} < R_\bot < 120$ kpc, for which the ionic ratios $\nOI/\nHI$ or $\nSiII/\nHI$ yield robust constraints or limits on the CGM metallicity (given our assumption that these ratios are less sensitive to ionization state in systems with $\nHI \gtrsim 10^{18.5}~\rm cm^{-2}$; e.g., \citealt{Crighton_2013_776L}).  These values range from a maximum of $\log Z/Z_{\odot} =-0.75$ (i.e., close to that observed in the ISM of LBGs) to an upper limit $\log Z/Z_{\odot} < -2.06$ (consistent with that observed in the IGM at this epoch).  These metallicities are consistent with or lower than those estimated for the associated DLA (using the ionic ratio $\nSiII/\nHI$) in five of these sightline pairs, and are higher than that of the associated DLA in two systems. Overall, we find no evidence for a correlation between the metallicities observed in the DLAs and their associated CGM.
       
    \item The $\Delta v_{90}$ velocity widths of both low-ionization transitions and the \ion{C}{4} $\lambda 1548$ transition in the DLA vs.\ CGM sightlines lie along a 1:1 relation and are correlated with high statistical significance. Moreover, the low-ionization $\Delta v_{90}$ values differ by ${<}40~\mkms$ in the preponderance (66\%) of sightline pairs. This suggests (1) that metal-line kinematic widths exhibit strong coherence over $R_\bot \lesssim 200$ kpc scales; and (2) that the $\Delta v_{90}$ kinematic measure is an effective indicator of the potential well of the host halo regardless of the location of the sightline relative to the halo center. We find that our DLA-CGM systems, along with several CGM systems drawn from the literature, lie along the same $\Delta v_{90}$-metallicity relation as that exhibited by DLAs themselves.
    
    \item Velocity centroids for \ion{C}{4} $\lambda 1548$ differ by ${>}100~\rm \mkms$ for nearly half (48$\%$) of velocity components across the sightline pairs.  However, the vast majority (32/35) of \ion{C}{4} and low-ion component centroids have line-of-sight velocities less than the escape velocity of a putative DLA host dark matter halo with $M_h \approx 10^{12}~M_\odot$.  If we instead assume a host halo mass of $M_h \approx 10^{11}~M_\odot$, 18 of these 35 components have radial velocities less than the implied escape velocity. 
\end{itemize}

The CGM, while crucial to our understanding of galaxy evolution, is diffuse and difficult to detect at $z\gtrsim2$. As demonstrated by the foregoing analysis, absorption line studies remain important tools for assessing the enrichment histories of the gas that feeds star formation during this critical epoch. Ongoing efforts to improve our understanding of the baryonic cycling through circumgalactic environments will include increasing the samples of known galaxy counterparts to strong absorbers at high redshift. One such effort is the MUSE Analysis of Gas around Galaxies (MAGG) survey, which uses VLT/MUSE to search for galaxy counterparts in emission in several bright quasar fields (see \citealt{Dupuis_2020,Lofthouse_2020_491,Fossati_2021_503,Lofthouse2023}). The increasing sample of confirmed galaxy/CGM sightline pairs will lead to a better understanding of the connection between the properties of host galaxy/halo centers (i.e., star-formation rates, halo mass) and those of extended halo gas. At the same time, comparison of measured CGM properties in these environments to the predictions of state-of-the-art cosmological zoom simulations (e.g., FIRE or FOGGIE; \citealt{Hopkins2018,Peeples2019,Stern2021}) will provide unique insight into the physics of the baryonic flows feeding such extended gas reservoirs. Ultimately these combined efforts will aid our understanding of the complex structure of the CGM and provide insight into how halo gas may be linked to different phases of galaxy growth.

\acknowledgments
We acknowledge Rob Simcoe for inspiring this work, and for many helpful discussions. We also appreciate illuminating conversations with Joe Burchett and Gwen Rudie.
This material is based upon work supported by the National Science Foundation under grant No.\ AST-1847909. 
KHRR acknowledges partial support from NSF grants AST-1715630 and AST-2009417.
E.N.K. gratefully acknowledges support from a Cottrell Scholar award administered by the Research Corporation for Science Advancement. This project has received funding from the European Research Council (ERC) under the European Union's Horizon 2020 research and innovation programme (grant agreement No.\ 757535). This work has also been supported by Fondazione Cariplo, grant No.\ 2018-2329.

The authors wish to recognize and acknowledge the very significant cultural role and reverence that the summit of Maunakea has always had within the indigenous Hawaiian community.  We are most fortunate to have the opportunity to conduct observations from this mountain.

\appendix

\section{Constraining Metallicities} \label{app:metallicities}

\subsubsection{Constraints on Ionization State from Ionic Ratios}

\begin{figure*}[hb]
	\centering
	\begin{minipage}{\linewidth}
		\centering
		\includegraphics[width=0.8\linewidth]{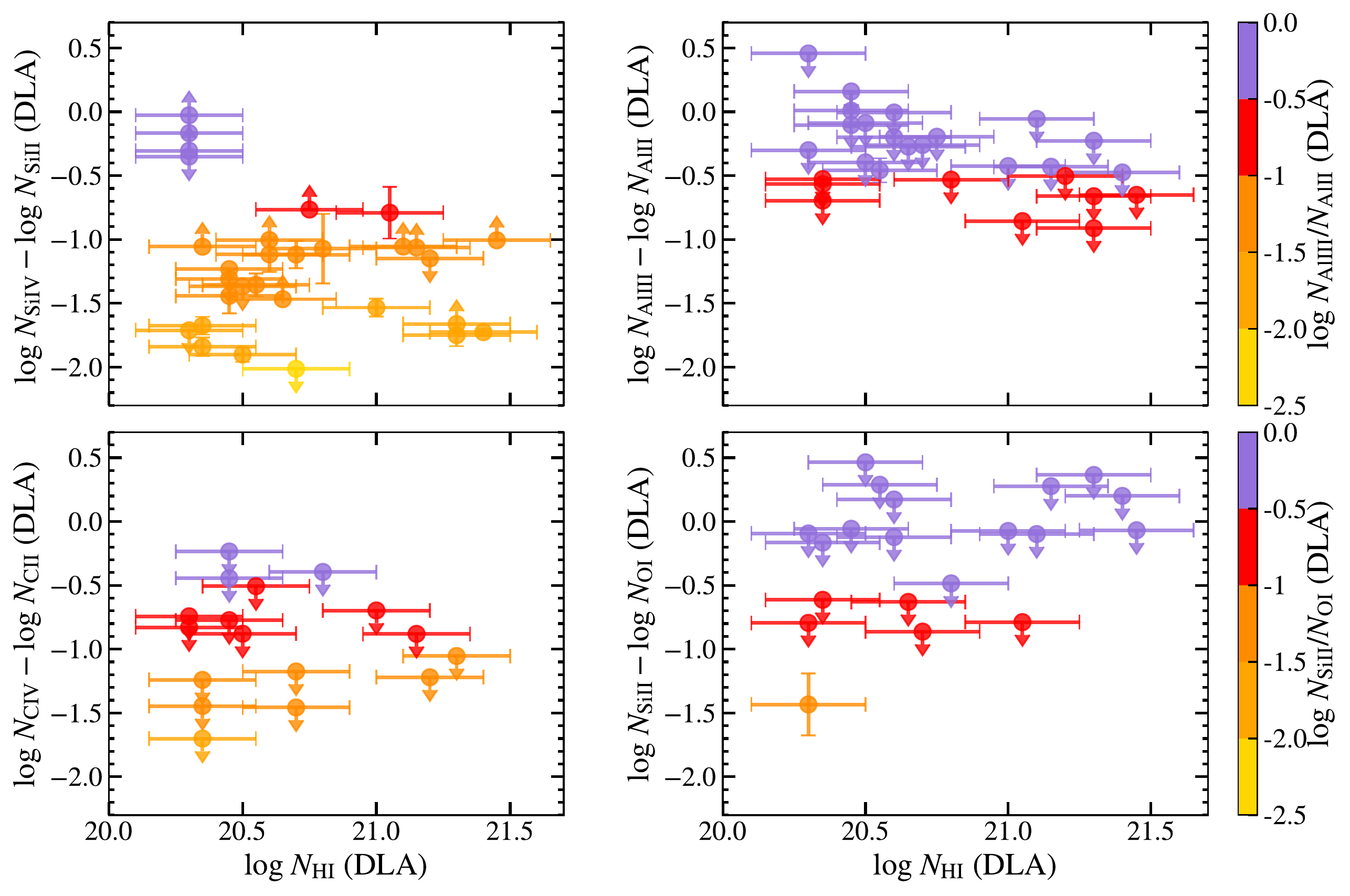} 
		\caption{Ionic ratios of $\nSiIV / \nSiII$ (top left), $\nCIV / \nCII$ (bottom left), $\nAlIII / \nAlII$ (top right), and $\nSiII / \nOI$ (bottom right) vs.\ $\nHI$ for the DLAs in our sample. The color bar reflects the y-axis value of each point in each panel, and will be used to color code these systems in Figure~\ref{fig:dlametalabund}.  Larger values are indicative of more highly ionized material, and imply larger uncertainties in metallicities estimated from ionic ratios.	Our constraints on $\nOI$, $\nAlII$, and $\nAlIII$ are upper limits in the majority of sightlines, resulting in the vast majority of ionic ratios including these species appearing here as limits. \ion{C}{2} is also typically saturated in DLA sightlines, such that many of our $\nCIV / \nCII$ values are upper limits. We therefore rely on ionic ratios calculated from $\nSiIV / \nSiII$ to indicate the ionization state of our DLA sightlines.  \label{fig:dlaionization}}
	\end{minipage}
\end{figure*}

 In order to estimate the systematic uncertainty in our metallicity measurements, to be described in Section \ref{secsub:metallicities}, we must first assess the ionization state along each sightline. We approach this by calculating the ratios of the column densities of high-ionization to low-ionization transitions of the same species.  Logarithmic ratios of $\nSiIV / \nSiII$, $\nCIV / \nCII$, and $\nAlIII / \nAlII$ vs.\ $\nHI$ for our DLA sample are shown in Figure \ref{fig:dlaionization}. We also include $\nSiII / \nOI$, as this ratio should yield $-1.2$ dex in neutral gas (assuming solar abundance ratios). If this ratio is above $-1.2$ dex, this suggests the material is highly ionized \citep{Prochaska_2015_221}.
 
 The measured ionic ratio constraints range from $-2.5$ dex to $+0.5$ dex; however, the vast majority of our constraints on $\nCIV / \nCII$, $\nAlIII / \nAlII$, and $\nSiII / \nOI$ are upper limits.  Our $\nSiIV / \nSiII$ estimates included the greatest number of direct measurements, and we therefore rely on this ionic ratio as our primary indicator of ionization fraction. Several of our DLAs exhibit $\log \nSiIV / \nSiII$ values ${\gtrsim} -0.5$ dex (shown in purple in Figure \ref{fig:dlaionization}), indicative of partially ionized conditions \citep{Prochaska_2015_221}.  However, the average $\log \nSiIV / \nSiII$ value for these systems is ${\approx} -1.1$ dex, consistent with the predominantly neutral conditions of most DLAs \citep[e.g.,][]{Prochaska_2000_533L}.

\begin{figure*}[ht]
	\centering
	\begin{minipage}{\linewidth}
		\centering
		\includegraphics[width=0.8\linewidth]{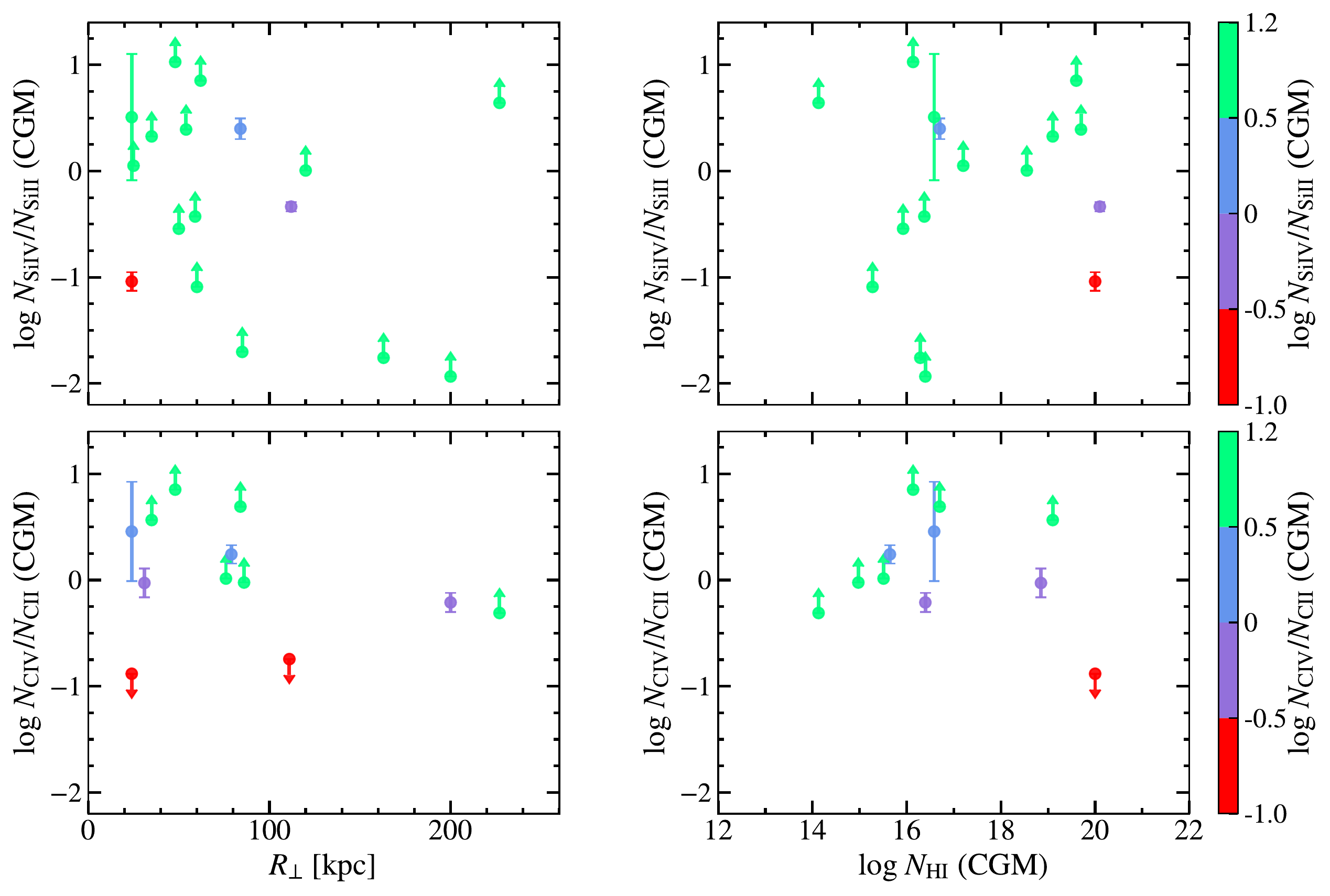}
		\caption{Ionic ratios for the CGM sightlines in this work.  $\log \nSiIV / \nSiII$ and $\log \nCIV / \nCII$ vs.\ $R_{\perp}$ constraints are shown at top left and bottom left, respectively.  The ionic ratios $\log \nSiIV / \nSiII$ and $\log \nCIV / \nCII$ vs.\ $N_\text{HI,CGM}$ are shown at top right and bottom right, respectively. Lower limits are shown in green because the value of the corresponding ionic ratio is ambiguous. As in our DLA sample, our $\log \nSiIV / \nSiII$ measurements yield a larger number of direct constraints than other ionic ratios, and we therefore rely on these measurements as our indicator of ionization state for the CGM.
		\label{fig:ionization}}
	\end{minipage}
\end{figure*}

We also calculate $\nSiIV / \nSiII$ and $\nCIV / \nCII$ ionic ratios for our CGM sightlines, shown in Figure~\ref{fig:ionization}. Here we color any lower limits as green.  As for our DLA sightlines, our $\nSiIV / \nSiII$ estimates include the greatest number of direct measurements. All except two sightlines exhibit ionic ratios ${\gtrsim} -0.5$ dex, indicative of predominantly ionized conditions. 

\subsubsection{Metallicity Constraints for the DLA Sample}\label{secsub:metallicities}

\begin{figure*}[ht]
	\centering
	\begin{minipage}{\linewidth}
		\centering
		\includegraphics[width=0.45\linewidth]{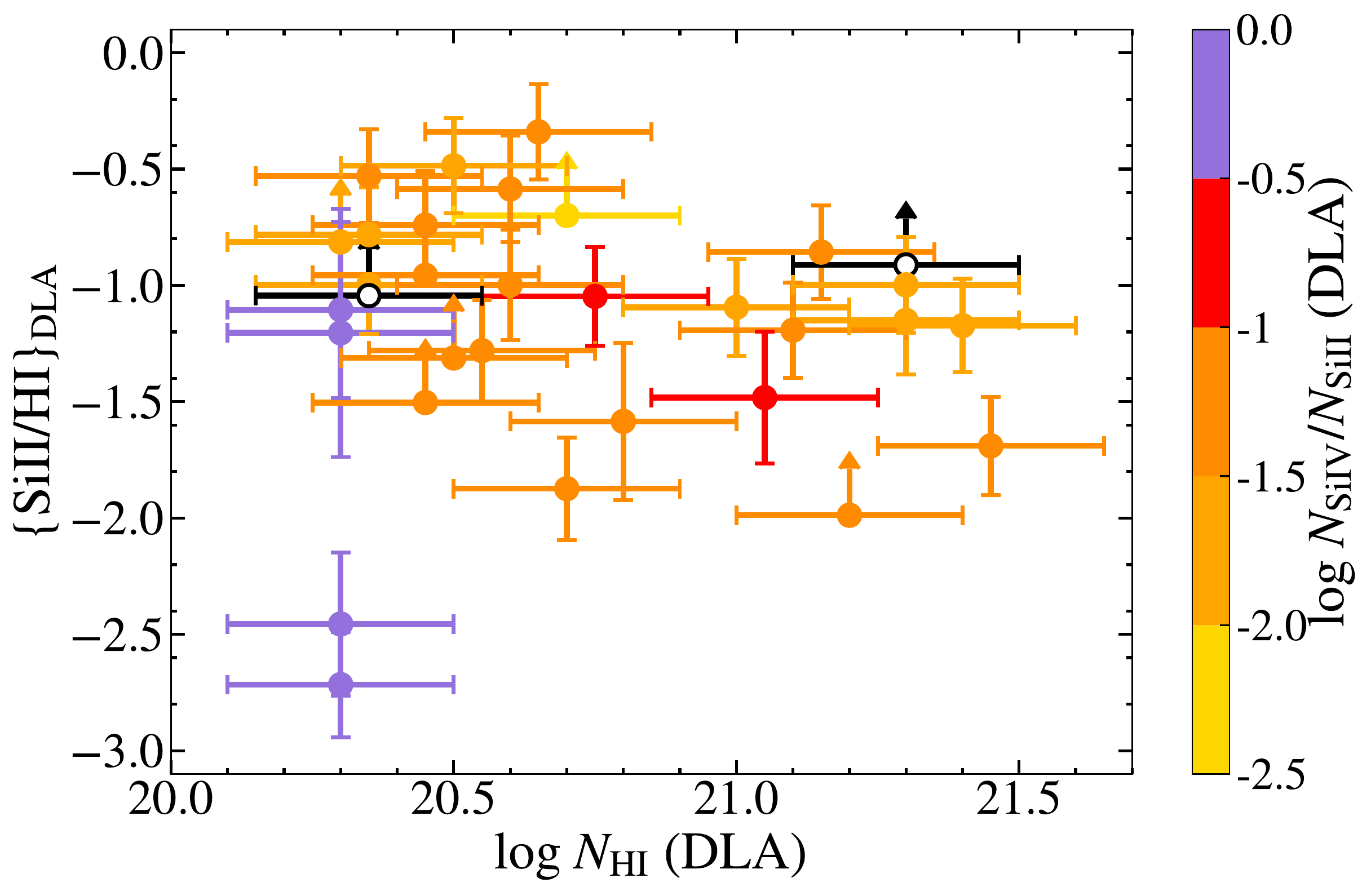}
		 \hspace{2mm}
		\includegraphics[width=0.45\linewidth]{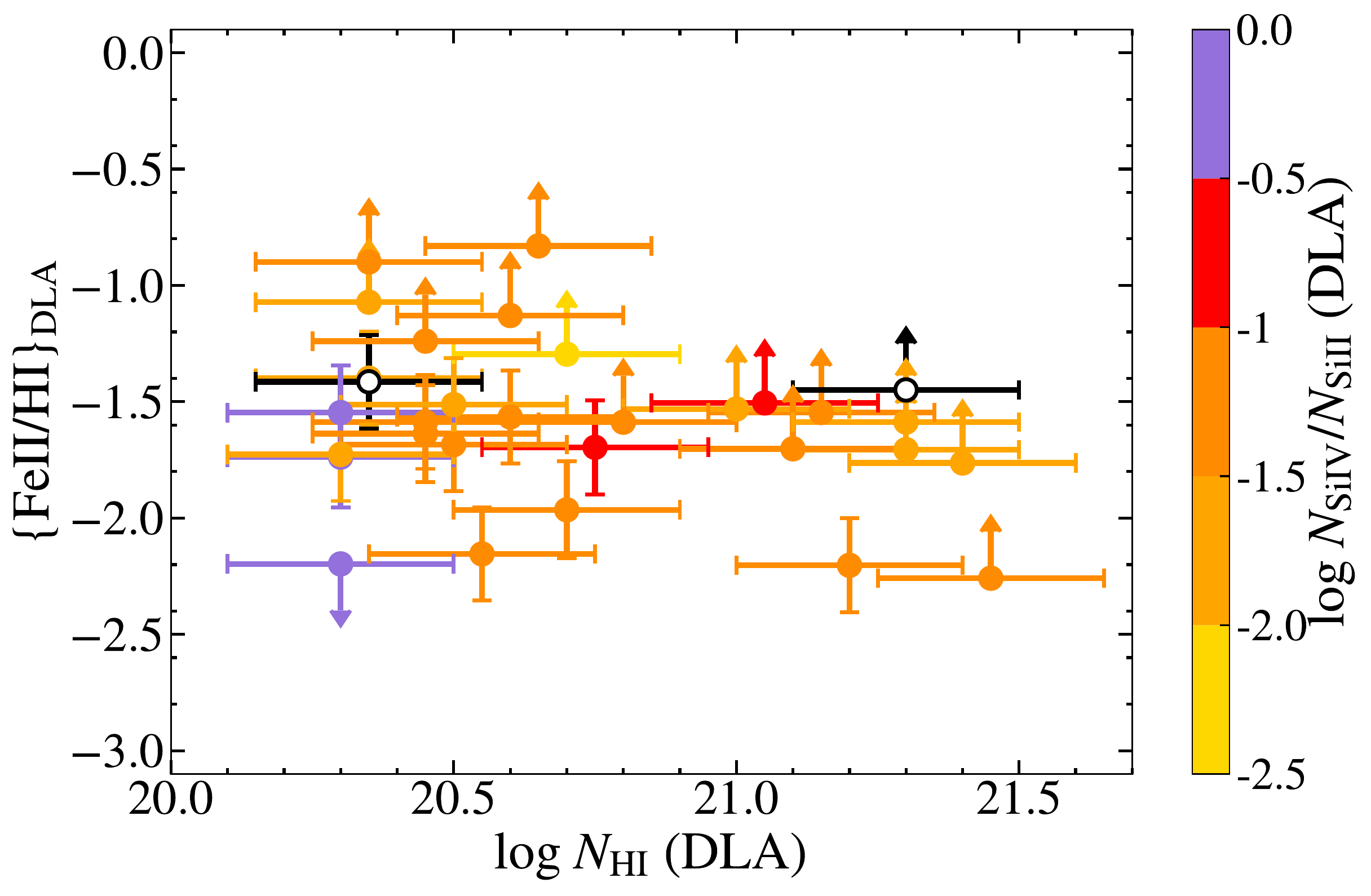}
		\hspace{2mm}
		\includegraphics[width=0.45\linewidth]{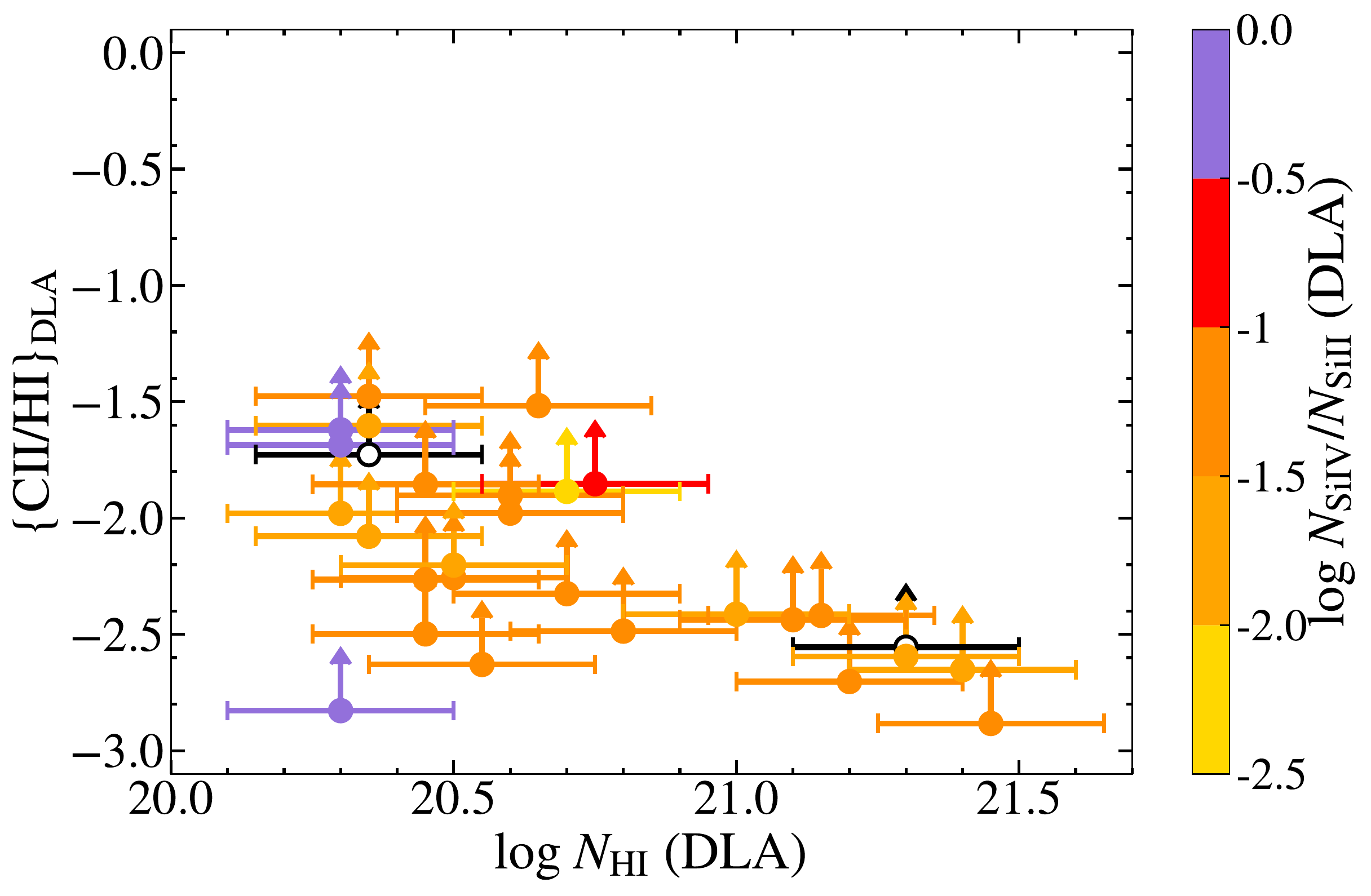}
		\hspace{2mm}
		\includegraphics[width=0.45\linewidth]{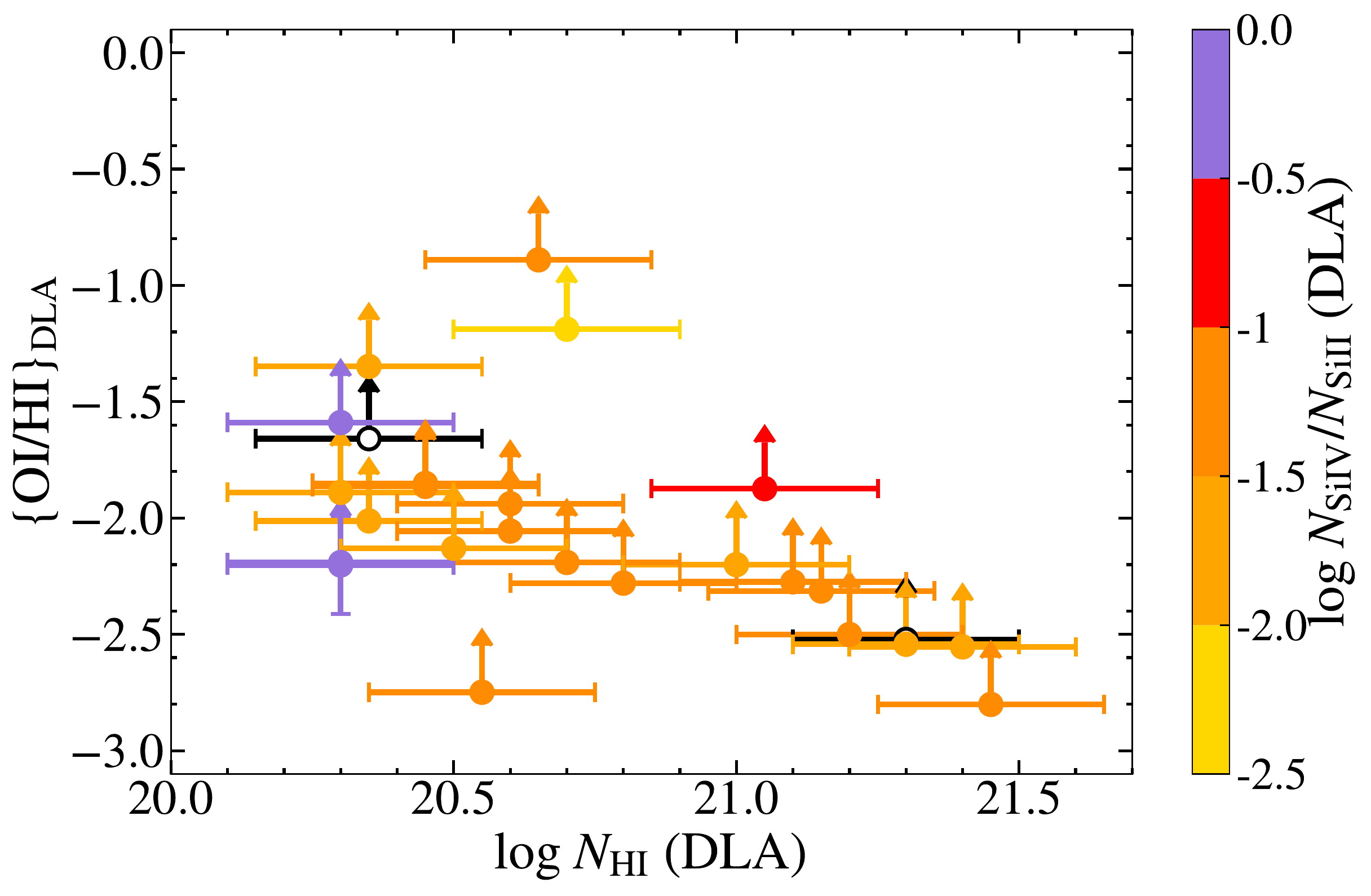}
		\caption{Ionic ratio metallicities estimated from \ion{Si}{2} (top left), \ion{Fe}{2} (top right), \ion{C}{2} (bottom left), and \ion{O}{1} (bottom right) column densities plotted versus \ion{H}{1} column density for our DLA sample. The colorbar shows the value of the ionic ratio $\log \nSiIV / \nSiII$, an indicator of the ionization fraction. Purple points mark systems with $\log \nSiIV / \nSiII > -0.5$ dex; i.e., systems which are predominantly ionized, and for which our ionic ratio metallicities are likely overestimated. The black open circles mark sightlines with an ambiguous ionization fraction due to heavy blending of the \ion{Si}{4} line profiles. In general, the measured DLA metallicities are consistent with the $\Delta v_{90}$-metallicity relation from \citet[][see Figure \ref{fig:Zwidth_relation}]{Neeleman_2013_769}.\label{fig:dlametalabund}}
	\end{minipage}
\end{figure*}

As explained in Section \ref{sec:metallicity}, to estimate the metallicity of our absorption systems, we make use of the quantity ($\{\mathrm{X}i/\mathrm{H}j\}$) introduced by \citet[][see Equation \ref{eq:solarmetals}]{Prochaska_2015_221}. In cases in which the ionization correction is small, such as for DLAs, we assume that $\{\mathrm{X}i/\mathrm{H}j\} =$ [X/H]. We adopt solar elemental abundances from \citet{Asplund_2009_47}.

In Figure \ref{fig:dlametalabund}, we show \{Si\textsc{ii}/H\textsc{i}\}, \{Fe\textsc{ii}/H\textsc{i}\}, \{C\textsc{ii}/H\textsc{i}\}, and \{O\textsc{i}/H\textsc{i}\} vs.\ $\nHI$ for our DLA sample. Points are color-coded by the value of $\log \nSiIV / \nSiII$ for the system (as in Figure \ref{fig:dlaionization}). This latter ionic ratio is low in the vast majority of these sightlines (${<}-0.5$ dex), implying that ionization corrections are small, and that the former ratios are indeed representative of the metallicity. Our values of $\{$Si\textsc{ii}/H\textsc{i}$\}$ range from ${\sim}-2.7$ to $-0.4$ dex and have an average of $-1.35$ dex.  The range of $\{$Fe\textsc{ii}/H\textsc{i}$\}$ values is similar (${\sim} -2.8$ to $-0.7$ dex, with an average of $-1.76$ dex).  These metallicities are typical of those exhibited by unbiased DLA samples at this epoch (see Figure \ref{fig:Zwidth_relation}; \citealt{Neeleman_2013_769}). 

\section{Comments on Individual Systems: Physical Origins } \label{sec:origin/fate}

To gain insight into the origins of the material detected along our CGM sightlines, we refer to the detailed analysis of the properties and origins of circumgalactic material in the FIRE-2 cosmological zoom simulations described in \citet{Hafen_2019_488}. This work used particle tracking to distinguish halo gas that has accreted from the IGM; that originates from an outflow from the host galaxy; or that originates from gas ejected from another galaxy. They performed this analysis for a suite of simulations of halos that evolve to have $M_h \sim 10^{10-12} M_{\odot}$ at low redshift, assessing the median metallicity as a function of radius for each of these three CGM components at both $z=2$ and $z=0.25$. Overall, these authors find that the material originating in winds on average exhibits only slight changes in metallicity with increasing radial distance; that wind material from the central galaxy on average has higher metallicity than winds from satellites; and that material accreting from the IGM exhibits metallicities $\lesssim 0.01Z_{\odot}$ at distances ${>} 0.5R_{\rm vir}$. We caution that these statements apply to the median metallicities as a function of radius, and therefore can only be used to indicate the most statistically likely origins of the gas in our sightlines.

\subsubsection{An Unusually Low-Metallicity CGM System: J0004-0844}

One of our CGM systems (in sightline pair J0004-0844 with $R_{\perp} = 35$ kpc and $z_\text{abs} = 2.75$), indicated with the purple square in Figure \ref{fig:ismigm}, has a metallicity $Z/Z_\odot < 10^{-2.09}$; i.e., within the range of metallicities measured in the $z\sim2$ IGM \citep{Schaye_2003_596,Simcoe_2004_606}, and consistent with the lowest-metallicity $z \sim2$ CGM sightlines probing distances $R_\bot > 100$ kpc from their galaxy counterparts \citep{Simcoe_2006_637,Mackenzie2019,Lofthouse_2020_491}. This sightline has a large \ion{H}{1} column density ($\nHI > 10^{19.1}~\rm cm^{-2}$), but exhibits no detectable absorption from low or intermediate metal ions. Our upper limit on the column density of \ion{O}{1} places a firm limit on the metallicity of this system at approximately two orders of magnitude lower than that measured for the associated DLA ($Z/Z_{\odot} = 10^{-0.1 \pm 0.24}$). The average velocity of this material as traced by \ion{C}{4} absorption is within ${<} 10 \ \mkms$ of that of the DLA, suggesting that this material is gravitationally bound even under the assumption that the system is hosted by a  $10^{10}M_\odot$ halo (see the purple square in Figure \ref{fig:vesc}). 

Given that almost none of the gas originating in winds in the FIRE-2 simulations at $z=2$ has a metallicity as low as our limit for this system (see Figure 17 in \citealt{Hafen_2019_488}), as well as the quiescent kinematics of the observed absorption profile, we posit that this system most likely arises from accreting IGM material. Moreover, we emphasize that our detection of such a low-metallicity system is likely only possible due to the absence of material originating in galactic winds along this sightline.

\subsubsection{A High-Metallicity, High Velocity-Offset CGM System: J2103+0646}

The CGM system detected toward sightline pair J2103+0646 (represented by the gold square in Figure \ref{fig:ismigm}) is one of the most metal-enriched in our sample, with an enrichment level near that of the ISM of the lensed LBG cB58 \citep{Pettini_2012_425}. The metallicity of this system is also an order of magnitude above the metallicity of the corresponding DLA despite being at $R_{\perp} = 31$ kpc. Considering that at this redshift ($z_\text{abs}=2.14$) the virial radius is likely $R_{\rm vir} < 90 ~\rm kpc$ for the host galaxy, this implies that this sightline probes gas at ${\gtrsim} 0.30R_{\rm vir}$. 
The \citet{Hafen_2019_488} analysis of $M_h \sim10^{12}M_{\odot}$ halos at such small radial distances predicts that both accreted IGM material and gas originating from winds may plausibly be enriched to this level in this inner CGM environment.

The CGM sightline has two velocity components: one close to the DLA \ion{H}{1} velocity centroid (the ``primary'') and one close to the CGM \ion{H}{1} velocity centroid (the ``secondary''). The primary component has a low $\nCIV$ compared to the secondary component, and therefore does not contribute significantly to the total metallicity of the sightline. The \ion{C}{4} profile of the secondary component is offset from the DLA redshift by 263 $\mkms$, and thus traces gas that would not be gravitationally bound to a halo with $M_h \le 10^{11} M_\odot$. However, the $\Delta v_{90}$ width of the \ion{C}{4} absorption is low (45 $\mkms$), suggesting this sightline does not probe a region of the CGM that is being actively enriched by outflows. The primary velocity component has a velocity centroid that is offset by only 32 $\mkms$ from the DLA redshift, and thus likely traces gas that remains gravitationally bound to its host halo.

\subsubsection{Systems that Must Be Wind-Enriched, But Which Have Quiescent Kinematics}

The CGM system observed toward sightline pair J0955-0123 (represented by the dark blue square in Figure~\ref{fig:ismigm}) has a redshift $z_\text{abs} = 2.73$ and a projected separation $R_{\perp} =89$ kpc. At this distance and redshift, this sightline likely probes gas near or beyond the virial radius of a galaxy in a ${\geq} 10^{12} M_\odot$ halo. This system has a high \ion{H}{1} column density ($\nHI = 10^{19.2}~\rm cm^{-2}$) and a metallicity $Z/Z_\odot = 10^{-1.43}$. This level of enrichment is for the most part only achieved in the FIRE-2 simulations by wind-enriched gas \citep{Hafen_2019_488}. 
The median metallicity of material accreted from the IGM and located at the virial radius is less than $3 \times 10^{-3} Z_\odot$ for all simulated halos, even with metal diffusion enriching the IGM particles; i.e., it is more than a factor of ten lower than that measured in the CGM of this system. We therefore suggest this gas likely probes wind-enriched material. The velocity centroid of the low-ionization gas along this sightline is close to that of the DLA ($\delta v_\text{weight} = -43 ~\rm \mkms$), and its $\Delta v_{90}$ width is narrow. The high-ionization gas traced by \ion{C}{4} is similarly near the DLA redshift ($\delta v_\text{weight} = -36 ~\rm \mkms$); however, it has a large $\Delta v_{90}$ width ($\Delta v_{90} = 178 ~\rm \mkms$). This gas is likely bound within halos with masses $\geq 10^{10}M_\odot$. 

The CGM system in our pair target J2146-0752, with a metallicity $Z/Z_\odot = 10^{-0.96}$,  is represented by the red square in Figure \ref{fig:ismigm}.  This sightline probes the CGM of a DLA at $z_\text{abs} = 1.85$ at an impact parameter $R_{\perp} = 120$ kpc. This particular sightline is therefore likely near or beyond the virial radius of typical DLA galaxies at this redshift. 
The median metallicity of IGM material at this redshift and proper distance as predicted by \citet[][$\leq 3 \times 10^{-3} Z_\odot$]{Hafen_2019_488} is 1.6 dex lower than what we measure along the sightline, suggesting that this gas likely originated from a wind of some type. The CGM absorption is this system is furthermore unusual in that the strengths of several metal transitions are comparable, and in some cases stronger, than those observed in the associated DLA sightline,  while at the same time it has a substantially lower \ion{H}{1} column density by approximately two orders of magnitude. However, its metallicity is only 0.2 dex below that of the DLA sightline. We observe velocity centroids of $\delta v_\text{weight} = 90~\mkms$ for low-ionization gas and $\delta v_\text{weight} = 59~\mkms$ for \ion{C}{4} along this sightline, along with $\Delta v_{90}$ widths of 87 $\mkms$ and 131 $\mkms$ for low- and high-ionization material, respectively. As with the J2103+0646 system discussed above, these quiescent kinematics suggest that although the metallicity of this system implies that it has been enriched by winds, our sightline is not probing the energetics associated with this enrichment.

\section{Additional Figures and Tables}
\label{chAppend: column}

Table~\ref{table: obs} lists all QSO sightlines analyzed in our study, along with the QSO redshifts, projected separations, the instrument used to observe each sightline, and the redshift and \ion{H}{1} column density of the corresponding foreground DLA. Table~\ref{table: metalcolm} lists metal-line column densities measured for each DLA and CGM absorption system. Table~\ref{table: kinmeasurements} lists kinematic quantities ($\delta v_{\rm weight}$ and $\Delta v_{90}$) measured for low-ionization and \ion{C}{4} 1548 absorption profiles in our DLA and CGM systems.  Table~\ref{table: Zmeasurements} lists ionic ratio measurements described in Section~\ref{secsub:metallicitiesmethods} along with our constraints on $N_{\rm H\textsc{i}}^{\rm CGM}$.  

\begin{figure*}[ht]
	 \centering 
	 \includegraphics[width=0.8\paperwidth]{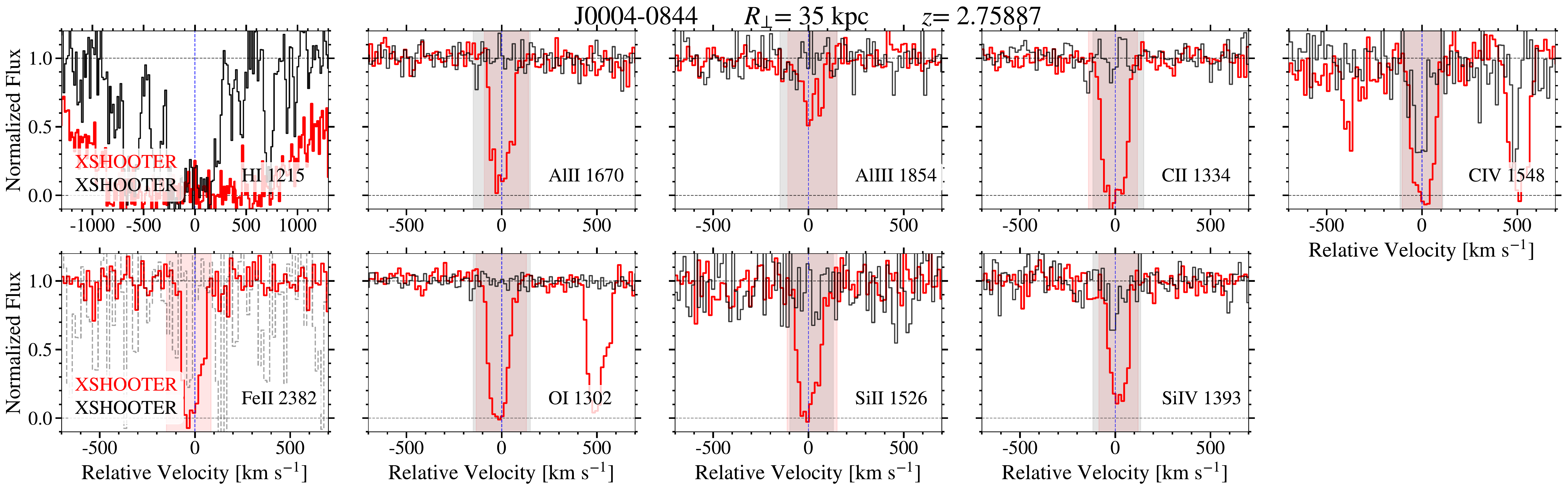} 
	 
	 \caption{\ion{H}{1} and metal-line absorption profiles for all 32 DLA-CGM sightline pairs in our sample. Each set of nine panels shows our coverage of
	 Ly$\alpha$, \ion{Al}{2}, \ion{Al}{3}, \ion{C}{2}, \ion{C}{4}, \ion{Fe}{2}, \ion{O}{1}, \ion{Si}{2} and \ion{Si}{4}  transitions associated with an individual DLA (red histogram) and the corresponding CGM system (black histogram).  
	 The QSO pair ID is shown at the top of each set of nine panels, along with the projected distance between the sightlines at the DLA redshift. The blue dotted line indicates the systemic velocity of the DLA. The shaded areas indicate the placement of the velocity windows used to measure metal-line absorption strength. In the case of profiles with multiple components, the shaded regions are marked `2' or `3' to indicate a second or third component. The instrument used for each sightline is labeled in the bottom left corner of each \ion{H}{1} panel. Transitions which are not used in this analysis due to extreme blending are shown with dotted histograms. The complete figure set (32 images) is available in the online journal. \label{fig:velstacks_appendix} }
\end{figure*}

\begin{figure*}[ht]
	\centering
	\begin{minipage}{\textwidth}
		\includegraphics[width=0.45\linewidth]{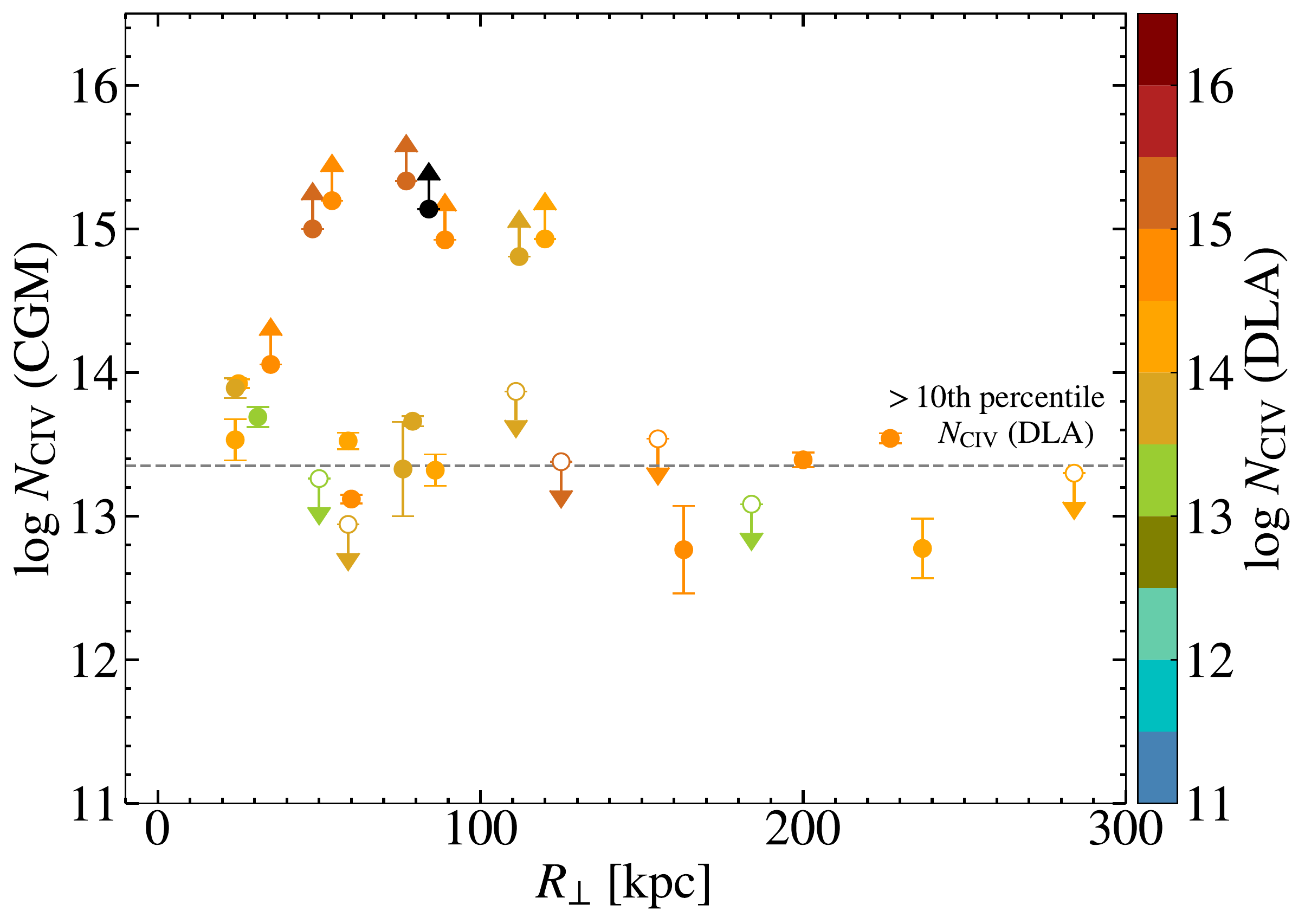}
		\includegraphics[width=0.45\linewidth]{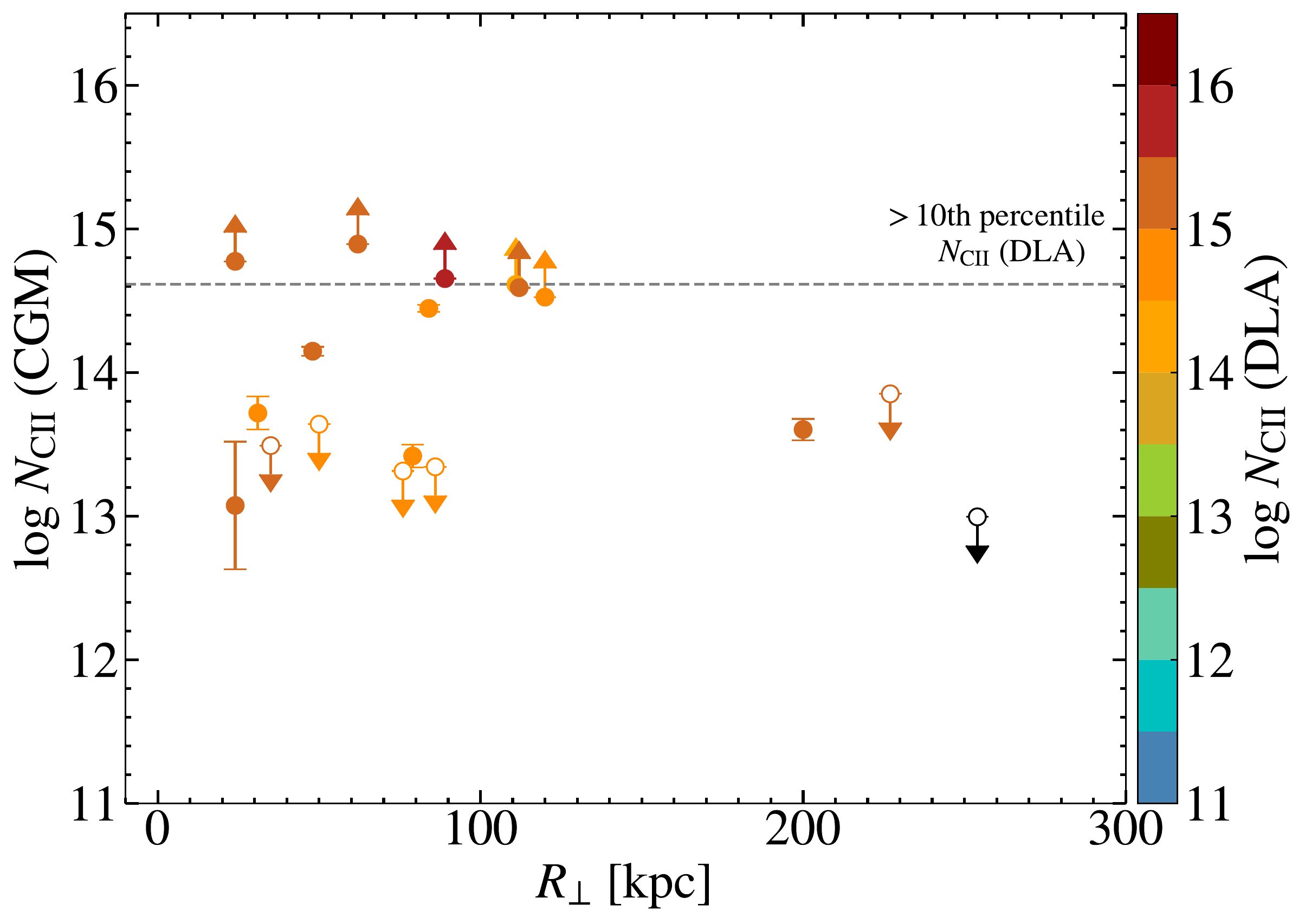}
		\includegraphics[width=0.45\linewidth]{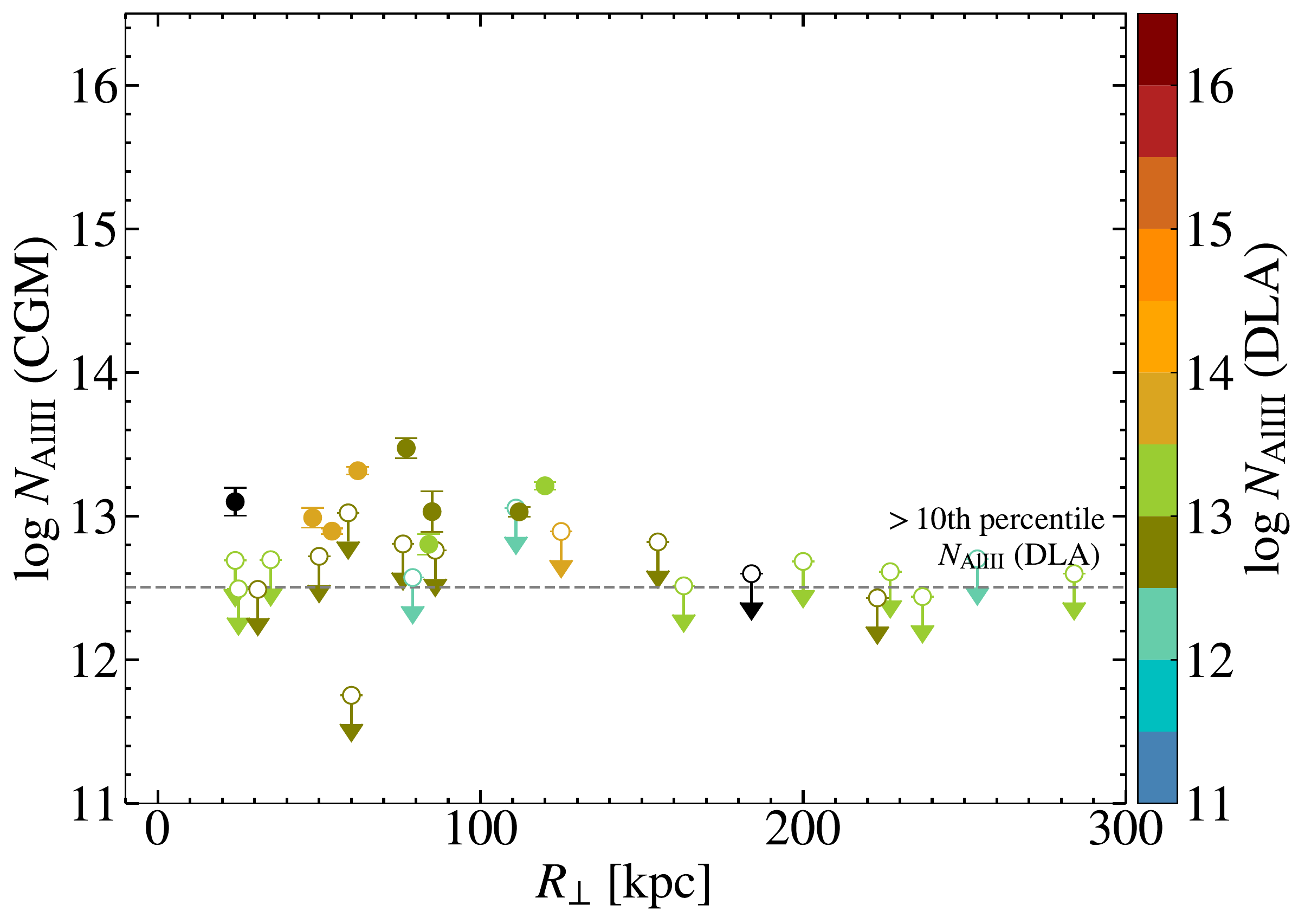}
		\includegraphics[width=0.45\linewidth]{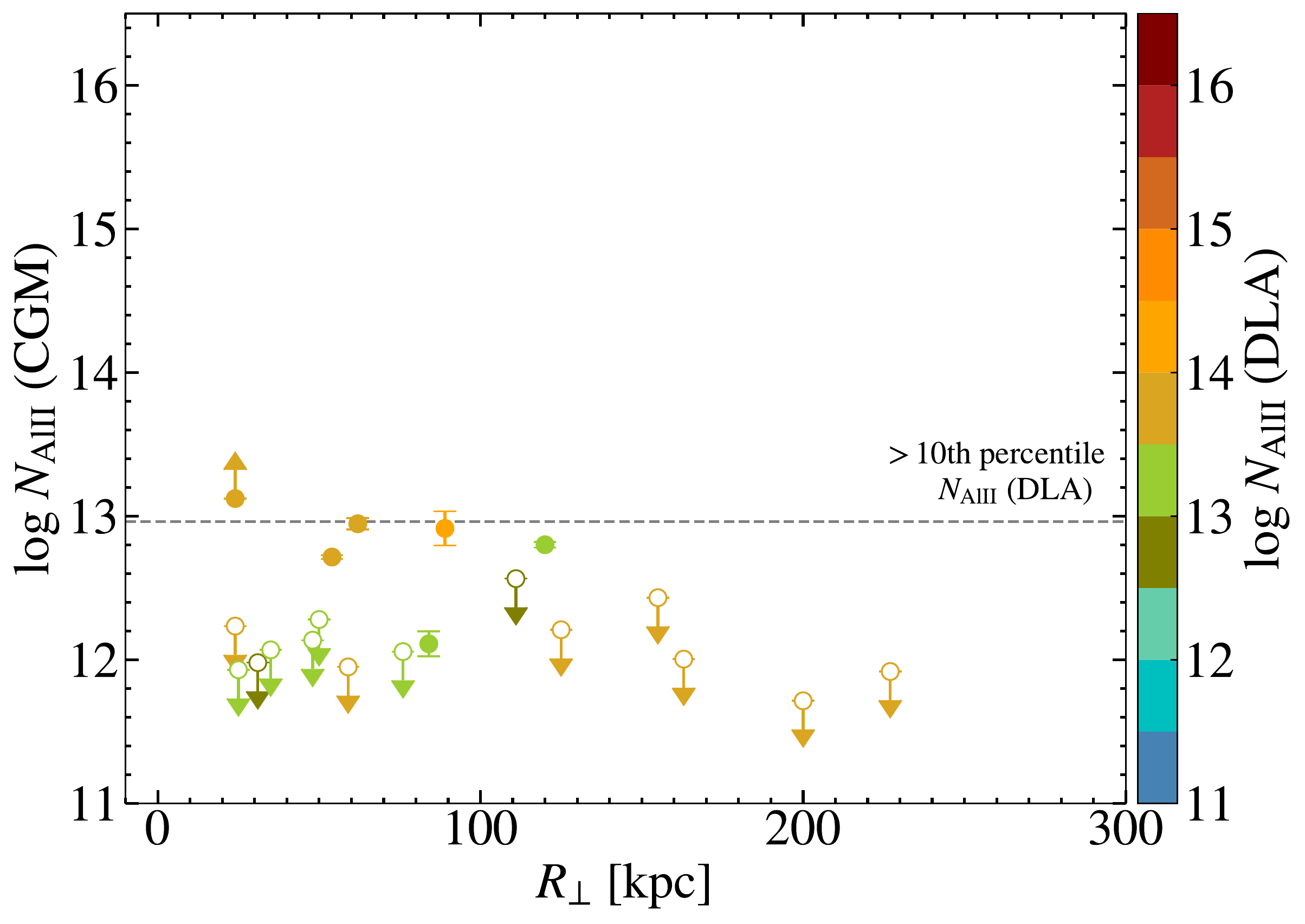}
		\includegraphics[width=0.45\linewidth]{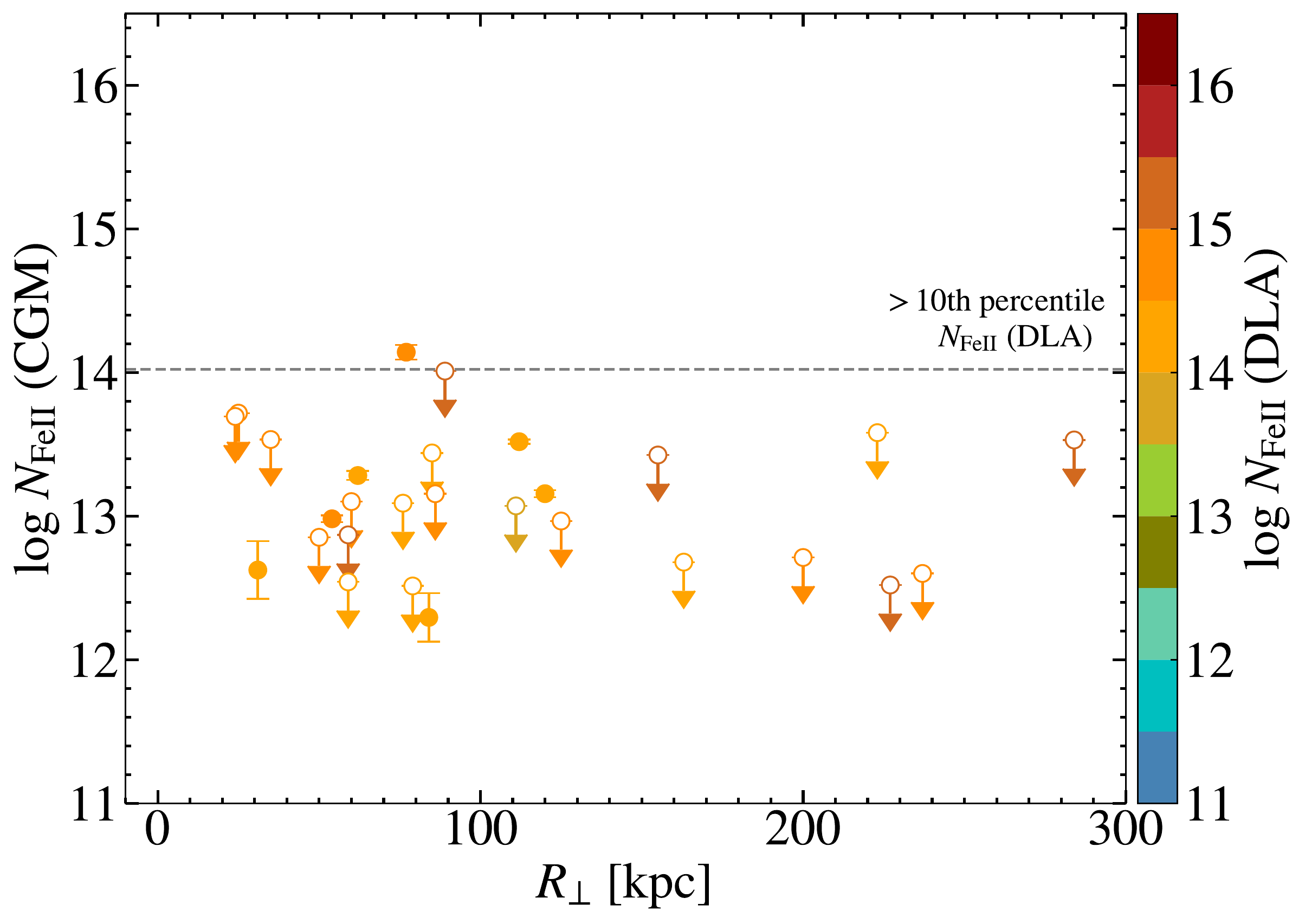} \hspace{1.7 cm}
		\includegraphics[width=0.45\linewidth]{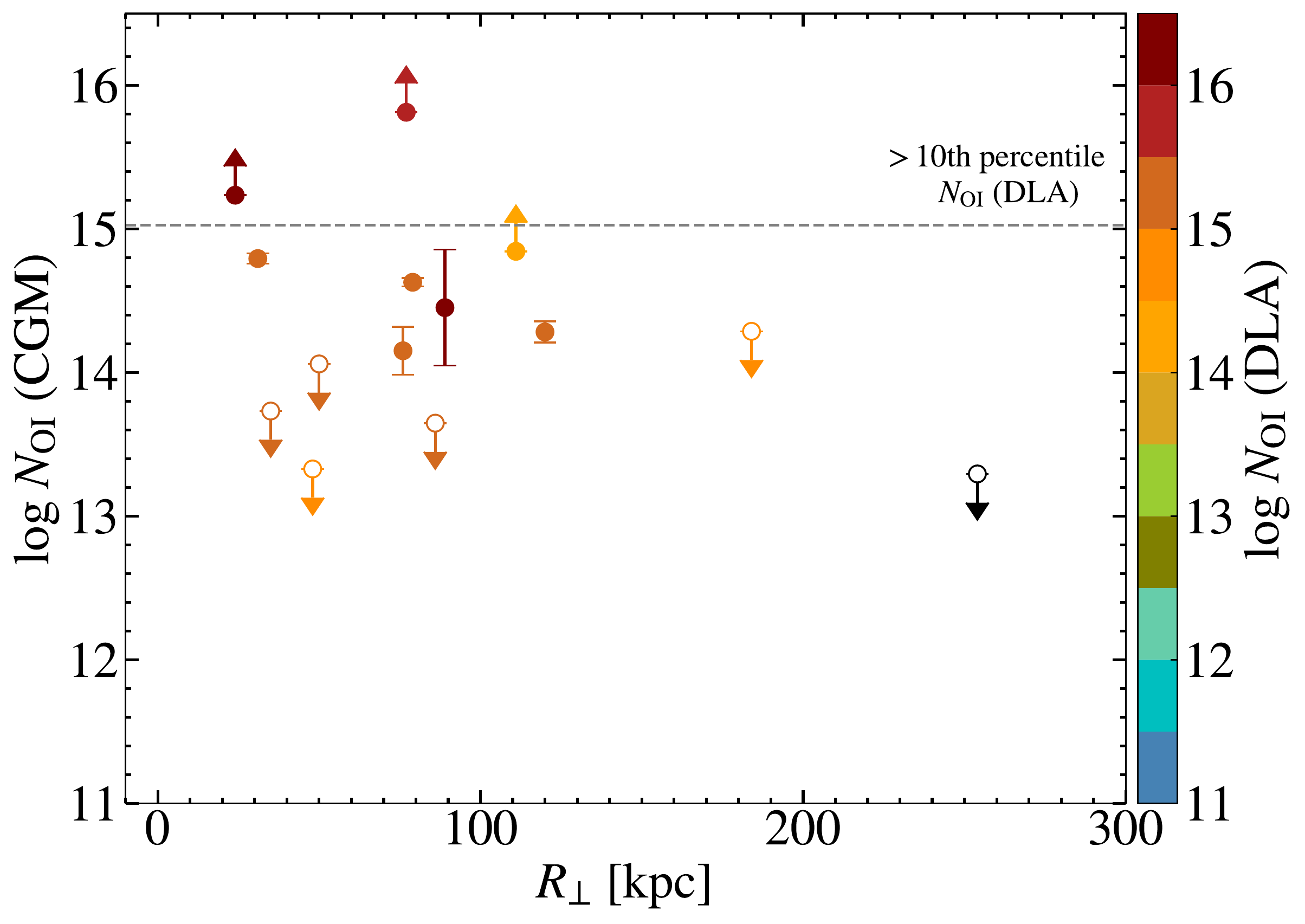}
		\caption{Column densities of \ion{C}{4} (top left panel), \ion{C}{2} (top right panel), \ion{Al}{3} (middle left panel), \ion{Al}{2} (middle right panel), \ion{Fe}{2} (bottom left panel), and \ion{O}{1} (bottom right panel) in our CGM sightlines vs.\ proper distance ($R_\bot$). Colors represent the corresponding DLA column density for that ion. Black points indicate ambiguous DLA column density values. Open symbols indicate that our constraint on the CGM sightline column is an upper limit. The horizontal dashed lines represent the threshold above which 90$\%$ of metal line column densities for the DLA sightlines fall. \label{fig:Rpropall}}
	\end{minipage}
\end{figure*}

We include a set of figures showing our spectroscopic coverage of Ly$\alpha$ and several metal-line transitions for all DLA and CGM sightlines.  An example is shown in Figure~\ref{fig:velstacks_appendix}. Figure \ref{fig:Rpropall} shows the column densities of several metal ions measured along our CGM sightlines vs.\ proper distance ($R_\bot$).

\begin{deluxetable*}{llccccccc}
	\tabletypesize{\footnotesize}
	\tablecolumns{9}
	\tablewidth{0pt}
	\tablecaption{QSO Observations and DLA Sample \label{table: obs}}
    \tablehead{
	    \colhead{Background QSO\tablenotemark{a}} & \colhead{Foreground QSO\tablenotemark{a}} & \colhead{$z_\text{bg}^\text{QSO}$} & \colhead{$z_\text{fg}^\text{QSO}$} & \colhead{$R_\bot$} & \colhead{Instrument\tablenotemark{b}} &  \colhead{Resolution} & \colhead{$z_\text{DLA}$} & \colhead{$N_\text{HI}^\text{DLA}$}  \\ 
	     \colhead{} & \colhead{}  & \colhead{} & \colhead{} & \colhead{kpc} & \colhead{bg/fg} &  \colhead{bg/fg} & \colhead{} & \colhead{$\log(\text{cm}^{-2}$)}   }
	\startdata
    J$000450.91-084452.0$ & J$000450.66-084449.6^*$ & 3.000 & 3.000 & 35 & XSHOOTER/XSHOOTER & 8000/8000 & 2.759 & 20.6 $\pm$ 0.2 \\ 
J$023317.54-054230.0$ & J$023316.29-054210.8^*$ & 2.959 & 2.629 & 223 & MagE/MagE & 5857/5857 & 2.375 & 20.45 $\pm$ 0.2 \\ 
J$025049.09-025631.7$ & J$025048.86-025640.7^*$ & 2.844 & 2.820 & 79 & ESI/ESI & 4545/4545 & 2.571 & 20.3 $\pm$ 0.2 \\ 
J$025836.62-044438.5$ & J$025837.57-044426.0^*$ & 2.540 & 2.428 & 163 & MagE/MagE & 5857/5857 & 1.879 & 20.3 $\pm$ 0.2 \\ 
J$083118.50+424728.8^*$ & J$083121.58+424722.3$ & 3.327 & 3.011 & 284 & ESI/ESI & 4545/4545 & 2.559 & 21.15 $\pm$ 0.2 \\ 
J$093225.60+092500.2$ & J$093226.34+092526.1^*$ & 2.602 & 2.410 & 237 & MagE/XSHOOTER & 5857/8000 & 2.252 & 20.45 $\pm$ 0.2 \\ 
J$093959.41+184757.1$ & J$093959.02+184801.7^*$ & 2.821 & 2.727 & 59 & ESI/ESI & 4545/4545 & 2.430 & 20.5 $\pm$ 0.2 \\ 
J$095543.67-012351.5$ & J$095544.29-012357.5^*$ & 2.844 & 2.833 & 89 & XSHOOTER/XSHOOTER & 8000/8000 & 2.727 & 20.65 $\pm$ 0.2 \\ 
J$095723.43+622322.9$ & J$095722.78+622335.2^*$ & 2.257 & 2.251 & 111 & ESI/ESI & 4545/4545 & 2.143 & 20.55 $\pm$ 0.2 \\ 
... & ... & ... & ... & ... &  .../BOSS &  .../2100 & ... & ... \\ 
J$102633.21+062909.5^*$ & J$102633.55+062901.5$ & 3.120 & 2.890 & 77 & MagE/MagE & 5857/5857 & 2.564 & 21.05 $\pm$ 0.2 \\ 
J$102633.21+062909.5$ & J$102633.55+062901.5^*$ & 3.120 & 2.890 & 76 & MagE/MagE & 5857/5857 & 2.782 & 21.2 $\pm$ 0.2 \\ 
J$105644.88-005933.4$ & J$105645.25-005938.1^*$ & 2.132 & 2.128 & 62 & MagE/MagE & 5857/5857 & 1.967 & 20.6 $\pm$ 0.2 \\ 
J$111610.69+411814.4$ & J$111611.74+411821.5^*$ & 3.000 & 3.000 & 112 & ESI/ESI & 4545/4545 & 2.662 & 20.35 $\pm$ 0.2 \\ 
J$114436.65+095904.9^*$ & J$114435.54+095921.7$ & 3.146 & 2.974 & 200 & MIKE-Blue/MagE & 28000/5857 & 2.093 & 20.35 $\pm$ 0.2 \\ 
J$114958.49+430048.4^*$ & J$114958.26+430041.3$ & 3.273 & 3.247 & 60 & ESI/ESI & 4545/4545 & 2.777 & 21.45 $\pm$ 0.2 \\ 
J$115031.14+045353.2^*$ & J$115031.54+045356.8$ & 2.521 & 2.517 & 59 & MagE/MagE & 5857/5857 & 2.000 & 21.3 $\pm$ 0.2 \\ 
J$123635.42+522057.3$ & J$123635.14+522059.0^*$ & 2.578 & 2.571 & 25 & ESI/ESI & 4545/4545 & 2.397 & 21.0 $\pm$ 0.2 \\ 
J$124025.15+432916.5$ & J$124024.93+432914.5^*$ & 3.264 & 3.249 & 24 & ESI/ESI & 4545/4545 & 2.979 & 21.3 $\pm$ 0.2 \\ 
J$124025.15+432916.5$ & J$124024.93+432914.5^*$ & 3.264 & 3.249 & 24 & ESI/ESI & 4545/4545 & 3.097 & 20.7 $\pm$ 0.2 \\ 
J$142758.74-012136.2$ & J$142758.89-012130.4^*$ & 2.352 & 2.279 & 54 & MIKE-Blue/MagE & 28000/4824 & 1.576 & 21.1 $\pm$ 0.2 \\ 
... & ... & ... & ... & ... & MagE/...  & 4824/... & ... & ... \\ 
J$142816.51+023229.2$ & J$142815.67+023243.5^*$ & 3.030 & 3.010 & 155 & XSHOOTER/XSHOOTER & 8000/8000 & 2.626 & 21.3 $\pm$ 0.2 \\ 
J$152928.37+231415.8^*$ & J$152929.03+231420.0$ & 2.637 & 2.492 & 85 & ESI/ESI & 4545/4545 & 2.077 & 20.35 $\pm$ 0.2 \\ 
... & ... & ... & ... & ... & BOSS/...  & 2100/... & ... & ... \\ 
J$154110.40+270231.2^*$ & J$154110.37+270224.8$ & 3.626 & 3.621 & 48 & ESI/ESI & 4545/4545 & 3.330 & 20.3 $\pm$ 0.2 \\ 
J$154225.81+173323.0$ & J$154226.90+173300.5^*$ & 3.261 & 2.782 & 227 & ESI/ESI & 4545/4545 & 2.423 & 21.4 $\pm$ 0.2 \\ 
J$161302.03+080814.3$ & J$161301.69+080806.0^*$ & 2.386 & 2.386 & 84 & MagE/MagE & 4824/4824 & 1.617 & 20.5 $\pm$ 0.2 \\ 
J$162737.25+460609.3^*$ & J$162738.63+460538.4$ & 4.110 & 3.813 & 254 & ESI/ESI & 4545/4545 & 3.550 & 20.3 $\pm$ 0.2 \\ 
J$163056.73+115250.3^*$ & J$163055.96+115229.4$ & 3.279 & 3.257 & 184 & MagE/MagE & 5857/5857 & 3.182 & 20.3 $\pm$ 0.2 \\ 
J$171946.66+254941.1$ & J$171945.87+254951.2^*$ & 2.172 & 2.170 & 125 & ESI/ESI & 4545/4545 & 2.019 & 20.75 $\pm$ 0.2 \\ 
 ... & ... & ... & ... & ... & GMOSN/GMOSN & 1872/1872 & ... & ... \\ 
J$172524.24+303801.0^*$ & J$172524.66+303803.9$ & 2.647 & 2.634 & 50 & ESI/ESI & 4545/4545 & 2.508 & 20.35 $\pm$ 0.2 \\ 
J$210329.25+064653.3$ & J$210329.37+064650.0^*$ & 2.572 & 2.551 & 31 & MagE/MagE & 5857/5857 & 2.139 & 20.7 $\pm$ 0.2 \\ 
J$214620.98-075303.8^*$ & J$214620.68-075250.6$ & 2.577 & 2.112 & 120 & MagE/MagE & 5857/5857 & 1.853 & 20.45 $\pm$ 0.2 \\ 
J$230044.52+015552.1^*$ & J$230044.36+015541.7$ & 2.951 & 2.910 & 86 & MagE/MagE & 5857/5857 & 2.730 & 20.8 $\pm$ 0.2 \\ 

	\enddata
	\tablenotetext{a}{The QSO marked with a ``$*$'' is the DLA sightline.}
	\tablenotetext{b}{Objects followed by ``...'' indicate  cases in which a low-resolution spectrum is used.}
\end{deluxetable*}

\clearpage 

\startlongtable
\begin{deluxetable*}{ccccccccccc}
	\tabletypesize{\footnotesize}
	\tablecolumns{11}
	\tablewidth{0pt}
	\tablecaption{Metal Line Column Density Measurements \label{table: metalcolm}}
    \tablehead{
	   \colhead{} & \colhead{} & \colhead{\hspace{15mm}} & \multicolumn{2}{c}{$\log N_\text{X}$ (cm$^{-2}$)}  & \colhead{} & \colhead{} &
	   \colhead{} & \colhead{}  & \colhead{} & \colhead{} \\
	   \colhead{ QSO Pair\tablenotemark{a} } & \colhead{ $z_\text{DLA}$}  & \colhead{\hspace{15mm}} & \colhead{AlII} & \colhead{AlIII}  & \colhead{CII} & \colhead{CIV} &
	   \colhead{FeII} & \colhead{SiII}  & \colhead{SiIV} & \colhead{OI}  }  
	\startdata
    J$0004-0844$ & 2.759 &  &  $>$13.47 & 13.27$\pm$0.03 &  $>$15.05 &  $>$14.79 &  $>$14.92 & 15.11$\pm$0.13 & 14.00$\pm$0.05 &  $>$15.23\\
 &   &  & $<$12.07 & $<$12.70 & $<$13.49 & $>$14.06 & $<$13.53 & $<$12.75 & 13.08$\pm$0.21 & $<$13.73\\
J$0233-0542$ & 2.375 &  & 12.89$\pm$0.03 & 12.90$\pm$0.13 &  $>$14.38 & 13.94$\pm$0.04 & 14.26$\pm$0.06 &  $>$14.46 & 13.22$\pm$0.09 & ...  \\
 &   &  & ...   & $<$12.43 & ...   & ...   & $<$13.58 & $<$13.18 & ...   & ...  \\
J$0250-0256$ & 2.571 &  & ...   &  $<$12.05 &  $>$14.75 & 13.92$\pm$0.01 & 14.02$\pm$0.01 &  $>$15.00 & 13.28$\pm$0.02 &  $>$15.10\\
 &   &  & ...   & $<$12.57 & 13.42$\pm$0.08 & 13.66$\pm$0.03 & $<$12.51 & $<$13.19 & $<$12.91 & 14.63$\pm$0.03\\
J$0258-0444$ & 1.879 &  &  $>$13.61 & 13.30$\pm$0.05 &  $>$15.04 &  $>$14.69 & 14.20$\pm$0.03 & 14.61$\pm$0.49 &  $>$14.30 &  $>$15.40\\
 &   &  & $<$12.00 & $<$12.52 & ...   & 12.77$\pm$0.30 & $<$12.68 & $<$14.85 & 13.09$\pm$0.16 & ...  \\
J$0831+4247$ & 2.559 &  &  $>$13.77 & 13.34$\pm$0.03 &  $>$15.16 & 14.28$\pm$0.01 &  $>$15.05 & 15.80$\pm$0.02 &  $>$14.74 &  $>$15.53\\
 &   &  & ...   & $<$12.60 & ...   & $<$13.30 & $<$13.53 & $<$12.89 & $<$12.77 & ...  \\
J$0932+0925$ & 2.252 &  &  $>$13.36 & 13.26$\pm$0.05 &  $>$15.02 & 14.25$\pm$0.01 &  $>$14.66 &  $>$15.00 & 13.69$\pm$0.02 &  $>$15.29\\
 &   &  & ...   & $<$12.44 & ...   & 12.78$\pm$0.21 & $<$12.60 & $<$14.75 & ...   & ...  \\
J$0939+1848$ & 2.430 &  &  $>$13.09 & 12.70$\pm$0.10 &  $>$14.67 & 13.79$\pm$0.02 & 14.27$\pm$0.01 &  $>$14.70 & 13.33$\pm$0.02 & ...  \\
 &   &  & ...   & $<$13.02 & ...   & $<$12.94 & $<$12.54 & $<$14.48 & $<$12.57 & ...  \\
J$0955-0123$ & 2.727 &  &  $>$14.03 & 13.76$\pm$0.04 &  $>$15.56 &  $>$14.77 &  $>$15.27 & 15.82$\pm$0.04 &  $>$14.35 &  $>$16.45\\
 &   &  & 12.91$\pm$0.12 & ...   & $>$14.65 & $>$14.92 & $<$14.01 & $>$14.44 & $>$14.07 & 14.45$\pm$0.40\\
J$0957+6223$ & 2.143 &  & 12.80$\pm$0.02 & 12.35$\pm$0.09 &  $>$14.35 & 13.84$\pm$0.02 & 13.85$\pm$0.01 & 14.78$\pm$0.09 & 13.42$\pm$0.02 &  $>$14.49\\
 &   &  & $<$12.57 & $<$13.06 & $>$14.61 & $<$13.87 & $<$13.07 & $>$14.56 & $>$14.42 & $>$14.84\\
J$1026+0629$ & 2.564 &  &  $>$13.59 & 12.74$\pm$0.22 & ...   &  $>$15.09 &  $>$14.99 & 15.08$\pm$0.20 & 14.29$\pm$0.02 &  $>$15.87\\
 &   &  & ...   & 13.47$\pm$0.07 & ...   & $>$15.33 & 14.14$\pm$0.05 & $>$14.74 & $>$14.62 & $>$15.81\\
J$1026+0629$ & 2.782 &  &  $>$13.47 &  $<$12.96 &  $>$14.93 & 13.71$\pm$0.15 & 14.45$\pm$0.03 &  $>$14.72 & 13.57$\pm$0.08 &  $>$15.39\\
 &   &  & $<$12.06 & $<$12.81 & $<$13.31 & 13.33$\pm$0.33 & $<$13.09 & ...   & $<$13.19 & 14.15$\pm$0.17\\
J$1056-0059$ & 1.967 &  &  $>$13.59 & 13.59$\pm$0.02 &  $>$15.13 &  $>$15.05 & 14.48$\pm$0.02 & 15.52$\pm$0.11 &  $>$14.52 &  $>$15.35\\
 &   &  & 12.95$\pm$0.04 & 13.32$\pm$0.03 & $>$14.89 & ...   & 13.28$\pm$0.03 & 13.67$\pm$0.16 & $>$14.52 & ...  \\
J$1116+4118$ & 2.662 &  & ...   & 12.92$\pm$0.02 &  $>$15.18 & 13.73$\pm$0.01 & 14.40$\pm$0.01 & 15.08$\pm$0.04 & 13.24$\pm$0.06 &  $>$15.69\\
 &   &  & ...   & 13.03$\pm$0.03 & $>$14.59 & $>$14.81 & 13.52$\pm$0.02 & 14.10$\pm$0.03 & 13.77$\pm$0.03 & ...  \\
J$1144+0959$ & 2.093 &  &  $>$14.08 & 13.61$\pm$0.00 &  $>$15.30 &  $>$14.97 &  $>$14.99 & 15.39$\pm$0.01 &  $>$14.44 & ...  \\
 &   &  & $<$11.71 & $<$12.68 & 13.60$\pm$0.07 & 13.39$\pm$0.05 & $<$12.71 & $<$14.65 & 12.72$\pm$0.17 & ...  \\
J$1149+4300$ & 2.777 &  &  $>$13.33 & 12.68$\pm$0.07 &  $>$15.00 &  $>$14.73 &  $>$14.64 & 15.27$\pm$0.07 &  $>$14.26 &  $>$15.34\\
 &   &  & ...   & $<$11.75 & ...   & 13.12$\pm$0.03 & $<$13.10 & $<$14.24 & 13.15$\pm$0.01 & ...  \\
J$1150+0453$ & 2.000 &  &  $>$13.60 & 13.38$\pm$0.04 & ...   & 14.31$\pm$0.05 &  $>$15.16 & 15.81$\pm$0.05 & 14.06$\pm$0.07 &  $>$15.45\\
 &   &  & $<$11.95 & ...   & ...   & 13.52$\pm$0.06 & $<$12.87 & $<$13.25 & 12.82$\pm$0.11 & ...  \\
J$1236+5220$ & 2.397 &  &  $>$13.48 & 13.06$\pm$0.05 &  $>$15.02 & 14.32$\pm$0.03 &  $>$14.92 & 15.42$\pm$0.06 & 13.88$\pm$0.04 &  $>$15.49\\
 &   &  & $<$11.93 & $<$12.49 & ...   & 13.92$\pm$0.03 & $<$13.72 & $<$13.10 & 13.15$\pm$0.08 & ...  \\
J$1240+4329$ & 2.979 &  &  $>$13.74 & 13.08$\pm$0.04 &  $>$15.18 & 14.12$\pm$0.02 &  $>$15.30 &  $>$15.90 &  $>$14.35 &  $>$15.47\\
 &   &  & $<$12.23 & $<$12.69 & 13.07$\pm$0.45 & 13.53$\pm$0.14 & ...   & 12.94$\pm$0.59 & 13.45$\pm$0.05 & ...  \\
J$1240+4329$ & 3.097 &  &  $>$13.59 & ...   &  $>$15.24 & 14.07$\pm$0.02 &  $>$14.85 &  $>$15.51 & 13.50$\pm$0.03 &  $>$16.20\\
 &   &  & $>$13.12 & 13.10$\pm$0.10 & $>$14.77 & 13.89$\pm$0.07 & $<$13.69 & 14.44$\pm$0.04 & 13.40$\pm$0.08 & $>$15.24\\
J$1427-0121$ & 1.576 &  &  $>$13.67 & 13.62$\pm$0.02 &  $>$15.09 &  $>$14.74 &  $>$14.85 & 15.42$\pm$0.04 &  $>$14.36 &  $>$15.52\\
 &   &  & 12.72$\pm$0.01 & 12.90$\pm$0.02 & ...   & $>$15.20 & 12.98$\pm$0.02 & 14.06$\pm$0.02 & $>$14.45 & ...  \\
J$1428+0232$ & 2.626 &  &  $>$13.63 & 12.72$\pm$0.35 &  $>$15.14 &  $>$14.52 &  $>$15.04 & 15.66$\pm$0.12 &  $>$14.00 & ...  \\
 &   &  & $<$12.43 & $<$12.82 & ...   & $<$13.54 & $<$13.43 & $<$13.04 & $<$13.00 & ...  \\
J$1529+2314$ & 2.077 &  &  $>$13.32 &  $<$12.75 &  $>$15.05 & 13.35$\pm$0.11 & 14.39$\pm$0.01 &  $>$14.82 & ...   &  $>$15.38\\
 &   &  & ...   & 13.03$\pm$0.14 & ...   & ...   & $<$13.44 & $<$15.31 & $>$13.61 & ...  \\
J$1541+2702$ & 3.330 &  &  $>$13.24 & 13.69$\pm$0.06 &  $>$15.11 &  $>$15.26 & 14.01$\pm$0.08 & 14.71$\pm$0.32 &  $>$14.68 &  $>$14.80\\
 &   &  & $<$12.14 & 12.99$\pm$0.07 & 14.15$\pm$0.03 & $>$15.00 & ...   & $<$13.27 & $>$14.30 & $<$13.33\\
J$1542+1733$ & 2.423 &  &  $>$13.72 & 13.24$\pm$0.03 &  $>$15.18 &  $>$14.57 &  $>$15.09 & 15.74$\pm$0.02 & 14.01$\pm$0.02 &  $>$15.54\\
 &   &  & $<$11.92 & $<$12.61 & $<$13.85 & 13.54$\pm$0.03 & $<$12.52 & $<$12.89 & 13.53$\pm$0.07 & ...  \\
J$1613+0808$ & 1.617 &  &  $>$13.15 & 13.07$\pm$0.05 &  $>$14.73 & ...   & 14.44$\pm$0.02 & 15.52$\pm$0.04 & 13.62$\pm$0.04 &  $>$15.06\\
 &   &  & 12.11$\pm$0.09 & 12.80$\pm$0.07 & 14.45$\pm$0.02 & $>$15.14 & 12.30$\pm$0.17 & 13.83$\pm$0.10 & 14.22$\pm$0.02 & ...  \\
J$1627+4606$ & 3.550 &  & ...   &  $<$12.23 & ...   & 13.34$\pm$0.04 &  $<$13.55 & 13.09$\pm$0.11 & 12.93$\pm$0.09 & ...  \\
 &   &  & ...   & $<$12.70 & $<$13.00 & ...   & ...   & $<$12.98 & $<$12.44 & $<$13.29\\
J$1630+1152$ & 3.182 &  &  $<$12.21 & ...   &  $>$13.90 & 13.16$\pm$0.32 & ...   & 13.35$\pm$0.23 &  $<$13.00 & 14.79$\pm$0.07\\
 &   &  & ...   & $<$12.60 & ...   & $<$13.08 & ...   & $<$13.23 & $<$12.77 & $<$14.29\\
J$1719+2549$ & 2.019 &  &  $>$13.71 & 13.51$\pm$0.02 &  $>$15.33 &  $>$15.27 & 14.50$\pm$0.03 & 15.21$\pm$0.07 &  $>$14.45 & ...  \\
 &   &  & $<$12.21 & $<$12.89 & ...   & $<$13.38 & $<$12.97 & $<$13.86 & $<$13.88 & ...  \\
J$1725+3038$ & 2.508 &  &  $>$13.20 & 12.51$\pm$0.07 &  $>$14.70 & 13.46$\pm$0.04 &  $>$14.73 & 14.86$\pm$0.07 & 13.19$\pm$0.02 &  $>$15.03\\
 &   &  & $<$12.28 & $<$12.72 & $<$13.64 & $<$13.26 & $<$12.85 & $<$13.51 & 12.96$\pm$0.22 & $<$14.06\\
J$2103+0646$ & 2.139 &  & 12.96$\pm$0.05 &  $<$12.70 &  $>$14.81 & 13.35$\pm$0.20 & 14.19$\pm$0.06 & 14.34$\pm$0.09 & 13.22$\pm$0.05 &  $>$15.20\\
 &   &  & $<$11.98 & $<$12.49 & 13.72$\pm$0.12 & 13.69$\pm$0.07 & 12.63$\pm$0.20 & ...   & $<$12.38 & 14.79$\pm$0.04\\
J$2146-0753$ & 1.853 &  &  $>$13.16 & 13.32$\pm$0.04 &  $>$14.62 & 14.38$\pm$0.02 & 14.31$\pm$0.02 & 15.22$\pm$0.12 & 13.78$\pm$0.07 &  $>$15.28\\
 &   &  & 12.80$\pm$0.02 & 13.21$\pm$0.03 & $>$14.53 & $>$14.93 & 13.16$\pm$0.02 & 14.30$\pm$0.05 & $>$14.30 & 14.28$\pm$0.07\\
J$2300+0155$ & 2.730 &  &  $>$13.44 &  $<$12.91 &  $>$14.74 & 14.35$\pm$0.04 &  $>$14.66 & 14.73$\pm$0.27 & 13.65$\pm$0.03 &  $>$15.21\\
 &   &  & ...   & $<$12.76 & $<$13.34 & 13.32$\pm$0.11 & $<$13.16 & $<$13.20 & $<$12.64 & $<$13.65\\

	\enddata
	\tablenotetext{a}{The first row of each pair lists the column densities for the DLA sightline. The second row of each pair lists the column densities for the CGM sightline. }
\end{deluxetable*}

\begin{longrotatetable}
\begin{deluxetable*}{ccccccccccc}
	\tabletypesize{\footnotesize}
	\tablecolumns{11}
	\tablewidth{0pt}
	\tablecaption{DLA and CGM Kinematic Measurements \label{table: kinmeasurements}}
    \tablehead{
	     \colhead{} & \colhead{} & \colhead{} \hspace{7mm}  & \colhead{} & 
      \colhead{DLA}
      & \colhead{} &  \colhead{}  \hspace{10 mm}  & \colhead{} & \colhead{CGM}  & \colhead{} & \colhead{} \\
	    \colhead{QSO Pair\tablenotemark{a}} & \colhead{$R_\bot$} & \colhead{$z_\text{DLA}$} \hspace{7mm} & \colhead{$\delta v_\text{weight}^\text{low-ion}$\tablenotemark{b},\tablenotemark{c}} & \colhead{$\Delta v_\text{90}^\text{low-ion}$\tablenotemark{b},\tablenotemark{d}} &  \colhead{$\delta v_\text{weight}^\text{1548 \AA}$\tablenotemark{c}} & \colhead{$\Delta v_\text{90}^{1548 \AA}$\tablenotemark{d}}  \hspace{10 mm} & \colhead{$\delta v_\text{weight}^\text{low-ion}$\tablenotemark{b},\tablenotemark{c}} & \colhead{$\Delta v_\text{90}^\text{low-ion}$\tablenotemark{b},\tablenotemark{d}} &  \colhead{$\delta v_\text{weight}^{1548 \AA}$\tablenotemark{c}} & \colhead{$\Delta v_\text{90}^{1548 \AA}$\tablenotemark{d}} \\ 
	     \colhead{} & \colhead{kpc} & \colhead{} \hspace{7mm} & \colhead{$\mkms$} & \colhead{$\mkms$} & \colhead{$\mkms$} &  \colhead{$\mkms$}  \hspace{10 mm} & \colhead{$\mkms$} & \colhead{$\mkms$} & \colhead{$\mkms$} & \colhead{$\mkms$}  }
	\startdata
    J$0004-0844$ & 35 & 2.759 \hspace{7mm}  & 5.4 & 118 & 6.7 & 88 $\hspace{10mm}$ & ...  & ...  & -9.5 & 88 \\ 
J$0233-0542$ & 223 & 2.375 \hspace{7mm}  & -1.6 & 87 & ...  & ...  $\hspace{10mm}$ & ...  & ...  & ...  & ...  \\ 
J$0250-0256$ & 79 & 2.571 \hspace{7mm}  & 5.4 & 111 & 62.7 & 81 $\hspace{10mm}$ & ...  & ...  & 243.7 & 81 \\ 
 & & $\hspace{7mm}$ & ... & ...  & ... & ... $\hspace{10mm}$  & ...  & ...  & 109.0 & 71 \\ 
J$0258-0444$ & 163 & 1.879 \hspace{7mm}  & 3.7 & 109 & 1.9 & 175 $\hspace{10mm}$ & ...  & ...  & -81.4 & 175 \\ 
J$0831+4247$ & 284 & 2.559 \hspace{7mm}  & 5.1 & 151 & ...  & ...  $\hspace{10mm}$ & ...  & ...  & ...  & ...  \\ 
J$0932+0925$ & 237 & 2.252 \hspace{7mm}  & ...  & ...  & 4.6 & 65 $\hspace{10mm}$ & ...  & ...  & -136.4 & 65 \\ 
J$0939+1848$ & 59 & 2.430 \hspace{7mm}  & 0.1 & 101 & ...  & ...  $\hspace{10mm}$ & ...  & ...  & ...  & ...  \\ 
J$0955-0123$ & 89 & 2.727 \hspace{7mm}  & 24.6 & 163 & -11.2 & 193 $\hspace{10mm}$ & ...  & ...  & -36.8 & 193 \\ 
J$0957+6223$ & 111 & 2.143 \hspace{7mm}  & -8.5 & 101 & ...  & ...  $\hspace{10mm}$ & ...  & ...  & ...  & ...  \\ 
J$1026+0629$ & 77 & 2.564 \hspace{7mm}  & ...  & ...  & 21.9 & 175 $\hspace{10mm}$ & ...  & ...  & 122.7 & 175 \\ 
 & & $\hspace{7mm}$  & ...  & ...  & 355.4 & 197 $\hspace{10mm}$ & ...  & ...  & 386.6 & 197 \\ 
J$1026+0629$ & 76 & 2.782 \hspace{7mm}  & -6.7 & 87 & -18.7 & 87 $\hspace{10mm}$ & ...  & ...  & -76.9 & 87 \\ 
J$1056-0059$ & 62 & 1.967 \hspace{7mm}  & 51.8 & 241 & ...  & ...  $\hspace{10mm}$ & -73.3 & 219 & ...  & ...  \\ 
J$1116+4118$ & 112 & 2.662 \hspace{7mm}  & -19.6 & 131 & 4.8 & 230 $\hspace{10mm}$ & 90.3 & 171 & 68.6 & 230 \\ 
J$1144+0959$ & 200 & 2.093 \hspace{7mm}  & -16.4 & 139 & ...  & ...  $\hspace{10mm}$ & ...  & ...  & ...  & ...  \\ 
J$1149+4300$ & 60 & 2.777 \hspace{7mm}  & ...  & ...  & 14.2 & 111 $\hspace{10mm}$ & ...  & ...  & 36.7 & 111 \\ 
J$1150+0453$ & 59 & 2.000 \hspace{7mm}  & 12.2 & 109 & 47.9 & 131 $\hspace{10mm}$ & -175.1 & 109 & -152.6 & 131 \\ 
J$1236+5220$ & 25 & 2.397 \hspace{7mm}  & ...  & ...  & -7.5 & 111 $\hspace{10mm}$ & ...  & ...  & 1.4 & 111 \\ 
J$1240+4329$ & 24 & 2.979 \hspace{7mm}  & ...  & ...  & -133.4 & 71 $\hspace{10mm}$ & ...  & ...  & -41.3 & 71 \\ 
 & & $\hspace{7mm}$ & ... & ...  & ... & ... $\hspace{10mm}$  & ...  & ...  & 143.3 & 81 \\ 
J$1240+4329$ & 24 & 3.097 \hspace{7mm}  & ...  & ...  & 72.7 & 171 $\hspace{10mm}$ & ...  & ...  & 75.0 & 171 \\ 
J$1427-0121$ & 54 & 1.576 \hspace{7mm}  & 12.9 & 82 & -30.4 & 250 $\hspace{10mm}$ & 181.3 & 103 & 150.2 & 250 \\ 
 & & $\hspace{7mm}$  & -280.5 & 104 & -298.4 & 88 $\hspace{10mm}$ & 972.1 & 58 & 946.8 & 88 \\ 
J$1428+0232$ & 155 & 2.626 \hspace{7mm}  & ...  & ...  & ...  & ...  $\hspace{10mm}$ & ...  & ...  & ...  & ...  \\ 
J$1529+2314$ & 85 & 2.077 \hspace{7mm}  & 21.7 & 211 & ...  & ...  $\hspace{10mm}$ & ...  & ...  & ...  & ...  \\ 
J$1541+2702$ & 48 & 3.330 \hspace{7mm}  & ...  & ...  & 0.7 & 250 $\hspace{10mm}$ & ...  & ...  & 131.2 & 250 \\ 
J$1542+1733$ & 227 & 2.423 \hspace{7mm}  & -8.4 & 101 & -18.9 & 71 $\hspace{10mm}$ & ...  & ...  & -226.0 & 71 \\ 
J$1613+0808$ & 84 & 1.617 \hspace{7mm}  & -0.4 & 82 & ...  & ...  $\hspace{10mm}$ & 70.8 & 82 & 78.5 & 148 \\ 
 & & $\hspace{7mm}$ & ... & ...  & ... & ... $\hspace{10mm}$  & 275.4 & 104 & 233.1 & 148 \\ 
J$1627+4606$ & 254 & 3.550 \hspace{7mm}  & ...  & ...  & ...  & ...  $\hspace{10mm}$ & ...  & ...  & ...  & ...  \\ 
J$1630+1152$ & 184 & 3.182 \hspace{7mm}  & ...  & ...  & ...  & ...  $\hspace{10mm}$ & ...  & ...  & ...  & ...  \\ 
J$1719+2549$ & 125 & 2.019 \hspace{7mm}  & 10.3 & 101 & ...  & ...  $\hspace{10mm}$ & ...  & ...  & ...  & ...  \\ 
J$1725+3038$ & 50 & 2.508 \hspace{7mm}  & ...  & ...  & ...  & ...  $\hspace{10mm}$ & ...  & ...  & ...  & ...  \\ 
J$2103+0646$ & 31 & 2.139 \hspace{7mm}  & -9.4 & 109 & -60.2 & 65 $\hspace{10mm}$ & ...  & ...  & 266.4 & 65 \\ 
 & & $\hspace{7mm}$  & ...  & ...  & 156.2 & 109 $\hspace{10mm}$ & ...  & ...  & 62.3 & 109 \\ 
J$2146-0753$ & 120 & 1.853 \hspace{7mm}  & 6.1 & 131 & 58.6 & 153 $\hspace{10mm}$ & 90.4 & 87 & 59.3 & 153 \\ 
J$2300+0155$ & 86 & 2.730 \hspace{7mm}  & ...  & ...  & -98.9 & 197 $\hspace{10mm}$ & ...  & ...  & 126.6 & 197 \\ 

	\enddata
	\tablenotetext{a}{The second row below a given QSO pair describes measurements of a secondary velocity component.}
	\tablenotetext{b}{The low-ion transition used for kinematic measurements is chosen to have the highest S/N at the peak of its optical depth profile without being saturated.}
	\tablenotetext{c}{We adopt a 1$\sigma$ uncertainty for $\delta v_\text{weight}$ of 15 $\mkms$.}
	\tablenotetext{d}{We adopt a 1$\sigma$ uncertainty for $\Delta v_\text{90}$ of 35 $\mkms$.}
\end{deluxetable*}
\end{longrotatetable}

\begin{longrotatetable}
\begin{deluxetable*}{cccccccccccc}
	\tabletypesize{\footnotesize}
	\tablecolumns{12}
	\tablewidth{0pt}
	\tablecaption{DLA and CGM Ionic Ratio Measurements \label{table: Zmeasurements}}
    \tablehead{
	    \colhead{QSO Pair} & \colhead{$R_\bot$} & \colhead{$z_\text{DLA}$} & \colhead{$\log \nSiIV/\nSiII^\text{DLA}$} & \colhead{\{Si\textsc{ii}/H\textsc{i}\}$^\text{DLA}$} & \colhead{\{Si\textsc{ii}/Fe\textsc{ii}\}$^\text{DLA}$} &  \colhead{$N_\text{HI}^\text{CGM}$} & \colhead{Method\tablenotemark{a}} & \colhead{$\log \nSiIV/\nSiII^\text{CGM}$} & \colhead{\{Si\textsc{ii}/H\textsc{i}\}$^\text{CGM}$} & \colhead{\{O\textsc{i}/H\textsc{i}\}$^\text{CGM}$} & \colhead{\{O\textsc{i}/Fe\textsc{ii}\}$^\text{CGM}$} \\ 
	     \colhead{} & \colhead{kpc} & \colhead{}  & \colhead{} & \colhead{} & \colhead{} &  \colhead{$\log(\text{cm}^{-2}$)} & \colhead{($N_\text{HI}^\text{CGM}$)} & \colhead{} & \colhead{} & \colhead{} & \colhead{}  }
	\startdata
    J$0004-0844$ & 35 & 2.759 & -1.12 $\pm$ 0.14 & -1.00 $\pm$ 0.24 & $<$ 0.13 & 19.10 $\pm$ 0.20 & 1  & $>$ 0.33 & $<$ -1.86 & $<$ -2.06 & ...  \\ 
J$0233-0542$ & 223 & 2.375 & $<$ -1.23 & $>$ -1.50 & $>$ 0.13 & 13.93 $\pm$ 0.10 & 5  & ...  & $<$ 3.74 & ...  & ...  \\ 
J$0250-0256$ & 79 & 2.571 & $<$ -1.71 & $>$ -0.81 & $>$ 0.91 & 15.64 $\pm$ 0.56 & 3  & ...  & $<$ 2.03 & 2.30 $\pm$ 0.56 & $>$ 0.87 \\ 
J$0258-0444$ & 163 & 1.879 & $>$ -0.30 & -1.20 $\pm$ 0.53 & 0.34 $\pm$ 0.50 & 16.29 $\pm$ 1.71 & 4  & $>$ -1.76 & $<$ 3.05 & ...  & ...  \\ 
J$0831+4247$ & 284 & 2.559 & $>$ -1.06 & -0.86 $\pm$ 0.20 & $<$ 0.69 & 14.33 $\pm$ 0.31 & 5  & ...  & $<$ 3.04 & ...  & ...  \\ 
J$0932+0925$ & 237 & 2.252 & $<$ -1.31 & $>$ -0.96 & ...  & 13.98 $\pm$ 0.13 & 5  & ...  & $<$ 5.26 & ...  & ...  \\ 
J$0939+1848$ & 59 & 2.430 & $<$ -1.37 & $>$ -1.31 & $>$ 0.37 & 16.22 $\pm$ 1.78 & 4  & ...  & $<$ 2.75 & ...  & ...  \\ 
J$0955-0123$ & 89 & 2.727 & $>$ -1.47 & -0.34 $\pm$ 0.20 & $<$ 0.49 & 19.20 $\pm$ 0.20 & 1  & ...  & $>$ -0.27 & -1.44 $\pm$ 0.45 & $>$ -0.80 \\ 
J$0957+6223$ & 111 & 2.143 & -1.36 $\pm$ 0.09 & -1.28 $\pm$ 0.22 & 0.87 $\pm$ 0.09 & ...  & ...  & ...  & ...  & ...  & $>$ 0.53 \\ 
J$1026+0629$ & 77 & 2.564 & -0.79 $\pm$ 0.20 & -1.48 $\pm$ 0.28 & $<$ 0.02 & 20.20 $\pm$ 0.20 & 1  & ...  & $>$ -0.97 & $>$ -1.08 & $>$ 0.43 \\ 
J$1026+0629$ & 76 & 2.782 & $<$ -1.15 & $>$ -1.99 & $>$ 0.22 & 15.51 $\pm$ 0.79 & 3  & ...  & ...  & 1.95 $\pm$ 0.81 & $>$ -0.18 \\ 
J$1056-0059$ & 62 & 1.967 & $>$ -1.00 & -0.59 $\pm$ 0.23 & 0.98 $\pm$ 0.11 & 19.60 $\pm$ 0.20 & 1  & $>$ 0.85 & -1.44 $\pm$ 0.26 & ...  & ...  \\ 
J$1116+4118$ & 112 & 2.662 & -1.84 $\pm$ 0.07 & -0.78 $\pm$ 0.20 & 0.62 $\pm$ 0.04 & 20.10 $\pm$ 0.20 & 1  & -0.33 $\pm$ 0.04 & -1.51 $\pm$ 0.20 & ...  & ...  \\ 
J$1144+0959$ & 200 & 2.093 & $>$ -0.94 & -0.47 $\pm$ 0.20 & $<$ 0.34 & 16.40 $\pm$ 1.60 & 4  & $>$ -1.93 & $<$ 2.75 & ...  & ...  \\ 
J$1149+4300$ & 60 & 2.777 & $>$ -1.01 & -1.69 $\pm$ 0.21 & $<$ 0.57 & 15.27 $\pm$ 0.68 & 3  & $>$ -1.09 & $<$ 3.45 & ...  & ...  \\ 
J$1150+0453$ & 59 & 2.000 & -1.75 $\pm$ 0.09 & -1.00 $\pm$ 0.21 & $<$ 0.59 & 16.37 $\pm$ 1.63 & 4  & $>$ -0.43 & $<$ 1.37 & ...  & ...  \\ 
J$1236+5220$ & 25 & 2.397 & -1.53 $\pm$ 0.07 & -1.09 $\pm$ 0.21 & $<$ 0.44 & 17.20 $\pm$ 0.20 & 2  & $>$ 0.05 & $<$ 0.39 & ...  & ...  \\ 
J$1240+4329$ & 24 & 2.979 & ...  & $>$ -0.91 & ...  & 16.58 $\pm$ 1.42 & 4  & 0.51 $\pm$ 0.60 & 0.85 $\pm$ 1.54 & ...  & ...  \\ 
J$1240+4329$ & 24 & 3.097 & $<$ -2.01 & $>$ -0.70 & ...  & 20.00 $\pm$ 0.20 & 1  & -1.04 $\pm$ 0.09 & -1.07 $\pm$ 0.20 & $>$ -1.45 & $>$ 0.30 \\ 
J$1427-0121$ & 54 & 1.576 & $>$ -1.06 & -1.19 $\pm$ 0.20 & $<$ 0.51 & 19.70 $\pm$ 0.20 & 1  & $>$ 0.39 & -1.15 $\pm$ 0.20 & ...  & ...  \\ 
J$1428+0232$ & 155 & 2.626 & $>$ -1.66 & -1.15 $\pm$ 0.23 & $<$ 0.56 & 15.46 $\pm$ 0.94 & 3  & ...  & $<$ 2.07 & ...  & ...  \\ 
J$1529+2314$ & 85 & 2.077 & ...  & $>$ -1.04 & $>$ 0.37 & ...  & ...  & $>$ -1.70 & ...  & ...  & ...  \\ 
J$1541+2702$ & 48 & 3.330 & $>$ -0.03 & -1.10 $\pm$ 0.38 & 0.63 $\pm$ 0.33 & 16.14 $\pm$ 0.51 & 3  & $>$ 1.03 & $<$ 1.63 & $<$ 0.50 & ...  \\ 
J$1542+1733$ & 227 & 2.423 & -1.73 $\pm$ 0.03 & -1.17 $\pm$ 0.20 & $<$ 0.59 & 14.13 $\pm$ 1.37 & 5  & $>$ 0.64 & $<$ 3.25 & ...  & ...  \\ 
J$1613+0808$ & 84 & 1.617 & -1.90 $\pm$ 0.06 & -0.49 $\pm$ 0.20 & 1.03 $\pm$ 0.05 & 16.70 $\pm$ 0.10 & 2  & 0.40 $\pm$ 0.10 & 1.62 $\pm$ 0.14 & ...  & ...  \\ 
J$1627+4606$ & 254 & 3.550 & -0.17 $\pm$ 0.14 & -2.72 $\pm$ 0.23 & $>$ -0.52 & 14.36 $\pm$ 0.03 & 5  & ...  & $<$ 3.10 & $<$ 2.24 & ...  \\ 
J$1630+1152$ & 184 & 3.182 & $<$ -0.35 & -2.46 $\pm$ 0.31 & ...  & 15.53 $\pm$ 1.02 & 3  & ...  & $<$ 2.19 & $<$ 2.07 & ...  \\ 
J$1719+2549$ & 125 & 2.019 & $>$ -0.77 & -1.05 $\pm$ 0.21 & 0.65 $\pm$ 0.08 & 16.19 $\pm$ 1.81 & 4  & ...  & $<$ 2.17 & ...  & ...  \\ 
J$1725+3038$ & 50 & 2.508 & -1.68 $\pm$ 0.07 & -1.00 $\pm$ 0.21 & $<$ 0.07 & 15.92 $\pm$ 1.08 & 3  & $>$ -0.54 & $<$ 2.07 & $<$ 1.45 & ...  \\ 
J$2103+0646$ & 31 & 2.139 & -1.12 $\pm$ 0.11 & -1.87 $\pm$ 0.22 & 0.09 $\pm$ 0.11 & 18.85 $\pm$ 0.20 & 1  & ...  & ...  & -0.75 $\pm$ 0.20 & 0.93 $\pm$ 0.20 \\ 
J$2146-0753$ & 120 & 1.853 & -1.44 $\pm$ 0.14 & -0.74 $\pm$ 0.23 & 0.85 $\pm$ 0.12 & 18.55 $\pm$ 0.20 & 1  & $>$ 0.01 & 0.24 $\pm$ 0.21 & -0.96 $\pm$ 0.21 & -0.11 $\pm$ 0.08 \\ 
J$2300+0155$ & 86 & 2.730 & -1.07 $\pm$ 0.27 & -1.58 $\pm$ 0.34 & $<$ 0.00 & 14.97 $\pm$ 0.48 & 3  & ...  & $<$ 2.72 & $<$ 1.98 & ...  \\ 

	\enddata
 	\tablenotetext{a}{Method used to constrain $N_\text{HI}$ for the CGM sightline as described in Section \ref{sec:NHI}. Method 1 corresponds to the fitting of Ly$\alpha$ damping wings; method 2 corresponds to model fitting of the Lyman limit covered in {\it{HST}} WFC3/UVIS grism spectroscopy; method 3 corresponds to model fitting of the Lyman limit covered in our optical medium-/high-resolution spectra; method 4 corresponds to bounds established from apparent optical depth measurements for saturated but undamped Ly$\alpha$ absorption; and method 5 corresponds to summed apparent optical depth column densities assuming Ly$\alpha$ is optically thin.}
\end{deluxetable*}
\end{longrotatetable}

\bibliographystyle{aasjournal}
\bibliography{thebib}

\end{document}